\documentclass[prd,twocolumn,showpacs,superscriptaddress,nofootinbib,floatfix,showkeys,10pt]{revtex4-1}

\usepackage{graphicx}
\usepackage{amsmath}
\usepackage{bm}
\usepackage{yhmath}
\usepackage{mathtools}
\usepackage{wasysym}
\usepackage[colorlinks,citecolor=blue,urlcolor=blue,linkcolor=blue]{hyperref}
\usepackage{subfigure}
\usepackage{color}
\usepackage{cases}
\usepackage{subfigure}
\usepackage{times}
\usepackage{dcolumn,booktabs,bm}
\usepackage{slashed}
\usepackage{amsfonts,amssymb,stmaryrd,latexsym,amsmath}
\usepackage{textcomp}
\usepackage{multirow}
\usepackage{cancel}
\usepackage{array}
\usepackage{enumerate,enumitem}
\usepackage{orcidlink}
\usepackage{dashrule}
\usepackage{multirow}
\usepackage{times}
\usepackage{mathtools}
\allowdisplaybreaks[4]
\newcounter{cnt}
\makeatletter
\let\oldhypertarget\hypertarget
\renewcommand{\hypertarget}[2]{%
  \oldhypertarget{#1}{#2}%
    \protected@write\@mainaux{}{%
        \string\expandafter\string\gdef
          \string\csname\string\detokenize{#1}\string\endcsname{#2}%
    }%
  }
\newcommand{\myhyperlink}[1]{%
  \hyperlink{#1}{\csname #1\endcsname}%
  }
\makeatother

\newcommand{\clabel}[2][]{#2}
\newcommand{\change}[1]{#1}
\def\Tcc{{T_{cc}^+}}
\def\xEFT{{$\chi$EFT}}
\def\N2LO{{N$^2$LO}}
\newcommand{\etal}{\textit{et al}.}

\begin{document}

%%%%%%%%%%%%%%%%open the reply mode%%%%%%%%%%%%%%%%%%
%\onecolumngrid
%\input{reply_b.tex}
%\newpage
%\setcounter{page}{0}
%%%%%%%%%%%%%%%%open the reply mode%%%%%%%%%%%%%%%%%%

\title{Revisiting the $DD^\ast$ chiral interactions with the local momentum-space regularization up to the third order and the nature of $\Tcc$}

\author{Bo Wang\,\orcidlink{0000-0003-0985-2958}}%\email{wangbo@hbu.edu.cn}
\affiliation{School of Physical Science and Technology, Hebei University, Baoding 071002, China}
\affiliation{Key Laboratory of High-precision Computation and Application of Quantum Field Theory of Hebei Province, Baoding 071002, China}
\affiliation{Research Center for Computational Physics of Hebei Province, Baoding, 071002, China}

\author{Lu Meng\,\orcidlink{0000-0001-9791-7138}}\email{lu.meng@rub.de}
\affiliation{Ruhr-Universit\"at Bochum, Fakult\"at f\"ur Physik und Astronomie, Institut f\"ur Theoretische Physik II, D-44780 Bochum, Germany}

\begin{abstract}
We revisit the $DD^\ast$ interactions in chiral effective field theory up to the third order for the first time. We deal with the pion-exchanged interactions via local momentum-space regularization, in which we focus on their long-range behaviors through demanding their contributions vanish at the origin in the coordinate space. The short-range contact interactions and subleading pion-charmed meson couplings are estimated with the phenomenological resonance saturation model. The subleading pion-charmed meson couplings are much weaker than those in the pion-nucleon system, thus the $DD^\ast$ binding mechanism is very different with that of the $NN$ system. We also obtain the analytic structure of the two-pion exchange interactions in the coordinate space, and we find that its asymptotic behavior at long distance is similar to but slightly different with the $NN$ interactions. We get the same asymptotic behavior of the two-pion exchange interaction with that from HAL QCD method but appearing in the longer distance rather than $1 <r<2\text{ fm}$. The binding solution only exists in the isoscalar channel. Our calculation supports the molecular interpretation of $\Tcc$.
\end{abstract}

%\pacs{12.39.Fe, 12.39.Hg, 14.40.Nd, 14.40.Rt}
\maketitle
\thispagestyle{empty}
\section{Introduction}\label{Introduction}

The interactions between a pair of heavy-light hadrons can be fairly regarded as the extension of the pattern of nuclear forces. The theoretical tools that designed for the nucleon systems shall also be generalized to the heavy-light systems via including the restriction of extra symmetries, such as the heavy quark symmetry. Meanwhile, in recent years, the observations of many near-threshold exotic states provide golden platforms to test and redevelop these tools~\cite{Chen:2016qju,Guo:2017jvc,Liu:2019zoy,Lebed:2016hpi,Esposito:2016noz,Brambilla:2019esw,Chen:2021ftn,Chen:2022asf,Meng:2022ozq,Albuquerque:2022weq,Albuquerque:2023rrf}, in which the successful generalizations of the effective field theories (EFTs), e.g., the pionless and pionful EFTs, is a epitome of the intimate connection between the nuclear physics and the hadron physics~\cite{Meng:2022ozq}.

Based on the instructive works of Weinberg~\cite{Weinberg:1990rz,Weinberg:1991um}, in the past decades, the modern framework of nuclear forces was constructed upon the chiral effective field theory (\xEFT)~\cite{Epelbaum:2008ga,Machleidt:2011zz}. In \xEFT, the short-range part of the nuclear forces is parameterized as the four-fermion contact interactions through integrating out the heavy particle exchanging (e.g., the vector meson $\rho$ and $\omega$, etc.), while the long- and intermediate-range parts are presented by the one-pion exchange (OPE) and multi-pion exchange interactions, respectively~\cite{Ordonez:1995rz,Kaiser:1997mw,Epelbaum:1999dj}. The latter can be derived from the chiral symmetry of QCD via a model-independent way. The study of nucleon-nucleon ($NN$) interactions indicates that the leading order (LO) two-pion exchange (TPE) potential is very weak and insufficient to provide the appropriate attractive force at the intermediate range, and which is in fact described by the subleading TPE potential with an insertion of the subleading pion-nucleon vertices~\cite{Epelbaum:2005pn,Epelbaum:2008ga,Machleidt:2011zz}. It was found that the large values of the low energy constants (LECs) in the subleading pion-nucleon Lagrangians leads to the attractive source. The values of these LECs can be quantitatively understood using the phenomenological resonance saturation model (RSM)~\cite{Bernard:1996gq}. It was shown that these large value LECs in the \xEFT~without explicit $\Delta$ resonance actually stem from the `high' (note that $m_\Delta-m_N\approx 2m_\pi<m_\rho$, where $m_\rho\sim770$ MeV is usually regarded as the truly high energy scale in chiral perturbation theory) energy scale $\Delta$ baryon as well as the pion-pion correlation (or the $\sigma$ meson)~\cite{Bernard:1996gq}.

Epelbaum \etal~noticed that the TPE loop diagrams calculated within the dimensional regularization accompanying with the large value LECs in subleading pion-nucleon vertices lead to unsatisfactory convergence of chiral expansion and uncertain consequences in few-nucleon systems, e.g., the unphysical deeply bound states in the low partial waves of isoscalar channel~\cite{Epelbaum:2002ji,Epelbaum:2003gr}. The expediency is to use the small value LECs, but this is not compatible with the pion-nucleon scattering data~\cite{Fettes:1998ud,Krebs:2007rh}. In order to cure this problem, Epelbaum \etal~argued that one needs to suppress the high-momentum modes of the exchanged pions, since they cannot be suitably handled in an EFT who only properly works in the soft scales. In order to solve this problem, the TPE loop diagrams using the cutoff regularization combining the spectral function representation scheme~\cite{Epelbaum:2003gr, Epelbaum:2003xx, Epelbaum:2004fk}, local regularization scheme~\cite{Epelbaum:2014efa, Epelbaum:2014sza} and semilocal regularization scheme~\cite{Reinert:2017usi} were proposed. One can see review~\cite{Epelbaum:2019kcf} for the state-of-the-art. This is analogous to the means for improving the convergence of chiral expansion in the SU(3) case~\cite{Donoghue:1998krd,Donoghue:1998bs}. In addition, it was shown that an covariant \xEFT~can moderate the TPE contribution~\cite{Xiao:2020ozd,Wang:2021kos,Lu:2021gsb} even using the dimensional regularization.

Obviously, one needs to consider the possible emergence of the above mentioned problem when generalizing the \xEFT~to the heavy-light systems. The application of \xEFT~in heavy-light systems for dealing with the hadronic molecules has achieved much progress in recent years~\cite{Meng:2022ozq}. In Ref.~\cite{Liu:2012vd}, Liu \etal~first calculated the $BB$ interactions with considering the leading TPE contributions. Along this line, Xu \etal~studied the $DD^\ast$ interactions and used the RSM to determine the contact LECs, in which they predicted a bound state in the isoscalar channel with $J^P$ quantum numbers $1^+$~\cite{Xu:2017tsr}. Four years latter, the LHCb Collaboration observed a state, the $\Tcc$ in $D^0D^0\pi^+$ invariant mass spectrum~\cite{LHCb:2021vvq,LHCb:2021auc}. The $\Tcc$ is below the $D^{\ast+}D^0$ threshold about $300$ keV, thus it is the very good candidate of $DD^\ast$ hadronic molecule. Similar to Ref.~\cite{Xu:2017tsr}, Wang \etal~studied the $B^{(\ast)}B^{(\ast)}$ interactions and predicted the possible bound states in the isoscalar $BB^\ast$ and $B^\ast B^\ast$ systems with $J^P=1^+$~\cite{Wang:2018atz}. The same framework was also adopted to investigate the LHCb pentaquarks $P_\psi^N(4312)$, $P_\psi^N(4440)$ and $P_\psi^N(4457)$~\cite{Meng:2019ilv,Wang:2019ato} (throughout this paper, we use the new naming scheme of the exotic states proposed by the LHCb~\cite{Gershon:2022xnn}), as well as to predict the existence of molecular pentaquarks with strangeness in $\Xi_c^{(\prime,\ast)}\bar{D}^{(\ast)}$ systems~\cite{Wang:2019nvm} (see also the recent experimental measurements for the $P_{\psi s}^\Lambda$ states near the $\Xi_c\bar{D}^\ast$~\cite{LHCb:2020jpq} and $\Xi_c\bar{D}$~\cite{LHCb:2022jad} thresholds), and the double-charm pentaquarks~\cite{Chen:2021htr}. For a review of this topic, we refer to Ref.~\cite{Meng:2022ozq}. In Ref.~\cite{Chen:2022iil}, the study of $\Sigma_c\Sigma_c$ interactions turns out that there results in bad convergence and unnaturally deep bound state in the lowest isospin channel if one calculates the leading TPE diagrams with dimensional regularization. This demands us to properly treat the TPE contributions for heavy-light systems as those in the $NN$ case.

Recently, the S-wave $DD^\ast$ potential in isoscalar channel was extracted from lattice QCD simulations near the physical pion mass within HAL QCD method~\cite{Lyu:2023xro}. It was shown that the potential favors the $e^{-2m_\pi r}/r^2$ behavior in the range $1 <r<2\text{ fm}$. Thus, it is worthwhile to investigate TPE interaction for $DD^\ast$ in the coordinate space to compare with that from HAL QCD.

In this work, we revisit the $DD^\ast$ interactions within \xEFT, and calculate the $DD^\ast$ interactions up to the third order [i.e. the next-to-next-to-leading order (N$^2$LO)] for the first time. We construct the subleading $\pi D^{(\ast)}$ Lagrangians and determine the corresponding LECs with the RSM. The TPE diagrams will be calculated with the cutoff regularization, but we use the fully local momentum-space regularization rather than the semi-local form  as those in Ref.~\cite{Reinert:2017usi}. The $DD^\ast$ interactions shall strongly correlate to the $\Tcc$ inner structures and its other properties. In contrast to the well-known $X(3872)$, there is no coupling with the charmonia for $\Tcc$. Thus it provides a clean environment for investigating the interactions between the charmed mesons. This is very similar to the $NN$ interactions.

The $\Tcc$~state has been intensively studied from various aspects, such as the decay behaviors~\cite{Meng:2021jnw,Agaev:2021vur,Ling:2021bir,Yan:2021wdl,Feijoo:2021ppq,Ren:2021dsi}, the mass spectra~\cite{Dong:2021bvy,Chen:2021vhg,Weng:2021hje,Xin:2021wcr,Chen:2021tnn,Chen:2021cfl,Deng:2021gnb,Ke:2021rxd,Padmanath:2022cvl,Lin:2022wmj,Kim:2022mpa,Cheng:2022qcm,Albuquerque:2022weq}\clabel[spectra]{~}, the productions~\cite{Qin:2020zlg,Huang:2021urd,Jin:2021cxj,Hu:2021gdg,Abreu:2022lfy,Braaten:2022elw}, the lineshapes~\cite{Dai:2021wxi,Fleming:2021wmk,Du:2021zzh}, and the magnetic moment~\cite{Azizi:2021aib}, etc. In order to pin down the inner configuration of $\Tcc$, a systematic study of the $DD^\ast$ interactions is very necessary.

This paper is organized as follows. The $DD^\ast$ effective potentials within the local momentum-space regularization are shown in Sec.~\ref{sec:effpot}. The analyses of effective potential and the pole trajectories of $DD^\ast$ bound state and related discussions are given in Sec.~\ref{sec:numsanddis}. A short summary is given in Sec.~\ref{sec:sum}. The estimations of LECs within the RSM are listed in the Appendix~\ref{sec:app}.

\section{Effective chiral potentials up to the third order}\label{sec:effpot}

The effective potential of $DD^\ast$ can be extracted from their scattering amplitude. In \xEFT, the scattering amplitude of $DD^\ast$ is expanded in powers of the ratio $\mathcal{Q}/\Lambda_{\rm{b}}$, where $\mathcal{Q}$ represents the soft scale, which could be the pion mass or the external momenta of $D^{(\ast)}$, while $\Lambda_{\rm{b}}$ denotes the hard scale at which the \xEFT~breaks down. The relative importance of the terms in the expansion is weighed by the power $\nu$ of $(\mathcal{Q}/\Lambda_{\rm{b}})^\nu$, this is known as the power counting scheme. According to the naive dimensional analysis~\cite{Weinberg:1990rz,Weinberg:1991um}, the power $\nu$ for a system with two matter fields (charmed mesons) is measured as
\begin{eqnarray}\label{eq:powercounting}
\nu=2L+\sum_i V_i \Delta_i,\qquad \Delta_i=d_i+\frac{n_i}{2}-2,
\end{eqnarray}
with $L$ the number of loops in a diagram, $V_i$ the number of vertices of type-$i$. The $d_i$ is the number of derivatives (or the pion-mass insertions), and $n_i$ is the number of charmed meson fields that involved in the vertex-$i$.

\begin{figure*}[!hptb]
\begin{centering}
%\vspace{0.3cm}
    %\setlength{\abovecaptionskip}{0.01cm}
    %\setlength{\belowcaptionskip}{-0.4cm}
    \scalebox{1.0}{\includegraphics[width=1.0\linewidth]{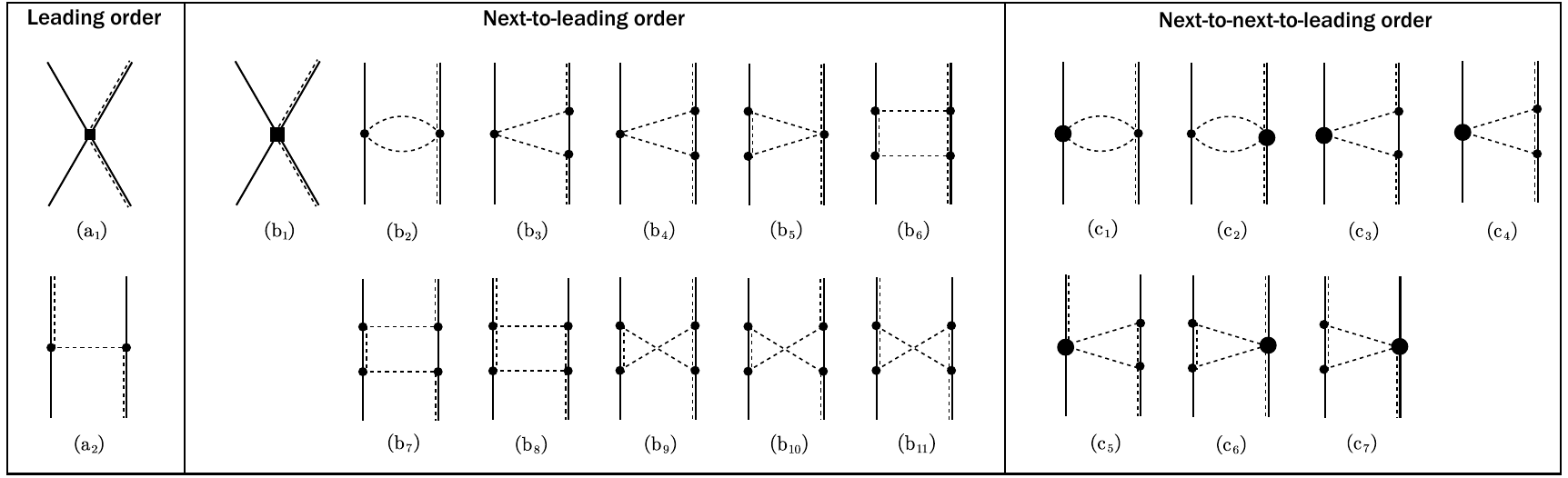}}
    \caption{The irreducible Feynmann diagrams for the $DD^\ast$ interactions at the LO, NLO and N$^2$LO. We use the single solid line, the double line (solid plus dashed) and the dashed line to denote the $D$, $D^\ast$ and pion, respectively. The small solid square [in (a$_1$)] and large solid square [in (b$_1$)] denote the LO ($\Delta_i=0$) and NLO ($\Delta_i=2$) $DD^\ast$ contact vertices, respectively. The small solid dot [in (a$_2$), (b$_{2,\dots,11}$), (c$_{1,\dots,7}$)] and large solid dot [in (c$_{1,\dots,7}$)] represent the LO ($\Delta_i=0$) and NLO ($\Delta_i=1$) $\pi D^{(\ast)}$ vertices, respectively.\label{fig:loopdiagrams}}
\end{centering}
\end{figure*}

The $DD^\ast$ interaction starts at $\nu=0$ (first order, the LO), and the higher orders come as $\nu=2$ [second order, the next-to-leading order (NLO)], $\nu=3$ (third order, the N$^2$LO), etc. At the given order, the number of the corresponding irreducible diagrams is limited. In Fig.~\ref{fig:loopdiagrams}, we show the pertinent Feynmann diagrams for the LO, NLO and N$^2$LO interactions of the $DD^\ast$ system. Then the effective potential of the $DD^\ast$ system can be written as
\begin{eqnarray}\label{eq:vctopetpe}
V_{\rm eff}=V_{\rm ct}+V_{1\pi}+V_{2\pi}+\dots,
\end{eqnarray}
with
\begin{eqnarray}\label{eq:vctvpi}
V_{\rm ct}&=&V_{\rm ct}^{(0)}+V_{\rm ct}^{(2)}+\dots,\nonumber\\
V_{2\pi}&=&V_{2\pi}^{(2)}+V_{2\pi}^{(3)}+\dots,
\end{eqnarray}
where $V_{\rm ct}$, $V_{1\pi}$ and $V_{2\pi}$ denote the contact, OPE and TPE potentials, respectively. The numbers in the parentheses of the superscripts represent the order $\nu$ [see Eq.~\eqref{eq:powercounting}]. Each piece of the right hand side of Eq.~\eqref{eq:vctopetpe} can be further decomposed into the following form,
\begin{eqnarray}\label{eq:vi}
V_i=\left[V_{i,c}+\bm{\tau}_{1}\cdot\bm{\tau}_{2}W_{i,c}\right]\mathcal{O}_1+\left[V_{i,t}+\bm{\tau}_{1}\cdot\bm{\tau}_{2}W_{i,t}\right]\mathcal{O}_2,
\end{eqnarray}
where $i=\mathrm{ct},1\pi,2\pi$, and $\bm{\tau}_{1}\cdot\bm{\tau}_{2}$ denotes the isospin-isospin interaction. The matrix element $\langle\bm{\tau}_{1}\cdot\bm{\tau}_{2}\rangle=-3$ and $1$ for the isoscalar and isovector channels, respectively. The operators $\mathcal{O}_1$ and $\mathcal{O}_2$ are given as
\begin{eqnarray}\label{eq:operatoro1o2}
\mathcal{O}_{1}=\bm{\varepsilon}^{\prime\dagger}\cdot\bm{\varepsilon},\qquad\qquad \mathcal{O}_{2}=(\bm{q}\cdot\bm{\varepsilon}^{\prime\dagger})(\bm{q}\cdot\bm{\varepsilon}),
\end{eqnarray}
where $\bm q=\bm p^\prime-\bm p$ ($\bm p$ and $\bm p^\prime$ denote the initial and final state momenta in the center of mass system, respectively) is the transferred momentum between $D$ and $D^\ast$, $\bm{\varepsilon}$ and $\bm{\varepsilon}^{\prime\dagger}$ denote the polarization vectors of the initial and final $D^\ast$ mesons, respectively. In the heavy quark limit, we will not consider the $1/m$ (with $m$ the mass of the charmed mesons) corrections of the charmed meson fields. Then only two pertinent operators survive in the effective potentials of $DD^\ast$ (for the $NN$ case, see Ref.~\cite{Machleidt:1987hj}), i.e., the $\mathcal{O}_1$ and $\mathcal{O}_2$.

In the following subsections, we will derive the $V_{\rm ct}$, $V_{1\pi}$ and $V_{2\pi}$, respectively.

\subsection{Short-range contact interactions}

The contact potentials of $DD^\ast$ system at the order $\nu=0,2,4$ can be respectively parameterized as
\begin{eqnarray}
V_{\mathrm{ct}}^{(0)}	&=&	\left(C_{1}+\bm{\tau}_{1}\cdot\bm{\tau}_{2}C_{2}\right)\mathcal{O}_{1},\label{eq:vcto0}\\
V_{\mathrm{ct}}^{(2)}	&=&	(C_{3}+\bm{\tau}_{1}\cdot\bm{\tau}_{2}C_{4})\bm{q}^{2}\mathcal{O}_{1}\nonumber\\
&&+\left(C_{5}+\bm{\tau}_{1}\cdot\bm{\tau}_{2}C_{6}\right)\mathcal{O}_{2},\label{eq:vcto2}\\
V_{\mathrm{ct}}^{(4)}	&=&	(C_{7}+\bm{\tau}_{1}\cdot\bm{\tau}_{2}C_{8})\bm{q}^{4}\mathcal{O}_{1}\nonumber\\
                        &&+(C_{9}+\bm{\tau}_{1}\cdot\bm{\tau}_{2}C_{10})\bm{q}^{2}\mathcal{O}_{2},\label{eq:vcto4}%\nonumber\\
\end{eqnarray}
where $C_{1,\dots,10}$ are the corresponding LECs. In the following calculations, we will take the $V_{\mathrm{ct}}^{(4)}$ to test the convergence of the expansion in different isospin channels. In Eqs.~\eqref{eq:vcto2} and~\eqref{eq:vcto4}, we ignore the pion-mass dependent terms for which are of irrelevance in our studies. In calculations, the local form Gaussian regulator $\exp{(-\bm q^2/\Lambda^2)}$ is multiplied to Eqs.~\eqref{eq:vcto0}-\eqref{eq:vcto4} to ensure the convergence when they are inserted into the Lippmann-Schwinger equations (LSEs).

In order to determine all the LECs in Eqs.~\eqref{eq:vcto0}-\eqref{eq:vcto4}, we resort to the phenomenological RSM~\cite{Ecker:1988te,Epelbaum:2001fm} (see also the applications in heavy-light systems~\cite{Xu:2017tsr,Du:2016tgp,Xu:2021vsi,Peng:2021hkr}). Within the RSM, we consider the exchanging of the scalar, pseudoscalar, vector and axial-vector mesons [the tensor exchanges (e.g., $a_2,f_2$ mesons) are not considered, since their contributions start at least at the fourth order~\cite{Du:2016tgp}]. The derivation details are given in appendix~\ref{sec:app1}. Their numerical values are listed in Table~\ref{tab:numsci}.

\begin{table*}[htbp]
%\centering
\centering
\renewcommand{\arraystretch}{1.5}
\caption{The numerical values of the LECs [see Eqs.~\eqref{eq:vcto0}-\eqref{eq:vcto4}] determined from the RSM. The $C_{1,2}$, $C_{3,\dots,6}$ and $C_{7,\dots,10}$ are in units of $\mathrm{GeV}^{-2}$, $\mathrm{GeV}^{-4}$ and $\mathrm{GeV}^{-6}$, respectively. The values for the isoscalar ($I=0$) and isovector ($I=1$) channels can be easily obtained via replacing the ${\rm sgn}$ with $(-1)^{I}$. \change{The errors come from the parameters in the Lagrangians that quoted in the appendix.}\label{tab:numsci}}
\setlength{\tabcolsep}{4.mm}
{
\begin{tabular}{ccccc}
\toprule[0.8pt]
$C_1$ & $C_2$ & $C_3$ & $C_4$ & $C_5$\\
$7.5\pm1.3$ & $10.9\pm0.2+0.3\mathrm{sgn}$ & $-26.0\pm9.1$ & $-18.8+(23.1\pm7.4)\mathrm{sgn}$ & $14.7\pm7.2$\\
$C_6$ & $C_7$ & $C_8$ & $C_9$ & $C_{10}$\\
$(-23.1\pm7.4)\mathrm{sgn}$ & $-5.2\pm30.5$ & $32.1-(40.6\pm12.9)\mathrm{sgn}$ & $-10.6\pm12.1$ & $(40.6\pm12.9)\mathrm{sgn}$\\
\bottomrule[0.8pt]
\end{tabular}
}
\end{table*}
\subsection{Long-range one-pion exchange interactions}

The $\Tcc$ was observed in the $D^0D^0\pi^+$ final state, and its signal is absent in the $D^+D^0\pi^+$ channel~\cite{LHCb:2021vvq,LHCb:2021auc}, which implies that the $\Tcc$ is an isoscalar state rather than the isovector one. The flavor wave function of $DD^\ast$ in the isoscalar and isovector channels read, respectively,
\begin{eqnarray}
|DD^\ast,I=0,I_{3}=0\rangle	&=&	\frac{1}{\sqrt{2}}\left[D^{0}D^{\ast+}-D^{\ast0}D^{+}\right],\\
|DD^\ast,I=1,I_{3}=0\rangle	&=&	\frac{1}{\sqrt{2}}\left[D^{0}D^{\ast+}+D^{\ast0}D^{+}\right].
\end{eqnarray}

We consider the explicit chiral dynamics from the light pion and relegate the heavy $\eta$ ($m_\eta\simeq4m_\pi$) contribution to the contact terms. In the following, we show the complete LO ($\Delta_i=0$) chiral Lagrangian of $\varphi D^{(\ast)}$ ($\varphi=\pi,\eta$) coupling~\cite{Wise:1992hn,Manohar:2000dt} for the latter convenience.
\begin{eqnarray}\label{eq:mesonlagsf}
\mathcal{L}_{\varphi\mathcal{H}}^{(0)}&=&i\langle \mathcal{H}v\cdot\mathcal{D}\bar{\mathcal{H}}\rangle-\frac{1}{8}\delta_b\langle\mathcal{H}\sigma^{\mu \nu}\bar{\mathcal{H}}\sigma_{\mu \nu}\rangle\nonumber\\
&&+g_\varphi\langle\mathcal{H} \slashed{u}\gamma_{5}\bar{\mathcal{H}}\rangle,
\end{eqnarray}
where $v=(1,\bm 0)$ denotes the four-velocity of heavy mesons, and $\mathcal{D}_\mu=\partial_\mu+\Gamma_\mu$, with $\Gamma_\mu=[\xi^{\dagger},\partial_{\mu}\xi]/2$ the chiral connection. $\delta_b=m_{D^\ast}-m_D\simeq 142$ MeV, and $g_\varphi=-0.59$. The axial-vector current $u_\mu$ is defined as $u_\mu=i\{\xi^{\dagger},\partial_{\mu}\xi\}/2$. Meanwhile, the $\xi^2=U=\exp(i\varphi/f_\varphi)$, and the matrix form of $\varphi$ reads
\begin{eqnarray}
\varphi=\left[\begin{array}{cc}
\pi^{0}+\frac{1}{\sqrt{3}}\eta & \sqrt{2}\pi^{+}\\
\sqrt{2}\pi^{-} & -\pi^{0}+\frac{1}{\sqrt{3}}\eta
\end{array}\right].
\end{eqnarray}
The $\mathcal{H}$ denotes the superfield of ($D,D^\ast$) doublet in the heavy quark symmetry, which reads
\begin{eqnarray}
\mathcal{H}&=&\frac{1+\slashed{v}}{2}\left(P_\mu^\ast\gamma^\mu+iP\gamma_5\right),\quad\bar{\mathcal{H}}=\gamma^0\mathcal{H}^\dag\gamma^0,
\end{eqnarray}
with $P=(D^0,D^+)^T$ and $P^\ast=(D^{\ast0},D^{\ast+})^T$.

With the OPE diagram in Fig.~\ref{fig:loopdiagrams} ($\mathrm{a}_2$) and the LO chiral Lagrangian in Eq.~\eqref{eq:mesonlagsf}, one can easily get the OPE potential, which reads
\begin{eqnarray}\label{eq:opeporig}
\mathcal{V}_{1\pi}(u_\pi,\bm{q})&=&\mathrm{sgn}(\bm{\tau}_{1}\cdot\bm{\tau}_{2})\frac{g_\varphi^{2}}{4f_\pi^{2}}\frac{\mathcal{O}_2}{\bm{q}^{2}-u_\pi^{2}-i\epsilon},
\end{eqnarray}
where $f_\pi=92.4$ MeV, $u_\pi=\sqrt{\delta_b^2-m_\pi^2}$ (with $m_\pi\simeq137$ MeV the pion mass), and $\mathrm{sgn}=(-1)^I$. Eq.~\eqref{eq:opeporig} contains two parts---the principle-value and the imaginary parts. Its principle-value corresponds to an oscillatory potential in the coordinate space, e.g., see Eq.~\eqref{eq:fourtrans}, while the imaginary part comes from the three-body ($DD\pi$) cut, it will contribute a finite width to the bound state of $DD^\ast$. We then separate the operator $\mathcal{O}_2$ into the `spin-spin' part and the tensor part via the equation
\begin{eqnarray}
\mathcal{O}_2=(\bm{q}\cdot\bm{\varepsilon}^{\prime\dagger})(\bm{q}\cdot\bm{\varepsilon})=\frac{1}{3}(\bm{\varepsilon}^{\prime\dagger}\cdot\bm{\varepsilon})\bm{q^{2}}+\bm{q}^{2}\mathcal{S}_{12},
\end{eqnarray}
where $\mathcal{S}_{12}=(\bm{\varepsilon}^{\prime\dagger}\cdot\hat{\bm{q}})(\bm{\varepsilon}\cdot\hat{\bm{q}})-\frac{1}{3}\bm{\varepsilon}^{\prime\dagger}\cdot\bm{\varepsilon}$, with $\hat{\bm{q}}=\bm q/|\bm q|$. Then the principle-value part of Eq.~\eqref{eq:opeporig} can be transformed into
\begin{eqnarray}
\mathcal{V}_{1\pi}(u_\pi,\bm{q})&=&\mathrm{sgn}(\bm{\tau}_{1}\cdot\bm{\tau}_{2})\frac{g_\varphi^{2}}{4f_\pi^{2}}\bigg(\frac{1}{3}\mathcal{O}_{1}+\frac{1}{3}\frac{u_\pi^{2}}{\bm{q}^{2}-u_\pi^{2}}\mathcal{O}_{1}\nonumber\\
&&+\frac{\bm{q}^{2}}{\bm{q}^{2}-u_\pi^{2}}\mathcal{S}_{12}\bigg),
\end{eqnarray}
in which the first term corresponds to a $\delta$-function in the coordinate space ($r$-space) after the Fourier transform. It is an artefact arising from the idealized point-like $\pi D^{(\ast)}$ coupling. In reality, the OPE dominates at the long-distance region, i.e., $r\gtrsim 2~\mathrm{fm}\simeq 1.5 m_\pi^{-1}$~\cite{Ericson:1988gk}. Therefore, it is better to subtract the unphysical $\delta$-function part from the OPE potential. In Ref.~\cite{Reinert:2017usi}, Reinert \etal~introduced a subtraction scheme for the $NN$ interaction with the following form
\begin{eqnarray}
V_{1\pi,\Lambda}(u_\pi,\bm{q})&=&\mathrm{sgn}(\bm{\tau}_{1}\cdot\bm{\tau}_{2})\frac{g_\varphi^{2}}{4f_\pi^{2}}\Bigg\{\frac{u_\pi^{2}}{3}\frac{1}{\bm{q}^{2}-u_\pi^{2}}\mathcal{O}_{1}\nonumber\\
&&+\left[\frac{1}{3}+\mathcal{C}(u_\pi,\Lambda)\right]\mathcal{O}_{1}+ \frac{\bm{q}^{2}}{\bm{q}^{2}-u_\pi^{2}} \mathcal{S}_{12}\Bigg\}\nonumber\\
&&\times\exp\Big(-\frac{\bm{q}^{2}-u_\pi^{2}}{\Lambda^{2}}\Big),
\end{eqnarray}
where an $u_\pi$-dependent term in the Gaussian regulator is introduced to ensure the strength of OPE potential remains unchanged at the pion pole~\cite{Epelbaum:2022cyo}. The subtraction term $\mathcal{C}(u_\pi,\Lambda)$ is determined by the requirement that the OPE potential vanishes at the origin, i.e., when $r\to0$. With the following relations of Fourier transform,
\begin{eqnarray}\label{eq:fourtrans}
&&\int\frac{d^{3}q}{(2\pi)^{3}}e^{i\bm{q}\cdot\bm{r}}\frac{1}{\bm{q}^{2}-u_\pi^{2}}=\frac{1}{4\pi r}\cos (u_\pi r),\\
&&\mathcal{U}_{\Lambda}(u_\pi,r)=\int\frac{d^{3}q}{(2\pi)^{3}}e^{i\bm{q}\cdot\bm{r}}\frac{1}{\bm{q}^{2}-u_\pi^{2}}\exp\Big(-\frac{\bm{q}^{2}-u_\pi^{2}}{\Lambda^{2}}\Big)\nonumber\\
&&~~~~~=\frac{\cos(u_\pi r)}{8\pi r}\left[\mathrm{erfc}\Big(\frac{u_\pi}{\Lambda}-\frac{\Lambda r}{2}\Big)-\mathrm{erfc}\Big(\frac{u_\pi}{\Lambda}+\frac{\Lambda r}{2}\Big)\right],\nonumber\\
\\
&&\int\frac{d^{3}q}{(2\pi)^{3}}e^{i\bm{q}\cdot\bm{r}}\exp\Big(-\frac{\bm{q}^{2}-u_\pi^{2}}{\Lambda^{2}}\Big)\nonumber\\
&&\qquad\qquad=\Big(\frac{\Lambda^{2}}{4\pi}\Big)^{3/2}\exp\Big(\frac{u_\pi^{2}}{\Lambda^{2}}-\frac{\Lambda^{2}r^{2}}{4}\Big),
\end{eqnarray}
one easily obtains
\begin{eqnarray}\label{eq:opesub}
&&V_{1\pi,\Lambda}(u_\pi,r)	=	\int\frac{d^{3}q}{(2\pi)^{3}}e^{i\bm{q}\cdot\bm{r}}V_{1\pi,\Lambda}(u_\pi,\bm{q})\nonumber\\
&&\quad=	-\mathrm{sgn}(\bm{\tau}_{1}\cdot\bm{\tau}_{2})\frac{g_\varphi^{2}}{4f_\pi^{2}}\Bigg\{ \mathcal{S}_{12}r\frac{\partial}{\partial r}\left(\frac{1}{r}\frac{\partial}{\partial r}\right)\mathcal{U}_{\Lambda}(u_\pi,r)\nonumber\\
&&\quad\quad-\mathcal{O}_{1}\bigg[\frac{u_\pi^{2}}{3}\mathcal{U}_{\Lambda}(u_\pi,r)+\Big(\frac{1}{3}+\mathcal{C}(u_\pi,\Lambda)\Big)\Big(\frac{\Lambda^{2}}{4\pi}\Big)^{3/2}\nonumber\\
&&\quad\quad\times\exp\Big(\frac{u_\pi^{2}}{\Lambda^{2}} -\frac{\Lambda^{2}r^{2}}{4}\Big)\bigg]\Bigg\}.
\end{eqnarray}
With the constraint $V_{1\pi,\Lambda}(u_\pi,r\to 0)=0$, we get
\begin{eqnarray}
\mathcal{C}(u_\pi,\Lambda)=-\Big(\frac{1}{3}+\frac{2u_\pi^{2}}{3\Lambda^{2}}e^{-\frac{2u_\pi^{2}}{\Lambda^{2}}}\Big).
\end{eqnarray}
Note that, in Eq.~\eqref{eq:fourtrans} the $\mathrm{erfc}(x)$ represents the complementary error function, i.e.,
\begin{eqnarray}
\mathrm{erfc}(x)=\frac{2}{\sqrt{\pi}}\int_{x}^{\infty}dt e^{-t^2}.
\end{eqnarray}
It should be stressed that the so-called OPE interactions in this work are in fact parts of their effects that cannot be compensated by the contact terms.

\begin{figure}[htb]
\begin{center}
\begin{minipage}[t]{0.51\linewidth}
\centering
\includegraphics[width=\columnwidth]{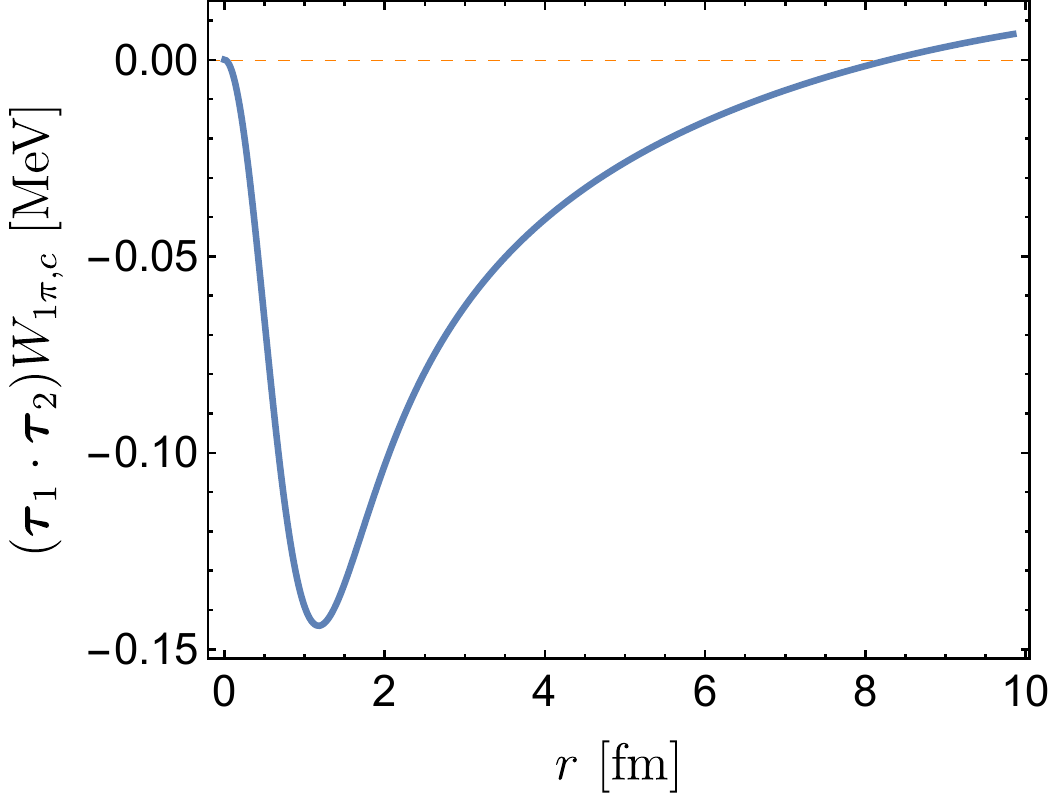}
\end{minipage}%
%\hspace{0.005in}
\begin{minipage}[t]{0.48\linewidth}
\centering
\includegraphics[width=\columnwidth]{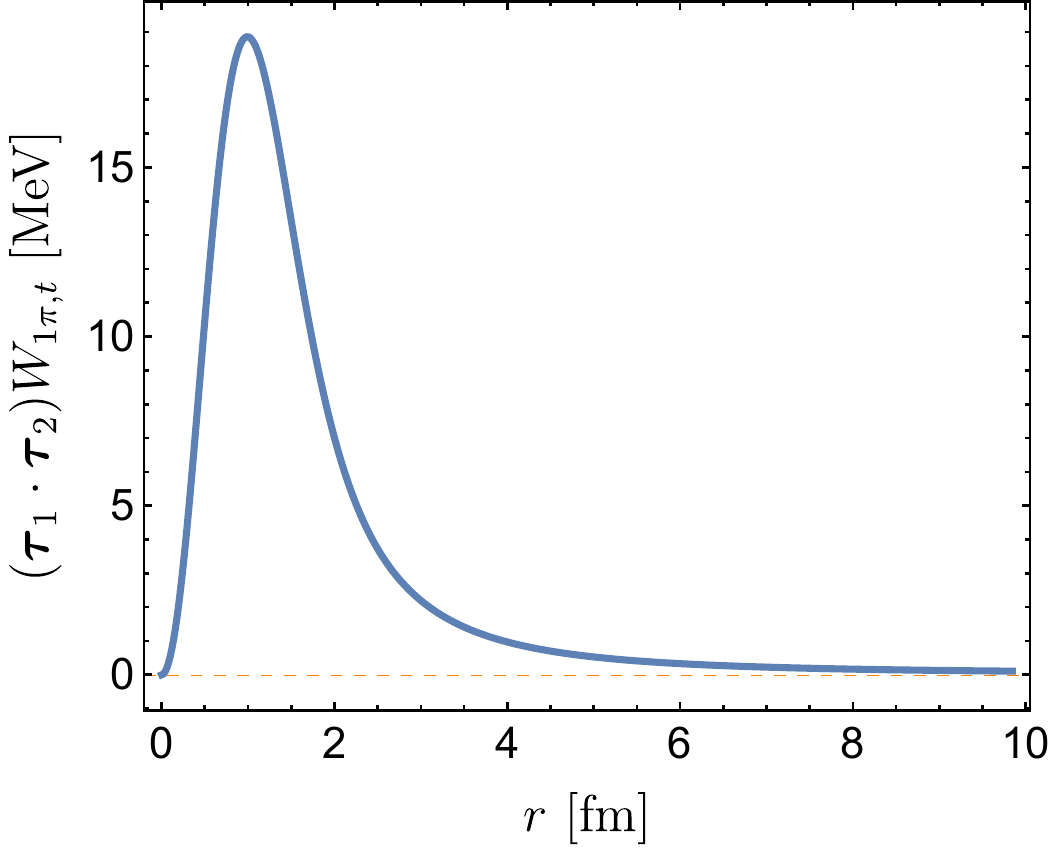}
\end{minipage}
%\hspace{0.068in}
\caption{The central part (left panel) and tensor part (right panel) of the subtracted OPE potential [see Eq.~\eqref{eq:opesub}] in isoscalar channel with the cutoff $\Lambda=0.5$ GeV.\label{fig:ope}}
\end{center}
\end{figure}

In Fig.~\ref{fig:ope}, we show the behaviors of the central part [$\mathcal{O}_1$ related term in Eq.~\eqref{eq:opesub}] and tensor part [$\mathcal{S}_{12}$ related term in Eq.~\eqref{eq:opesub}] of the $\delta$-function subtracted OPE potential for the $I=0$ case (the behaviors of the $I=1$ case are similar). One can see that both the central and tensor potentials vanish for $r\to 0$, and the strength of the central potential is much weaker than that of the tensor potential. Therefore, though the central potential is attractive, it is too weak to form bound state. However, if one does not subtract the $\delta$-function, then there would result in very attractive central potential once using a large cutoff when making the Fourier transform. This may also lead to the bound state, but it is unreasonable. One also sees that the central potential can extend to large distances since the effective mass $u_\pi$ in the pion propagator is much smaller than the $m_\pi$, this is a very typical feature of the $DD^\ast$ system.
%Besides, because the central potential is proportional to $\cos(u_\pi r)$, thus it is oscillatory in coordinate space. One may notice that the vanishing point of the central potential is around $8~\mathrm{fm}$, which is of similar size as the size of $\Tcc$,
%\begin{eqnarray}
%r_{T_{cc}}=\frac{1}{\sqrt{2\mu_{D^{\ast+}D^0}|\delta m^U|}}=7.5\pm0.4~\mathrm{fm},
%\end{eqnarray}
%where $\mu_{D^{\ast+}D^0}$ is the reduced mass of $D^{\ast+}D^0$ and $\delta m^U=-361\pm40~\mathrm{keV}$~\cite{LHCb:2021auc}.

\subsection{Intermediate-range two-pion exchange interactions}
%\subsubsection{The leading two-pion exchange potential}
We first show the LO ($\nu=2$) TPE contributions, which come from the diagrams in Figs.~\ref{fig:loopdiagrams} ($\text{b}_2$)-($\text{b}_{11}$). They can be obtained using the Lagrangian~\eqref{eq:mesonlagsf} and calculating the loop integrals. We adopt the spectral function representation for the TPE interactions. The long-range part of the TPE interactions is determined by the non-analytic terms in momentum-space. They have the following forms within the dimensional regularization,
\begin{widetext}
\begin{eqnarray}
\mathcal{V}_{2\pi,c}^{(2)}&=&\frac{g_\varphi^{4}}{512\pi\delta_{b}f_{\pi}^{4}}\left[6\bm{q}^2\mathrm{sgn}(4m_{\pi}^{2}-4\delta_{b}^{2}+\bm{q}^2)A^{\prime}(q)-3(8m_{\pi}^{4}+10m_{\pi}^{2}\bm{q}^2+3\bm{q}^4)A(q)- \frac{8\delta_{b}}{\pi}(5\delta_{b}^{2}+3\bm{q}^2\mathrm{sgn})L(q)\right],\label{eq:v2pic}\\
\mathcal{W}_{2\pi,c}^{(2)}&=&\frac{g_{\varphi}^{4}}{256\pi\delta_{b}f_{\pi}^{4}}\bigg\{\frac{1}{g_{\varphi}^{2}}(4m_{\pi}^{2}-4\delta_{b}^{2}+3\bm{q}^2)\left[2\delta_{b}^{2}+g_{\varphi}^{2}(-2\delta_{b}^{2}+2m_{\pi}^{2}+\bm{q}^2)\right]A^{\prime}(q) -2\mathrm{sgn}(4m_{\pi}^{2}+\bm{q}^2)\bm{q}^2A(q)\nonumber\\
&&+\frac{2\delta_{b}}{3\pi g_{\varphi}^{4}}\left[16\delta_{b}^{2}g_{\varphi}^{2}(5g_{\varphi}^{2}-3)+4(-5g_{\varphi}^{4}+4g_{\varphi}^{2}+1)m_{\pi}^{2}+(-23g_{\varphi}^{4}+10g_{\varphi}^{2}+1)\bm{q}^2\right]L(q)\bigg\}, \\
\mathcal{V}_{2\pi,t}^{(2)}&=&\frac{g_\varphi^{4}}{512\pi\delta_{b}f_{\pi}^{4}}\left[\frac{3}{\bm{q}^2}(-8m_{\pi}^{4}-2m_{\pi}^{2}\bm{q}^2+\bm{q}^4)A(q)-6\mathrm{sgn}(-4\delta_{b}^{2}+4m_{\pi}^{2}+\bm{q}^2)A^{\prime}(q)+ \frac{24\delta_{b}}{\pi}\mathrm{sgn}L(q)\right],\\
\mathcal{W}_{2\pi,t}^{(2)}&=&\frac{g_{\varphi}^{4}}{256\pi\delta_{b}f_{\pi}^{4}}\left\{ 2\mathrm{sgn}(4m_{\pi}^{2}+\bm{q}^2)A(q)-\frac{1}{g_{\varphi}^{2}\bm{q}^2}(4\delta_{b}^{2}-4m_{\pi}^{2}+\bm{q}^2)\left[2\delta_{b}^{2}+g_{\varphi}^{2}(-2\delta_{b}^{2}+2m_{\pi}^{2}+\bm{q}^2)\right]A^{\prime}(q)\right\},\label{eq:w2pit}
\end{eqnarray}
\end{widetext}
where the three non-analytic functions $A(q)$, $A^\prime(q)$ and $L(q)$ respectively read
\begin{eqnarray}
A(q)&=&\frac{1}{2q}\arctan\frac{q}{2m_\pi},\\
A^{\prime}(q)&=&\frac{1}{2q}\arctan\frac{q}{2m^{\prime}},\\
L(q)&=&\frac{\varpi}{q}\ln\frac{q+\varpi}{2m_\pi},
\end{eqnarray}
with $q=|\bm q|$, $m^{\prime}=[m_\pi^{2}-\delta_b^{2}-i\epsilon]^{1/2}$, and $\varpi=[\bm{q}^{2}+4m_\pi^{2}-i\epsilon]^{1/2}$.
The terms containing the non-analytic functions $\mathcal{F}(q,n,i)=\int_{-1/2}^{1/2}\left(y^n\zeta^i\arctan\frac{\delta_b}{\zeta}\right)dy$ (with $\zeta=q\sqrt{a^2-y^2}$, $a=\frac{\varpi^\prime}{2q}$, and $\varpi^\prime=[\bm{q}^{2}+4(m_\pi^{2}-\delta_b^{2})-i\epsilon]^{1/2}$) and their derivatives with respect to $\delta_b$ are ignored for simplicity since we noticed that their contributions are much smaller than those in Eqs.~\eqref{eq:v2pic}-\eqref{eq:w2pit}.

In order to obtain the subleading ($\nu=3$) TPE potential (see the diagrams in the third column of Fig.~\ref{fig:loopdiagrams}), one needs an insertion of the subleading ($\Delta_i=1$) $\pi D^{(\ast)}$ Lagrangians. The Lagrangians read~\cite{Meng:2022ozq}
\begin{eqnarray}\label{eq:nlopid}
\mathcal{L}_{\varphi\mathcal{H}}^{(1)}&=&\tilde{c}_{1}\langle\mathcal{H}\bar{\mathcal{H}}\rangle\mathrm{Tr}(\chi_{+})+\tilde{c}_{2}\langle\mathcal{H}v\cdot uv\cdot u\bar{\mathcal{H}}\rangle+\tilde{c}_{3}\langle\mathcal{H}u\cdot u\bar{\mathcal{H}}\rangle\nonumber\\
&&+i\tilde{c}_{4}\langle\mathcal{H}[u_{\mu},u_{\nu}]\sigma^{\mu\nu}\bar{\mathcal{H}}\rangle+\tilde{c}_{5}\langle\mathcal{H}\hat{\chi}_{+}\bar{\mathcal{H}}\rangle,
\end{eqnarray}
where $\chi_{+}=\xi^{\dagger}\chi \xi^{\dagger}+ \xi\chi^{\dagger}\xi$, with $\chi=2B_0\mathrm{diag}(m_u,m_d)$, and $\hat{\chi}_{+}=\chi_{+}-\frac{1}{2}\mathbf{\mathrm{Tr}}(\chi_{+})$. One can see that the structure of $\mathcal{L}_{\varphi\mathcal{H}}^{(1)}$ is very similar to the ones of $\pi N$ Lagrangians~\cite{Bernard:1995dp}. 

\begin{table}[htbp]
%\centering
\centering
\renewcommand{\arraystretch}{1.5}
\caption{The numerical values of the LECs in Eq.~\eqref{eq:nlopid} determined from the RSM (in units of GeV$^{-1}$). \change{The errors come from the parameters in the Lagrangians that quoted in the appendix.}\label{tab:numscii}}
\setlength{\tabcolsep}{1.1mm}
{
\begin{tabular}{ccccc}
\toprule[0.8pt]
$\tilde{c}_1$ & $\tilde{c}_2$ & $\tilde{c}_3$ & $\tilde{c}_4$ & $\tilde{c}_5$\\
$-0.21\pm0.07$ & $-0.83\pm0.16$ & $-0.55\pm0.18$ & $0.61\pm0.10$ & $0.26$\\
\bottomrule[0.8pt]
\end{tabular}
}
\end{table}

In literature, only the LECs in partial terms in Eq.~\eqref{eq:nlopid} were determined for certain problems (see Ref.~\cite{Meng:2022ozq}). Here, we again use the RSM to estimate the $\tilde{c}_i$. One can consult appendix~\ref{sec:app2} for details. The numerical values of the LECs $\tilde{c}_i~(i=1,\dots,5)$ in Eq.~\eqref{eq:nlopid} are summarized in Table~\ref{tab:numscii}. From Table~\ref{tab:numscii} one can see that the couplings of the subleading $\pi D^{(\ast)}$ vertices are of natural size and are much smaller than those of the $\pi N$ system~\cite{Fettes:1998ud,Krebs:2007rh}. In contrast to the $NN$ system, this makes the main contribution for the binding forces of $DD^\ast$ come from the short-range contact interactions.

The non-analytic terms of the subleading TPE potentials read

\begin{widetext}
\begin{eqnarray}
\mathcal{V}_{2\pi,c}^{(3)}&=&\frac{g_{\varphi}^{2}}{512\pi f_{\pi}^{4}}\bigg\{ (4m_{\pi}^{2}+3\bm{q}^2)\left[48\tilde{c}_{1}m_{\pi}^{2}-3\tilde{c}_{3}(2m_{\pi}^{2}+\bm{q}^2)-8\tilde{c}_{5}\delta_{i}^{2}\right]A(q)+\big\{ 3(-12\delta_{b}^{2}+12m_{\pi}^{2}+5\bm{q}^2)\nonumber\\
&&\times\left[16\tilde{c}_{1}m_{\pi}^{2}-2\tilde{c}_{2}\delta_{b}^{2}-\tilde{c}_{3}(2m_{\pi}^{2}+\bm{q}^2)\right]-8\tilde{c}_{5}\delta_{i}^{2}(4m_{\pi}^{2}-4\delta_{b}^{2}+\bm{q}^2)\big\} A^{\prime}(q)\nonumber\\
&&-\frac{8\delta_{b}}{\pi}\left[48\tilde{c}_{1}m_{\pi}^{2}+\tilde{c}_{2}(m_{\pi}^{2}-6\delta_{b}^{2}+\bm{q}^2)+4\tilde{c}_{3}\delta_{b}^{2}-3\tilde{c}_{3}\bm{q}^2+4\tilde{c}_{5}\delta_{i}^{2}\right]L(q)\bigg\},\label{eq:v2pic3}\\
\mathcal{W}_{2\pi,c}^{(3)}&=&\frac{g_{\varphi}^{2}}{32\pi f_{\pi}^{4}}\left\{ \tilde{c}_{4}\mathrm{sgn}(4m_{\pi}^{2}+\bm{q}^2)\bm{q}^2A(q)-2\tilde{c}_{5}\delta_{i}^{2}(2m_{\pi}^{2}-2\delta_{b}^{2}+\bm{q}^2)A^{\prime}(q)+\frac{6}{\pi}\tilde{c}_{5}\delta_{b}\delta_{i}^{2}L(q)\right\},\\
\mathcal{V}_{2\pi,t}^{(3)}&=&\frac{g_{\varphi}^{2}}{512\pi f_{\pi}^{4}\bm{q}^2}\bigg\{ (\bm{q}^2-4m_{\pi}^{2})\left[-48\tilde{c}_{1}m_{\pi}^{2}+3\tilde{c}_{3}(2m_{\pi}^{2}+\bm{q}^2)+8\tilde{c}_{5}\delta_{i}^{2}\right]A(q)-(4\delta_{b}^{2}-4m_{\pi}^{2}+\bm{q}^2)\nonumber\\
&&\times\left[-48\tilde{c}_{1}m_{\pi}^{2}+ 6\tilde{c}_{2}\delta_{b}^{2}+3\tilde{c}_{3}(2m_{\pi}^{2}+\bm{q}^2)+8\tilde{c}_{5}\delta_{i}^{2}\right]A^{\prime}(q)+\frac{4\delta_{b}}{\pi}\left[\tilde{c}_{2}\bm{q}^2-2m_{\pi}^{2}(\tilde{c}_{2}-6\tilde{c}_{3})\right]L(q)\bigg\},\\
\mathcal{W}_{2\pi,t}^{(3)}&=&\frac{g_{\varphi}^{2}}{32\pi f_{\pi}^{4}}\left[-\tilde{c}_{4}\mathrm{sgn}(4m_{\pi}^{2}+\bm{q}^2)A(q)\right],\label{eq:w2pit3}
\end{eqnarray}
where $\delta_i^2=m_{K^+}^2-m_{K^0}^2$ denotes the $u,d$ quark mass difference that stems from the $\hat{\chi}_+$ term in Eq.~\eqref{eq:nlopid}.

The LO and subleading TPE potentials can be obtained from the spectral function representation associating with the local momentum-space regularization~\cite{Reinert:2017usi},
\begin{eqnarray}
V_{2\pi,i}^{(\nu)}&=&\frac{2}{\pi}\exp\left(-\frac{\bm q^2}{2\Lambda^2}\right)\int_0^\infty d\mu\left[\mu\rho_{2\pi,i}^{(\nu)}(\mu)\right]\left[\frac{1}{\mu^2+\bm q^2}+\mathcal{C}_1(\mu,\Lambda)+\mathcal{C}_2(\mu,\Lambda)\bm{q}^2\right]\exp\left(-\frac{\mu^2}{2\Lambda^2}\right),\\
W_{2\pi,i}^{(\nu)}&=&\frac{2}{\pi}\exp\left(-\frac{\bm q^2}{2\Lambda^2}\right)\int_0^\infty d\mu\left[\mu\eta_{2\pi,i}^{(\nu)}(\mu)\right]\left[\frac{1}{\mu^2+\bm q^2}+\mathcal{C}_1(\mu,\Lambda)+\mathcal{C}_2(\mu,\Lambda)\bm{q}^2\right]\exp\left(-\frac{\mu^2}{2\Lambda^2}\right),~\label{eq:w2pi}
\end{eqnarray}
where $i=c,t$ denotes the central and tensor parts, respectively, while the $\nu$ is the chiral order defined in Eq.~\eqref{eq:powercounting}. The spectral functions $\rho_{2\pi,i}^{(\nu)}(\mu)$ and $\eta_{2\pi,i}^{(\nu)}(\mu)$ respectively read
\begin{eqnarray}\label{eq:spfun}
\rho_{2\pi,i}^{(\nu)}(\mu)&=&\Im\left[\mathcal{V}_{2\pi,i}^{(\nu)}(0^+-i\mu)\right],\qquad\qquad
\eta_{2\pi,i}^{(\nu)}(\mu)=\Im\left[\mathcal{W}_{2\pi,i}^{(\nu)}(0^+-i\mu)\right].
\end{eqnarray}
In order to get $\rho_{2\pi,i}^{(\nu)}(\mu)$ and $\eta_{2\pi,i}^{(\nu)}(\mu)$, one also needs the following quantities~\cite{Epelbaum:2002ji},
\begin{eqnarray}
\Im A(0^+-i\mu)&=&\frac{\pi}{4\mu}\Theta(\mu-2m_\pi),\\
\Im A^\prime(0^+-i\mu)&=&
\begin{cases}
\frac{\pi}{4\mu}\Theta(\mu-2m^\prime)&m_\pi>\delta_b\\
\frac{1}{2\mu}\arctan\frac{\mu}{2\sqrt{\delta_b^2-m_\pi^2}}&m_\pi<\delta_b\\
\end{cases},\label{eq:ImAp}\\
\Im L(0^+-i\mu)&=&-\pi\frac{\sqrt{\mu^2-4m_\pi^2}}{2\mu}\Theta(\mu-2m_\pi),
\end{eqnarray}
with $\Theta$ the Heaviside step function.

The subtraction terms $\mathcal{C}_1$ and $\mathcal{C}_2$ are introduced to minimize the mixture of the long- and short-range forces in TPE interactions. They are determined by the following requirements~\cite{Reinert:2017usi}
\begin{eqnarray}
V_{2\pi,i}^{(\nu)}(r)\big|_{r\to0}=\frac{d^2}{dr^2}V_{2\pi,i}^{(\nu)}(r)\big|_{r\to0}=W_{2\pi,i}^{(\nu)}(r)\big|_{r\to0}=\frac{d^2}{dr^2}W_{2\pi,i}^{(\nu)}(r)\big|_{r\to0}=0,
\end{eqnarray}
where $V_{2\pi,i}^{(\nu)}(r)$ and $W_{2\pi,i}^{(\nu)}(r)$ represent the corresponding potentials in $r$-space. They are obtained with the following Fourier transform
\begin{eqnarray}
V_{2\pi,i}^{(\nu)}(r)=\frac{1}{2\pi^2}\int dq q^2 j_0(qr)V_{2\pi,i}^{(\nu)}(q),\label{eq:Vcoordinate}
\end{eqnarray}
where the similar form holds for the $W_{2\pi,i}^{(\nu)}$, and $j_0(qr)$ represents the spherical Bessel function of the first kind. The expressions of $\mathcal{C}_1$ and $\mathcal{C}_2$ are given as
\begin{eqnarray}
\mathcal{C}_1(\mu,\Lambda)&=&\frac{\sqrt{2\pi}\mu \exp\left(\frac{\mu^{2}}{2\Lambda^{2}}\right)\left(5\Lambda^{2}+\mu^{2}\right)\mathrm{erfc}\left(\frac{\mu}{\sqrt{2}\Lambda}\right)-2\Lambda\left(4\Lambda^{2}+\mu^{2}\right)}{4\Lambda^{5}},\\
\mathcal{C}_2(\mu,\Lambda)&=&\frac{2\Lambda\left(2\Lambda^{2}+\mu^{2}\right)-\sqrt{2\pi}\mu \exp\left(\frac{\mu^{2}}{2\Lambda^{2}}\right)\left(3\Lambda^{2}+\mu^{2}\right)\mathrm{erfc}\left(\frac{\mu}{\sqrt{2}\Lambda}\right)}{12\Lambda^{7}}.
\end{eqnarray}
\end{widetext}

\section{Numerical results and discussions}\label{sec:numsanddis}

In this section, we will first work out the TPE potentials in the coordinate space. We will compare their asymptotic behaviors at long distance $r$ with that from lattice QCD, specifically, HAL QCD method (see Sec.~\ref{sec:iiia}). We will also analyze the contributions from the contact, OPE and TPE interactions at each order (see Sec.~\ref{sec:ana}). In Sec.~\ref{sec:poletraj}, we will study the pole trajectories of the $DD^\ast$ bound state in two cases.

\subsection{TPE potentials in the coordinate space}\label{sec:iiia}

In the following, we first analyze the analytic structures of the TPE potentials in $r$-space. We will take the elements $W_{2\pi,c}^{(2)}$ and $W_{2\pi,c}^{(3)}$ as examples. Here, we resort to the inverse Fourier transform, in which the $r$-space potential is represented with a continuous superposition of Yukawa functions~\cite{Ericson:1988gk}. It can be formulated as
\begin{eqnarray}
    \tilde{V}(r)=\frac{2}{\pi}\int_{2m_{\pi}}^{\infty}\mu\rho(\mu)d\mu\int\frac{d^{3}q}{(2\pi)^{3}}\frac{e^{i\bm{q}\cdot\bm{r}}}{\mu^{2}+\bm{q}^{2}},
\end{eqnarray}
where $\rho(\mu)=\Im\left[\mathcal{V}(0^+-i\mu)\right]$ denotes the spectral function of the corresponding potential, e.g., see Eq.~\eqref{eq:spfun}. 

Since we are more interested in the long-range behavior of the potentials, we neglect the regulators and short-range subtractions in Eq.~\eqref{eq:w2pi} at first, which have no effects on the asymptotic behaviors. We call it the unregularized spectral method. In order to distinguish the potentials with those in the local regularization, we use the overhead tilde to denote the $r$-space potentials from unregularized spectral method. With the residue theorem, one can get the following form for the central potential,
\begin{eqnarray}\label{eq:Vrsp}
    \tilde{V}_c(r)=\frac{1}{2\pi^2r}\int_{2m_\pi}^\infty \mu e^{-\mu r}\rho(\mu)d\mu.
\end{eqnarray}
Inserting the $\eta_{2\pi,c}^{(\nu)}$ [in Eq.~\eqref{eq:spfun}] into Eq.~\eqref{eq:Vrsp} and integrating over $\mu$ with some assists of the integral representation of the modified Bessel function\footnote{$K_{n}(y)=\frac{\sqrt{\pi}(y/2)^{n}}{\Gamma(n+1/2)}\int_{0}^{\infty}(\sinh t)^{2n}\exp(-y\cosh t)dt,\quad y>0.$}, one then obtains
\begin{widetext}
\begin{eqnarray}
    \tilde{W}_{2\pi,c}^{(2)}(r)&=&\frac{1}{384\pi^{2}f_{\pi}^{4}}\Bigg\{ \frac{e^{-2x}}{r^{6}}\frac{3g_{\varphi}^{4}}{2\delta_{b}}(2x^{4}+4x^{3}+6x^{2}+6x+3)+\frac{e^{-2x}}{r^{2}}\frac{3g_{\varphi}^{2}}{2\delta_{b}}(m_{\pi}^{2}-\delta_{b}^{2})\left[\delta_{b}^{2}+g_{\varphi}^{2}(m_{\pi}^{2}-\delta_{b}^{2})\right]\nonumber\\
    &&+\frac{e^{-2x}}{r^{4}}\frac{1}{2}(x^{2}+x+\frac{1}{2})\left[g_{\varphi}^{4}(15\delta_{b}-3m_{\pi}^{2}/\delta_{b})-9\delta_{b}g_{\varphi}^{2}\right]-\frac{1}{r}\left[\frac{3K_{2}(2x)}{4x^{2}}+\frac{K_{1}(2x)}{2x}\right]\frac{4m_{\pi}^{4}}{\pi}(23g_{\varphi}^{4}-10g_{\varphi}^{2}-1)\nonumber\\
    &&-\frac{1}{r^{2}}K_{1}(2x)\frac{2m_{\pi}}{\pi}\left[4\delta_{b}^{2}g_{\varphi}^{2}(5g_{\varphi}^{2}-3)+(-5g_{\varphi}^{4}+4g_{\varphi}^{2}+1)m_{\pi}^{2}\right]\Bigg\},\label{eq:w2picr2}\\ 
    \tilde{W}_{2\pi,c}^{(3)}(r)&=&\frac{1}{32\pi^{2}f_{\pi}^{4}}\Bigg[\frac{e^{-2x}}{r^{6}}\tilde{c}_{4}g_{\varphi}^{2}(2x^{4}+4x^{3}+6x^{2}+6x+3)-\frac{e^{-2x}}{r^{2}}\tilde{c}_{5}\delta_{i}^{2}g_{\varphi}^{2}(m_{\pi}^{2}-\delta_{b}^{2})/2\nonumber\\
    &&-\frac{e^{-2x}}{r^{4}}g_{\varphi}^{2}\left(x^{2}+x+\frac{1}{2}\right)(2\tilde{c}_{4}m_{\pi}^{2}-\tilde{c}_{5}\delta_{i}^{2})-\frac{K_{1}(2x)}{r^{2}}3\tilde{c}_{5}\delta_{b}\delta_{i}^{2}g_{\varphi}^{2}m_{\pi}/\pi\Bigg]\label{eq:w2picr3},
\end{eqnarray}
\end{widetext}
where $x=m_\pi r$, and $K_n(y)$ are the modified Bessel function of the second kind. Note that in order to detour the complicate integrals involving the arctangent function in Eq.~\eqref{eq:ImAp} when $\delta_b>m_\pi$, we have used the expression for $\delta_b<m_\pi$ in deriving the Eqs.~\eqref{eq:w2picr2} and \eqref{eq:w2picr3}, which become true at unphysical pion mass used by HAL QCD simulation~\cite{Lyu:2022imf,Lyu:2023xro}. The nonphysical hadron masses used in the lattice QCD simulations read,
\begin{eqnarray}
      &m_\pi=146.4\text{ MeV},\;m_D=1878.2\text{ MeV}, \nonumber\\
      &m_{D^\ast}=2018.1\text{ MeV} . 
\end{eqnarray}

We focus on the range of $1<r<2\text{ fm}$ where were stressed in Ref.~\cite{Lyu:2023xro}. In this range, for the typical dimensionless variable $2x=2m_\pi r$ in Eqs.~\eqref{eq:w2picr2} and \eqref{eq:w2picr3}, there is $2m_\pi r\in (1.5,3)$.  The Eqs.~\eqref{eq:w2picr2} and \eqref{eq:w2picr3} can be generally written as
\begin{eqnarray}\label{eq:w2pic23ab}
    \tilde{W}_{2\pi,c}^{(\nu)}(r)=\frac{e^{-2x}}{(2x)^2}\left[\sum_{i=0}^4a_i^{(\nu)}\frac{1}{(2x)^i}+\sum_{j=1/2}^\infty b_j^{(\nu)}\frac{1}{(2x)^j}\right],
\end{eqnarray}
where $a_i^{(\nu)},b_j^{(\nu)}$ are the corresponding constants that can be deduced from Eqs.~\eqref{eq:w2picr2} and \eqref{eq:w2picr3}. We have used the following expansions for $K_{1,2}(y)$ for $y>1$
\begin{eqnarray}
    K_{1}(y)&=&\sqrt{\frac{\pi}{2}}e^{-y}\left(\frac{1}{y^{1/2}}+\frac{3}{8}\frac{1}{y^{3/2}}+\dots\right),\\
K_{2}(y)&=&\sqrt{\frac{\pi}{2}}e^{-y}\left(\frac{1}{y^{1/2}}+\frac{15}{8}\frac{1}{y^{3/2}}+\dots\right).
\end{eqnarray}

If ideally the $r$ is so large that $2x\gg1$, then one obtains the following asymptotic behavior
\begin{eqnarray}\label{eq:w2pic23ab1}
    \tilde{W}_{2\pi,c}^{(\nu)}(r)\propto \frac{e^{-2m_\pi r}}{r^2}.~\label{eq:r2TPE}
\end{eqnarray}
It is the same with that of lattice QCD result in the range $1 <r<2\text{ fm}$. It should be noticed that the asymptotic behavior at large distance is slightly different with that of $NN$, which has been found to be $e^{-2m_\pi r}/r^{3\over 2}$ long ago~\cite{Kaiser:1997mw}. This difference arises from the box diagrams. For the $NN$ system, the box diagrams with $NN$ as intermediate states are subtracted. For the $DD^*$ scattering, although the contribution of $DD^*$ in the box diagram is subtracted, the box diagram with $DD$, $DD^*$ and $DD\pi$ as intermediate states are kept, which have no counter part in $NN$ case. 

\begin{figure}[htb]
\begin{center}
\includegraphics[width=\columnwidth]{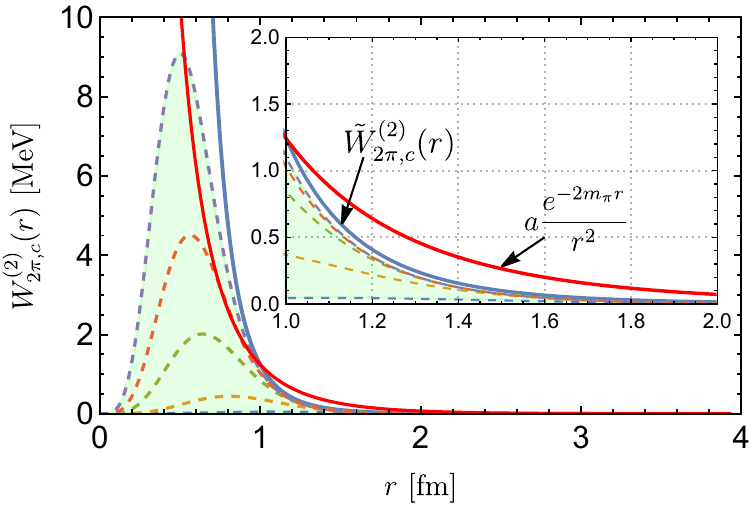}
\caption{The potentials $W_{2\pi,c}^{(2)}(r)$ and $\tilde{W}_{2\pi,c}^{(2)}(r)$ in the coordinate space. The blue and red solid lines denote the results of $\tilde{W}_{2\pi,c}^{(2)}(r)$ and the asymptotic function $ae^{-2m_\pi r}/r^2$ ($a\simeq 5.4$ MeV$\cdot$fm$^2$), respectively. The band represents the result from the local momentum-space regularization with the cutoff $\Lambda\in[400,900]$ MeV. The dashed lines by the up direction denote the results with $\Lambda=400,550,700,800,900$ MeV, in order.\label{fig:dl}}
\end{center}
\end{figure}

However, the $1.5<2x<3.0$  are not large enough to neglect the subleading effect in Eq.~\eqref{eq:w2pic23ab}. In Fig.~\ref{fig:dl}, we present the TPE potential from unregularized spectral method. As a comparison, we also present a $ae^{-2m_\pi r}/r^2$ function with $a$ making the function cross with the potential from unregularized spectral method at $r=1$ fm. We also try to vary $a$ but fail to use the $ae^{-2m_\pi r}/r^2$ function to depict the TPE potential from the unregularized spectral method. The terms with higher power of $r$ in the denominator, i.e. $e^{-2m_\pi r}/r^n$ with $n>2$ could be also important. Therefore, the $\chi$EFT calculations supports the significance of $ae^{-2m_\pi r}/r^2$ behavior of the long-range TPE interaction, but this behavior is not dominated in the range $1 <r<2$\text{ fm}.

 In Fig.~\ref{fig:dl}, we also present TPE potentials from local momentum-space regularization . One can see that the line shape of $W_{2\pi,c}^{(2)}(r)$ ($r>0$) in local momentum-space regularization will gradually approach that in dimensional regularization with the increasing of cutoff. This is because the potentials in these two regularization schemes differ from each other by an infinite series of higher-order contact interactions, e.g., see more discussions in Ref.~\cite{Epelbaum:2005pn}. It should be noticed that in local momentum-space regularization, the TPE behaviors in the range $1 <r<2$\text{ fm} will be distorted by the regulators and depend on the cutoff $\Lambda$. 

\subsection{Analyses of the contact, OPE and TPE contributions}\label{sec:ana}

We quantitatively analyze the behaviors of the contact, OPE and TPE interactions, respectively. We take the behavior of the quantity $W_{i,c}^{(\nu)}$ in coordinate space as an example. The element $W_{i,c}^{(\nu)}$ in Eq.~\eqref{eq:vi} is multiplied by the isospin factor $\bm{\tau}_1\cdot\bm{\tau}_2$, and $\langle\bm{\tau}_1\cdot\bm{\tau}_2\rangle=-3$ for $I=0$.
 It should be stressed that the analysis will depend on the scheme to separate the contact interaction and pion-exchange interactions.  The requirement $W_{i,c}^{(\nu)}(r)|_{r\to 0}=0$ in local momentum space regularization minimizes the mixing of the (intermediate) long- and short-range forces. Thus, the so-called OPE and TPE interactions in local momentum space regularization are in fact parts of their effects that can not be compensated by the contact terms. 

\begin{figure}[htb]
\begin{center}
\begin{minipage}[t]{0.505\linewidth}
\centering
\includegraphics[width=\columnwidth]{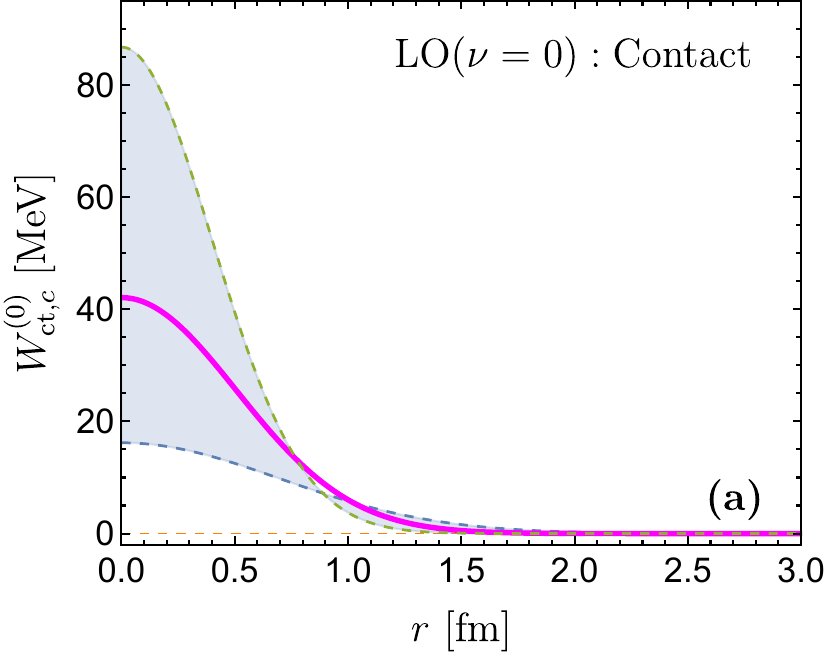}
\end{minipage}%
%\hspace{-0.005in}
\begin{minipage}[t]{0.525\linewidth}
\centering
\includegraphics[width=\columnwidth]{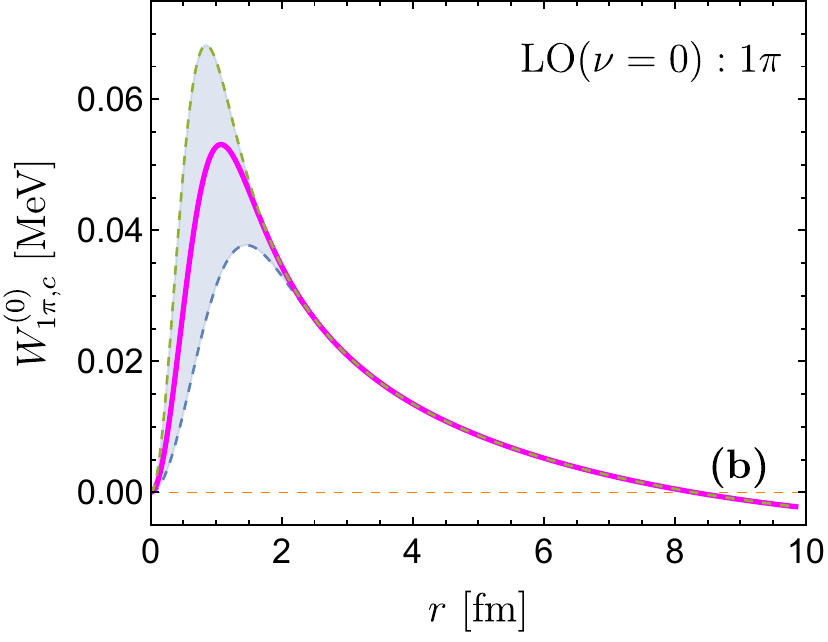}
\end{minipage}
\\
\begin{minipage}[t]{0.51\linewidth}
\centering
\includegraphics[width=\columnwidth]{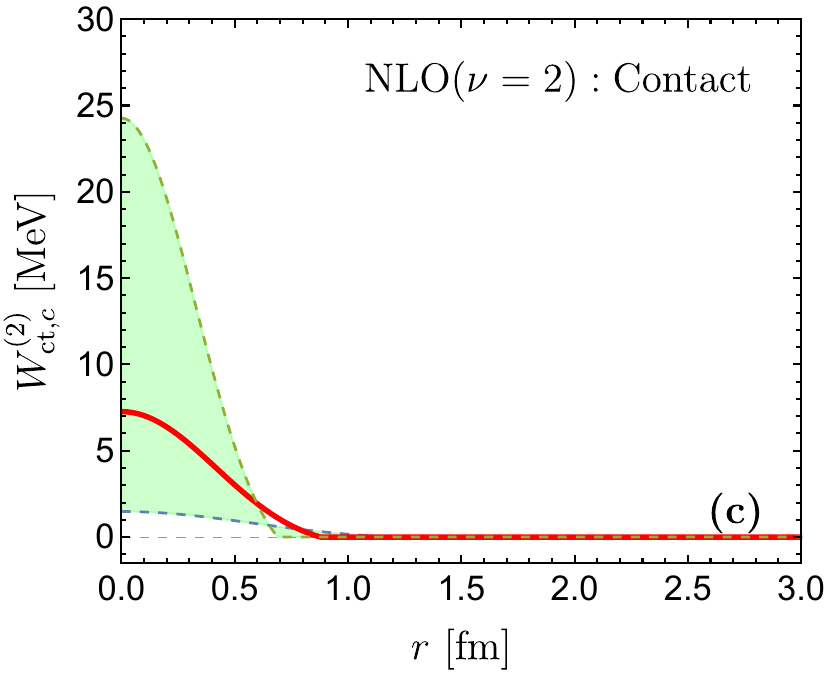}
\end{minipage}%
%\hspace{0.005in}
\begin{minipage}[t]{0.51\linewidth}
\centering
\includegraphics[width=\columnwidth]{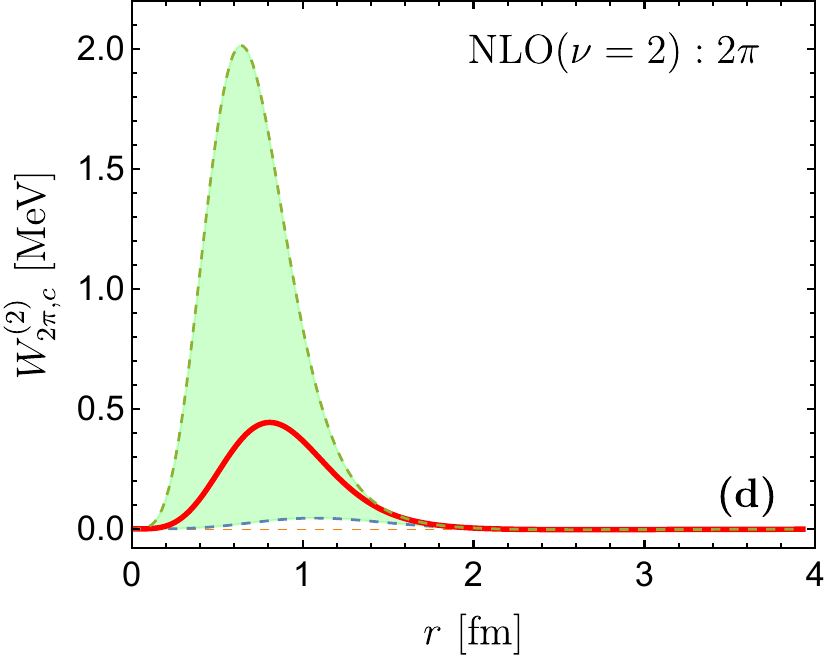}
\end{minipage}
%\hspace{0.068in}
\\
\begin{minipage}[t]{0.515\linewidth}
\centering
\includegraphics[width=\columnwidth]{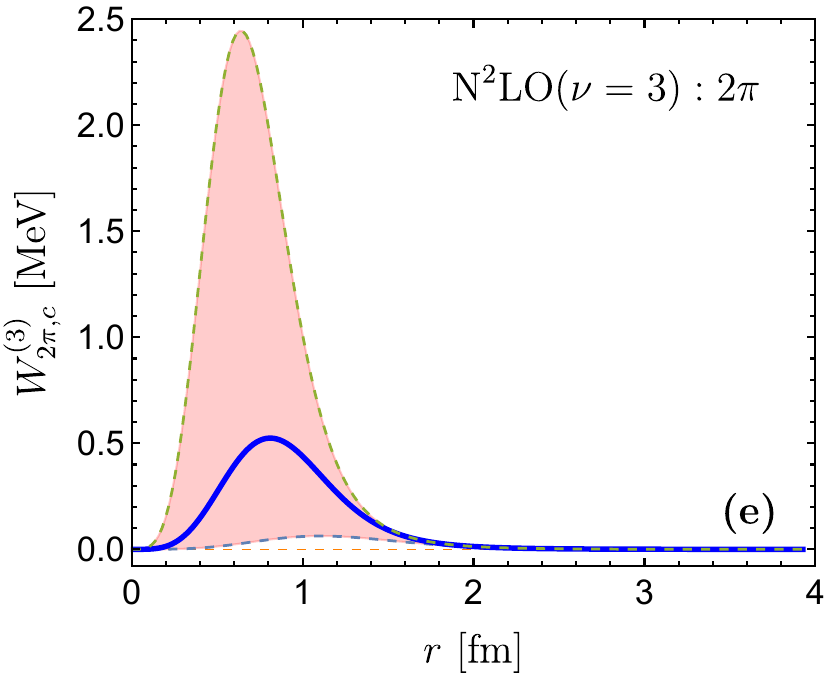}
\end{minipage}%
%\hspace{0.005in}
\begin{minipage}[t]{0.51\linewidth}
\centering
\includegraphics[width=\columnwidth]{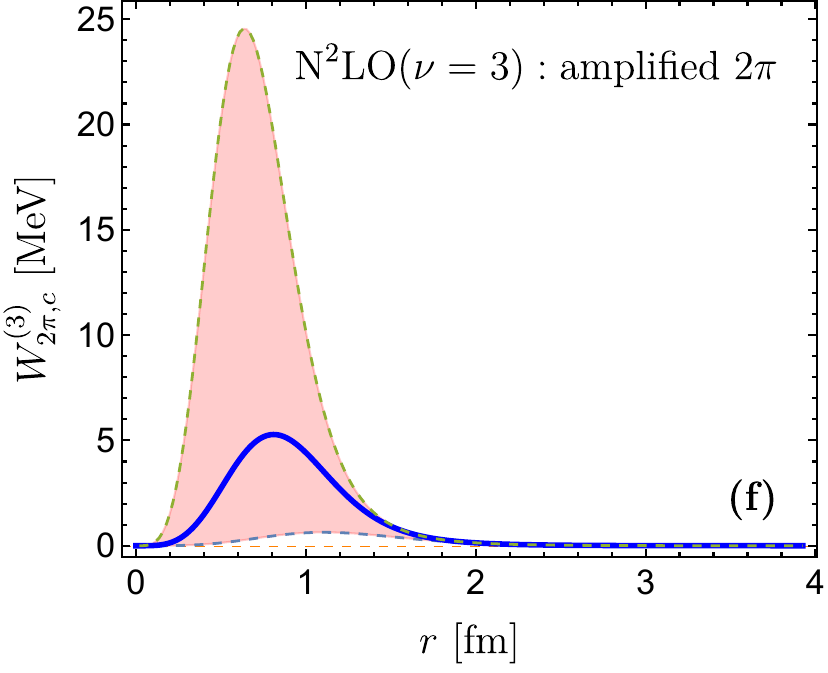}
\end{minipage}
\caption{The behaviors of the $W_{i,c}^{(\nu)}~(i=\mathrm{ct},1\pi,2\pi)$ in coordinate space. The first, second, and third rows denote the LO, NLO and N$^2$LO contributions, respectively. The shaded areas represent the regions where the cutoff is taken in the range $[400,700]$ MeV, while the solid lines stand for the results with $\Lambda=550$ MeV. Figure (f) represents the results that the LECs $\tilde{c}_i~(i=1,\dots,4)$ are artificially magnified by a factor of ten. \label{fig:amppot}}
\end{center}
\end{figure}

In Fig.~\ref{fig:amppot}, we show the behaviors of the $W_{i,c}^{(\nu)}(r)$ with the cutoff ranging in $[400,700]$ MeV:
\begin{enumerate}[label=(\arabic*)]
    \item In Figs.~\ref{fig:amppot} (a) and (b), we show the LO contact and OPE parts, respectively. One can see that as expected the short-range ($r\lesssim 1~\mathrm{fm}$) and long-range ($r\gtrsim 2~\mathrm{fm}$) behaviors of $DD^\ast$ potential are dominated by the contact and OPE interactions respectively, but the $W_{1\pi,c}^{(0)}(r)$ is much weaker than the $W_{\mathrm{ct},c}^{(0)}(r)$. One important reason is that at least for the S-wave, the long-range part central interaction in Eq.~\eqref{eq:opesub} is suppressed by the minor value of the effective mass $u_\pi$. This suppression directly leads to a new expansion of the $D\bar{D}^\ast/DD^\ast$ interactions~\cite{Fleming:2007rp} with perturbative OPE interaction. 
    
    \item In Figs.~\ref{fig:amppot} (c) and (d), we display the $W_{\mathrm{ct},c}^{(2)}(r)$ and $W_{2\pi,c}^{(2)}(r)$ contributions at NLO. One can notice that the contact and TPE interactions dominate the short-range ($r\lesssim 1~\mathrm{fm}$) and intermediate-range ($1\lesssim r\lesssim 2~\mathrm{fm}$) forces, respectively. Similarly, the strength of $W_{2\pi,c}^{(2)}(r)$ is weaker than that of the $W_{\mathrm{ct},c}^{(2)}(r)$.
    
    \item In Fig.~\ref{fig:amppot} (e), we illustrate the subleading $W_{2\pi,c}^{(3)}(r)$ contribution. One sees that its size and behavior are very similar to the $W_{2\pi,c}^{(2)}(r)$. This is because the LECs in Eq.~\eqref{eq:nlopid} determined from the RSM are of natural size (see Table~\ref{tab:numscii}) and we only focus on the medium-long character of the TPE interactions within the local momentum-space regularization. The values of $\tilde{c}_i$ are about one order of magnitude smaller than those of the $\pi N$ system, which makes the subleading TPE contributions in $DD^\ast$ system much moderate.
    
    \item In Fig.~\ref{fig:amppot} (f), we also illustrate the $W_{2\pi,c}^{(3)}(r)$ with the artificially magnified (by a factor of ten) LECs $\tilde{c}_i$. One sees that, in this case, the contrived $W_{2\pi,c}^{(3)}(r)$ is also amplified about ten times, and its magnitude is comparable with the $W_{\mathrm{ct},c}^{(2)}(r)$. This corresponds to the unnatural case in the $NN$ system.
\end{enumerate}

An overview of the contents in Fig.~\ref{fig:amppot} can be summarized as the following aspects:
\begin{itemize}
  \item Long-range force ($r\gtrsim 2~\mathrm{fm}$) is dominated by the OPE;
  \item Intermediate-range force ($1\lesssim r\lesssim 2~\mathrm{fm}$) is dominated by the TPE;
  \item Short-range force ($r\lesssim 1~\mathrm{fm}$) is dominated by the contact interaction;
  \item The short-range force is much stronger than the long- and intermediate-range ones;
  \item The OPE interaction is the weakest one, this may answer the question raised in Ref.~\cite{Lyu:2023xro}: {\it why the theoretically possible one-pion exchange contribution cannot been seen in the lattice data}.
  \item The requirement $W_{i,c}^{(\nu)}(r)|_{r\to 0}=0$ in local momentum space regularization minimizes the mixing of the (intermediate) long- and short-range forces.
\end{itemize}

\begin{figure*}[htb]
\begin{center}
\begin{minipage}[t]{0.333\linewidth}
\centering
\includegraphics[width=\columnwidth]{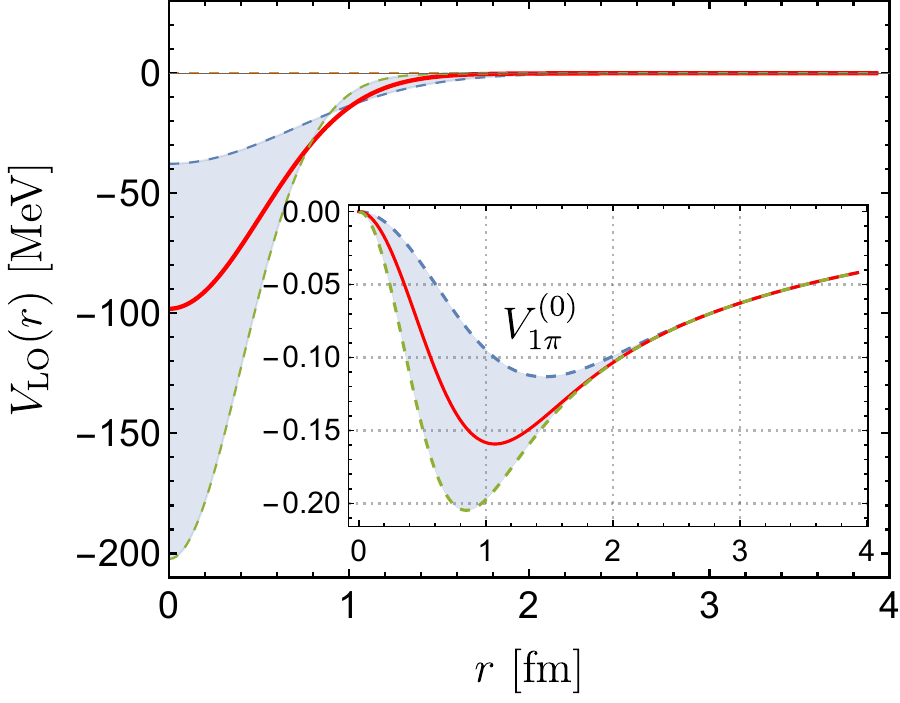}
\end{minipage}%
%\hspace{-0.005in}
\begin{minipage}[t]{0.325\linewidth}
\centering
\includegraphics[width=\columnwidth]{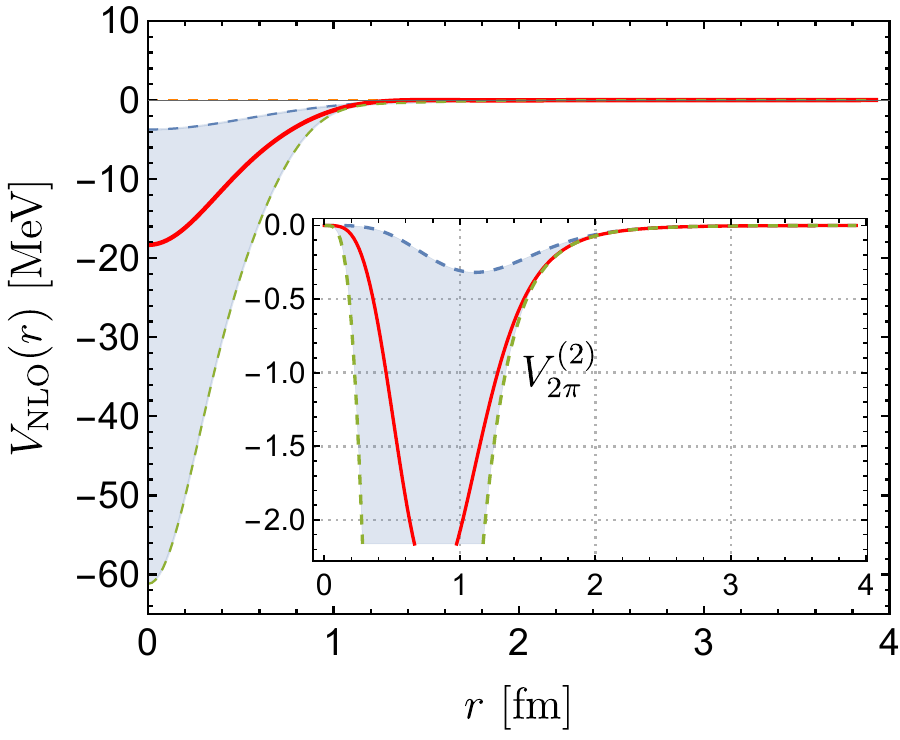}
\end{minipage}
\begin{minipage}[t]{0.325\linewidth}
\centering
\includegraphics[width=\columnwidth]{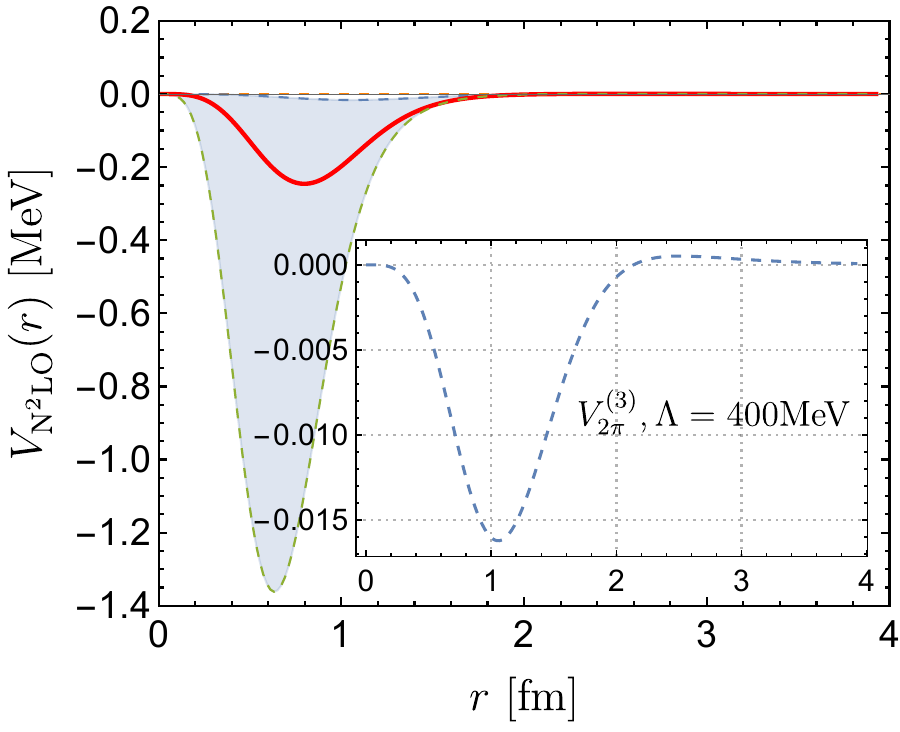}
\end{minipage}%
\caption{The panels from left to right show the LO ($V_{\rm ct}^{(0)}+V_{1\pi}^{(0)}$), NLO ($V_{\rm ct}^{(2)}+V_{2\pi}^{(2)}$), and N$^2$LO ($V_{2\pi}^{(3)}$) potentials for the isoscalar channel, in order. The bands represent the regions where the cutoff is taken in the range $[400,700]$ MeV, while
the solid lines stand for the results with $\Lambda=550$ MeV. The potentials become deeper with the increasing of cutoff. \label{fig:VLONLONNLO}}
\end{center}
\end{figure*}

In Fig.~\eqref{fig:VLONLONNLO}, we show the effective potentials of the isoscalar channel at each chiral order. The cutoff is taken in the range $[400,700]$ MeV. One sees that the potential becomes weaker with the increasing of chiral order. This implies that the chiral expansion works well in our study. The potentials at LO, NLO and N$^2$LO are all attractive, and the attraction mainly comes from the short-range forces (contact interactions). The attractive force in the isoscalar channel may lead to bound state. Thus, the next subsection is devoted to studying the pole trajectory of the $DD^\ast$ bound state.

%The scarce scattering data in $DD^\ast$ system is not enough to justify which regularization scheme is more suitable to date. For example, the binding energy alone cannot fairly serve as a criterion to check this, because it can be easily reproduced via introducing an appropriate potential. A lesson from the $NN$ interaction tells that it is better to study the $DD^\ast$ scattering phase shifts in the lower and peripheral partial waves (there have been some calculations for the S-wave from the lattice QCD~\cite{Padmanath:2022cvl,Chen:2022vpo,Lyu:2023xro}, but the data points are still not enough to make a thorough fitting). Though the local momentum-space regularization and dimensional regularization schemes might give similar results in $DD^\ast$ system (because the contributions from the subleading TPE are moderate), we will stick to the former in this study considering its virtue in dealing with the $NN$ interactions~\cite{Reinert:2017usi}.

\subsection{Pole trajectory of the $DD^\ast$ bound state}\label{sec:poletraj}

We use the isospin average mass for $DD^\ast$ system in our calculations. The threshold of $DD^\ast$ and the experimentally measured mass and width of $\Tcc$~\cite{LHCb:2021auc} are given as
\begin{eqnarray}\label{eq:nums}
m_{\mathrm{th}}&=&3875.8~\mathrm{MeV},\nonumber\\
m_{\mathrm{exp}}&\simeq&(m_{D^{\ast+}}+m_{D^0})-0.36=3874.7~\mathrm{MeV},\nonumber\\
\Gamma_{\mathrm{exp}}&=&48\pm2^{+0}_{-14}~\mathrm{keV}.
\end{eqnarray}

With the effective potentials in local momentum-space regularization, we solve the LSEs to analyze the pole distributions in the physical Riemann sheet. The LSE in the partial wave basis $|\ell sj\rangle$ reads
\begin{eqnarray}\label{eq:lses}
T_{\ell^\prime\ell sj}(p^\prime,p)&=&V_{\ell^\prime\ell sj}(p^\prime,p)
+\sum_{\ell^{\prime\prime}}\int\frac{k^2dk}{(2\pi)^3}V_{\ell^\prime\ell^{\prime\prime} sj}(p^\prime,k)\nonumber\\
&&\times\frac{2\mu_{DD^\ast}}{p^2-k^2+i\epsilon}T_{\ell^{\prime\prime}\ell sj}(k,p),
\end{eqnarray}
where $\mu_{DD^\ast}$ denotes the reduced mass of $DD^\ast$, and $p^2=2\mu_{DD^\ast}(E-m_{\rm th})$, with $E$ the total energy of $DD^\ast$ system. 
$V_{\ell^\prime\ell sj}(p^\prime,p)=\langle \ell^\prime sj|V(\bm q)|\ell sj\rangle$ can be easily obtained with the approach in Ref.~\cite{Golak:2009ri}. The S-D wave coupling is considered in our calculations. Thus, the $V_{\ell^\prime\ell sj}(p^\prime,p)$ is given with the $2\times2$ matrix in the coupled-channel $|\ell sj\rangle$ basis.

The finite width of $D^\ast$ meson will be considered in $m_{\rm th}$ via using a complex mass, i.e., $m_{D^\ast}-i\Gamma_{D^\ast}^{\rm eff}/2$. The width of $D^\ast$ is about several tens of keV, such as $\Gamma_{D^{\ast+}}\simeq 83.4$ keV~\cite{Workman:2022ynf} and $\Gamma_{D^{\ast0}}\simeq 55.6$ keV~\cite{Meng:2022ozq}. The width of $\Tcc$ is narrow $\sim50$ keV, thus its width should strongly depend on the $\Gamma_{D^\ast}$ if it is indeed the molecule of $DD^\ast$. In principle, the $\Gamma_{D^\ast}$ is a distribution with respect to the energy $E$, e.g., see Ref.~\cite{Du:2021zzh}. Here, we use an effective width of $D^\ast$, the $\Gamma_{D^\ast}^{\rm eff}$ in our calculations. We then tune the $\Gamma_{D^\ast}^{\rm eff}$ to reproduce the width of $\Tcc$. The value of $\Gamma_{D^\ast}^{\rm eff}$ should be close to the $\Gamma_{D^{\ast+,0}}$ as naively expected.

\begin{figure}[!htbp]
\begin{center}
\includegraphics[width=\columnwidth]{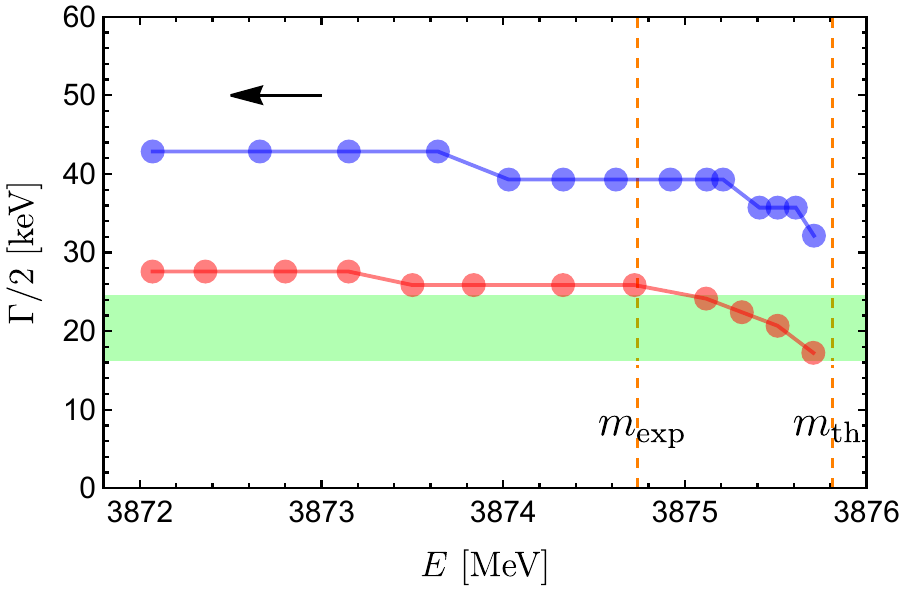}
\caption{Pole trajectory of the $DD^\ast$ bound state in the physical Riemann sheet with the change of cutoff $\Lambda\in[520,600]$ MeV. The arrow denotes the direction that the cutoff becomes larger. The two dashed vertical lines from left to right represent the experimentally measured mass of $T_{cc}^+$ and the threshold of $DD^\ast$ in order. The band denotes the measured width of $\Tcc$~\cite{LHCb:2021auc}.\label{fig:pole}}
\end{center}
\end{figure}

We first study the pole trajectory of $DD^\ast$ scattering T-matrix in the isoscalar channel with the contact interactions being kept up to NLO [including the Eqs.~\eqref{eq:vcto0}-\eqref{eq:vcto2} in effective potentials], and the cutoff is in the range $520-600$ MeV. We notice that the binding solution begins to appear when $\Lambda\approx600$ MeV in this case. The pole trajectory is very similar to the case in which the contact interaction is kept up to N$^2$LO [including the Eqs.~\eqref{eq:vcto0}-\eqref{eq:vcto4} in effective potentials], and the result in this case is shown in Fig.~\ref{fig:pole}. In this case, the pole appears when $\Lambda\approx520$ MeV, and the pole mass approaches the experimental value when $\Lambda\simeq560$ MeV. From Fig.~\eqref{fig:pole}, one also sees that the binding becomes deeper with the increasing of cutoff, while the half-width is not very insensitive to the cutoff.

The $DD^{\ast}$ lies above the threshold of the three-body $DD\pi$, thus there are two types of three-body cuts (see discussions in Ref.~\cite{Meng:2022ozq}): one comes from the OPE and another one comes from the self-energy correction of $D^\ast$ (this will contribute a finite width to $D^\ast$ meson). The first one is accounted for in Eq.~\eqref{eq:opeporig} using the static OPE potential. Our calculation contains two cases:
\begin{enumerate}[label=(\arabic*)]
\item We do not consider the width of $D^\ast$ in the propagator of Eq.~\eqref{eq:lses}, i.e., setting $\Gamma_{D^\ast}^{\rm eff}=0$ keV. The result in this case is shown as the blue dots in Fig.~\ref{fig:pole}. One can see that the half-width of $DD^\ast$ bound state is about two times larger than the $\Gamma_{\rm exp}/2$ [see Eq.~\eqref{eq:nums}]. This is somewhat not consistent with the experimental data~\cite{LHCb:2021auc} as well as the theoretical calculations~\cite{Meng:2021jnw,Ling:2021bir,Yan:2021wdl}.
\item We take the effective contribution of $D^\ast$ width into account. We noticed that we can reproduce the $\Tcc$ width when taking $\Gamma_{D^\ast}^{\rm eff}\sim30$ keV, and the result is shown as the red dots in Fig.~\ref{fig:pole}. The value of $\Gamma_{D^\ast}^{\rm eff}$ is close to the $\Gamma_{D^{\ast+,0}}$ as priorly expected.
\end{enumerate}
The calculations indicate that one has to consider the complete three-body effect for the $DD^\ast$ scattering. One can consult Ref.~\cite{Du:2021zzh} for a more formal formulation of the three-body dynamics in $\Tcc$.

In addition, we also investigated the situation in isovector channel, but we did not find binding solutions in this channel. This is consistent with the experimental facts---there are no structures in the $D^+D^0\pi^+$ invariant mass spectrum.

\section{Summary}\label{sec:sum}

We revisit the $DD^\ast$ interactions within the \xEFT~up to the third order. The pion-exchanged interactions are carefully treated with the local momentum-space regularization, in which their short-range components are subtracted via demanding the pion-exchanged contributions vanish at the origin in the coordinate space. This is consistent with the new developments of nuclear forces in Ref.~\cite{Reinert:2017usi}.

The contact interactions and the subleading $\pi D^{(\ast)}$ vertices are ascribed to the heavier meson exchanging, and consequently the LECs are estimated with the resonance saturation model. We notice that the subleading $\pi D^{(\ast)}$ couplings are much smaller that those in the $\pi N$ system, which makes the binding force of $DD^\ast$ mainly come from the short-range part. This is very different with that of the $NN$ system.

We study the analytic expression of the TPE interactions in coordinate space, and we find that its asymptotic behavior at long distance is similar but slightly different with that of the $NN$ forces. Along this line,  we get the asymptotic behavior $ae^{-2m_\pi r}/r^2$ of the TPE interaction in the long-range limit. However, for the range $1 <r<2$\text{ fm} where HAL QCD obtained the above behavior, our calculations imply that $ae^{-2m_\pi r}/r^n$ with $n>2$ behavior are also very important. We also analyze the contributions of the contact, OPE and TPE interactions at each order by defining pion interactions vanishing at origin. We notice that the contact interaction is much stronger than the OPE and TPE, which means the medium- and long-range parts of pion-exchange interaction are weak. 

We investigate the pole trajectory of $DD^\ast$ scattering T-matrix without and with considering the complete three-body effects, respectively. The width of $DD^\ast$ bound state would be two times larger than that of $\Tcc$ if not considering the effective contribution of $D^\ast$ width, while the width of $\Tcc$ can be reproduced once the complete three-body effects is considered. The binding solution only exists in the isoscalar channel, and this is consistent with the experimental data. Our calculation favors the molecular explanation of $\Tcc$.

\section*{Acknowledgement}
B.W is very grateful to Prof. Shi-Lin Zhu for helpful discussions and carefully reading the manuscript. This work is supported by the National Natural Science Foundation of China under Grants No. 12105072, the Youth Funds of Hebei Province (No. A2021201027) and the Start-up Funds for Young Talents of Hebei University (No. 521100221021). This project is also funded by the Deutsche Forschungsgemeinschaft (DFG, German Research Foundation, Project ID 196253076-TRR 110).

\begin{appendix}
\section{Estimations of the LECs}\label{sec:app}
\subsection{The LECs in contact interactions}\label{sec:app1}
In what follows, we list the coupling Lagrangians of the resonances with the ($D,D^\ast$) doublet under the heavy quark symmetry, and estimate the LECs in Eqs.~\eqref{eq:vcto0}-\eqref{eq:vcto4}.
\begin{itemize}

  \item Scalar exchange---$\sigma,a_0,f_0$ mesons

  The corresponding Lagrangians read
  \begin{eqnarray}
  \mathcal{L}_{\sigma\mathcal{H}}&=&g_{\sigma}\langle\mathcal{H}\bar{\mathcal{H}}\rangle\sigma,\label{eq:laghsigma}\\
  \mathcal{L}_{S\mathcal{H}}&=&g_{s}\langle\mathcal{H}S\bar{\mathcal{H}}\rangle,
  \end{eqnarray}
  where $g_\sigma$ and $g_s$ are the corresponding coupling constants. In the SU(2) case, $g_s=\sqrt{2}g_\sigma$ in the large-$N_c$ limit~\cite{Ecker:1988te}. The matrix form of $S$ is given as
  \begin{eqnarray}
  S&=&\left[\begin{array}{cc}
\frac{a_{0}^{0}}{\sqrt{2}}+\frac{f_{0}}{\sqrt{2}} & a_{0}^{+}\\
a_{0}^{-} & -\frac{a_{0}^{0}}{\sqrt{2}}+\frac{f_{0}}{\sqrt{2}}
\end{array}\right].
\end{eqnarray}

Within the parity-doubling model~\cite{Bardeen:2003kt}, the $g_\sigma$ reads
\begin{eqnarray}\label{eq:gsgp}
g_\sigma=-\frac{g_{\pi}}{2\sqrt{6}},\qquad g_{\pi}=\frac{\Delta}{f_{\pi}},
\end{eqnarray}
where $f_\pi=92.4$ MeV is the pion decay constant, and $\Delta=m_{0^+}-m_{0^-}$ denotes the mass difference of the $J^P=0^+$ and $0^-$ charmed mesons. In most previous studies, such as the one-boson exchange model, the $\Delta =m_{D_{s0}^\ast(2317)}-m_{D_s}\approx350$ MeV was usually used as the original work~\cite{Bardeen:2003kt}. In the SU(2) case in this study, we chose $\Delta=m_{D_{0}^\ast(2300)}-m_{D}$. The nature of the $D_{0}^\ast(2300)$ is still in controversy (one can consult the recent review~\cite{Meng:2022ozq} for more details). The analyses in~\cite{Moir:2016srx,Du:2020pui,Gayer:2021xzv} shown that the pole mass of $D_{0}^\ast(2300)$ is lower than the value in Review of Particle Physics (RPP)~\cite{Workman:2022ynf}. Here, we adopt the value $m_{D_{0}^\ast(2300)}=2196\pm64$ MeV from the lattice calculation at pion mass $m_\pi=239$ MeV~\cite{Gayer:2021xzv}. Then we have $\Delta\simeq330\pm64$ MeV. Feeding this $\Delta$ into Eq.~\eqref{eq:gsgp} one obtains
\begin{equation}\label{eq:gsigma}
g_\sigma=-0.73\pm 0.14.
\end{equation}
For the mass of $\sigma$ meson, we adopt the value that was determined in Refs.~\cite{Caprini:2005zr,Dai:2014zta} with the model-independent ways (\clabel[refsig]{one can also consult the similar results in Refs.~\cite{Yndurain:2007qm,Mennessier:2008kk,Mennessier:2010xg}}), which reads
\begin{equation}\label{eq:msigma}
m_\sigma=441~\rm{MeV}.
\end{equation}
Meanwhile, for the masses of the $a_0$ and $f_0$ mesons, we ignore their mass differences and use~\cite{Workman:2022ynf}
\begin{equation}\label{eq:ma0f0}
m_{a_0}=m_{f_0}\simeq 980~\rm{MeV}.
\end{equation}

\item Pseudoscalar exchange---$\eta$ meson

For the $\eta$ meson, its decay constant is $f_\eta=116$ MeV and the mass $m_\eta=548$ MeV. The eta-exchange contribution to the contact interaction is associated with the term in the second line of Eq.~\eqref{eq:mesonlagsf}.

\item Vector exchange---$\rho,\omega$ mesons

For the vector exchange form, we use the Lagrangians from the local hidden-gauge formalism~\cite{Casalbuoni:1996pg}, which read
\begin{eqnarray}\label{eq:vectorcoup}
\mathcal{L}_{V\mathcal{H}}&=&i\beta\langle\mathcal{H}v^{\mu}(\Gamma_{\mu}-\rho_{\mu})\bar{\mathcal{H}}\rangle+i\lambda\langle\mathcal{H}\sigma^{\mu\nu}F_{\mu\nu}\bar{\mathcal{H}}\rangle,
\end{eqnarray}
where $\rho_{\mu}=i\frac{g_{v}}{\sqrt{2}}V_{\mu}$ and $F_{\mu\nu}=\partial_{\mu}\rho_{\nu}-\partial_{\nu}\rho_{\mu}+[\rho_{\mu},\rho_{\nu}]$. The matrix form of $V_\mu$ is given as
\begin{eqnarray}
V_{\mu}	=	\left[\begin{array}{cc}
\frac{\rho^{0}}{\sqrt{2}}+\frac{\omega}{\sqrt{2}} & \rho^{+}\\
\rho^{-} & -\frac{\rho^{0}}{\sqrt{2}}+\frac{\omega}{\sqrt{2}}
\end{array}\right]_{\mu}.
\end{eqnarray}
The coupling constants $\beta,\lambda,g_v$~\cite{Casalbuoni:1992gi,Casalbuoni:1996pg} and the masses of $\rho,\omega$~\cite{Workman:2022ynf} read
\begin{eqnarray}\label{eq:consvec}
&&\beta=0.9,\quad\lambda=-0.63\pm0.1~\mathrm{GeV}^{-1},\quad g_v=5.8.\nonumber\\
&&m_\rho=770~\mathrm{MeV},\quad m_\omega=782~\mathrm{MeV}.
\end{eqnarray}

\item Axial-vector exchange---$a_1,f_1$ mesons

For involving the possible contribution of the heavier axial-vector mesons, we construct the following effective Lagrangians,
\begin{eqnarray}
\mathcal{L}_{A\mathcal{H}}&=&g_{a}\langle\mathcal{H}\gamma_{\mu}\gamma^{5}A^{\mu}\bar{\mathcal{H}}\rangle,
\end{eqnarray}
in which we use the ideal mixing for the axial-vector quartet in the SU(2) case,
\begin{eqnarray}
A^{\mu}&=&\left[\begin{array}{cc}
\frac{a_{1}^{0}}{\sqrt{2}}+\frac{f_{1}}{\sqrt{2}} & a_{1}^{+}\\
a_{1}^{-} & -\frac{a_{1}^{0}}{\sqrt{2}}+\frac{f_{1}}{\sqrt{2}}
\end{array}\right]^{\mu}.
\end{eqnarray}
It is hard to determine the value of $g_a$ through a reliable way presently. In Ref.~\cite{Yan:2021tcp}, Yan \etal~roughly estimated the $g_a$ via introducing the field of $a_1$ in the axial-vector current $u_\mu$, and they obtained the $g_a$ is about one order of magnitude larger than the $g$ in Eq.~\eqref{eq:mesonlagsf}. Here, we assume the coupling satisfies the naturalness, which amounts to setting the order of $g_a$ to be unity. We naively use $g_a=1$ in this study. For the masses of $a_1$ and $f_1$, we use~\cite{Workman:2022ynf}
\begin{eqnarray}
m_{a_1}=1230~\mathrm{MeV},\quad m_{f_1}=1282~\mathrm{MeV}.
\end{eqnarray}
\end{itemize}

In order to obtain the LECs $C_i$ in Eqs.~\eqref{eq:vcto0}-\eqref{eq:vcto4}, one needs to sum up the contributions from the scalar-, pseudoscalar-, vector- and axial-vector-exchange interactions and use the expansion
\begin{eqnarray}
\frac{g_i^2\mathcal{O}_j}{\bm q^2+m_i^2}=\frac{g_i^2\mathcal{O}_j}{m_i^2}\left(1-\frac{\bm q^2}{m_i^2}+\frac{\bm q^4}{m_i^4}+\dots\right),
\end{eqnarray}
where $i=\sigma,s,\varphi,v,a$ and $j=1,2$. The $m_i$ denotes either the mass of the exchanged particle or the effective mass $\sqrt{m_i^2-\delta_b^2}$. A matching with the terms in Eqs.~\eqref{eq:vcto0}-\eqref{eq:vcto4}, one gets
\begin{eqnarray}%\renewcommand{\arraystretch}{3.0}
C_{1}	&=&	-\frac{g_{a}^{2}}{2u_{f_{1}}^{2}}-\frac{g_{s}^{2}}{2m_{f_{0}}^{2}}-\frac{g_{\ensuremath{\sigma}}^{2}}{m_{\ensuremath{\sigma}}^{2}}+\frac{\beta^{2}g_{v}^{2}}{4m_{\omega}^{2}},\\[0.06cm]
C_{2}	&=&	\mathrm{sgn}\frac{g_{a}^{2}}{2u_{a_{1}}^{2}}-\frac{g_{s}^{2}}{2m_{a_{0}}^{2}}+\frac{\beta^{2}g_{v}^{2}}{4m_{\rho}^{2}},\\
C_{3}	&=&	\frac{g_{a}^{2}}{2u_{f_{1}}^{4}}+\frac{g_{s}^{2}}{2m_{f_{0}}^{4}}+\frac{g_{\sigma}^{2}}{m_{\sigma}^{4}}-\frac{\beta^{2}g_{v}^{2}}{4m_{\omega}^{4}}-\frac{g_{v}^{2}\lambda^{2}}{u_{\omega}^{2}},\\
C_{4}	&=&	-\mathrm{sgn}\frac{g_{a}^{2}}{2u_{a_{1}}^{4}}+\frac{g_{s}^{2}}{2m_{a_{0}}^{4}}-\frac{\beta^{2}g_{v}^{2}}{4m_{\rho}^{4}}+\mathrm{sgn}\frac{g_{v}^{2}\lambda^{2}}{u_{\rho}^{2}},\\
C_{5}	&=&	-\frac{g_\varphi^{2}}{12f_\eta^{2}u_{\eta}^{2}}-\frac{g_{a}^{2}}{2m_{f_{1}}^{2}u_{f_{1}}^{2}}+\frac{g_{v}^{2}\lambda^{2}}{u_{\omega}^{2}},\\
C_{6}	&=&	\mathrm{sgn}\frac{g_{a}^{2}}{2m_{a_{1}}^{2}u_{a_{1}}^{2}}-\mathrm{sgn}\frac{g_{v}^{2}\lambda^{2}}{u_{\rho}^{2}},\\
C_{7}	&=&	-\frac{g_{a}^{2}}{2u_{f_{1}}^{6}}-\frac{g_{s}^{2}}{2m_{f_{0}}^{6}}-\frac{g_{\ensuremath{\sigma}}^{2}}{m_{\ensuremath{\sigma}}^{6}}+\frac{\beta^{2}g_{v}^{2}}{4m_{\ensuremath{\omega}}^{6}}+\frac{g_{v}^{2}\lambda^{2}}{u_{\omega}^{4}},\\
C_{8}	&=&	\mathrm{sgn}\frac{g_{a}^{2}}{2u_{a_{1}}^{6}}-\frac{g_{s}^{2}}{2m_{a_{0}}^{6}}+\frac{\beta^{2}g_{v}^{2}}{4m_{\ensuremath{\rho}}^{6}}-\mathrm{sgn}\frac{g_{v}^{2}\lambda^{2}}{u_{\ensuremath{\rho}}^{4}},\\
C_{9}	&=&	\frac{g_\varphi^{2}}{12f_\eta^{2}u_{\eta}^{4}}+\frac{g_{a}^{2}}{2m_{f_{1}}^{2}u_{f_{1}}^{4}}-\frac{g_{v}^{2}\lambda^{2}}{u_{\omega}^{4}},\\
C_{10}	&=&	\mathrm{sgn}\frac{g_{v}^{2}\lambda^{2}}{u_{\ensuremath{\rho}}^{4}}-\mathrm{sgn}\frac{g_{a}^{2}}{2m_{a_{1}}^{2}u_{a_{1}}^{4}},
\end{eqnarray}
where ${\rm sgn}=(-1)^{I}$ (with $I$ the total isospin of $DD^\ast$), and $u_{x}=\sqrt{m_{x}^{2}-\delta_b^{2}}$.

\subsection{The LECs in subleading $\pi D^{(\ast)}$ Lagrangians}\label{sec:app2}
In the following, we estimate the LECs in Lagrangian~\eqref{eq:nlopid} with the RSM.
\begin{itemize}
  \item $\tilde{c}_1$ and $\tilde{c}_3$---with the $\sigma$-exchange

  One easily sees that the $\tilde{c}_1$ and $\tilde{c}_3$ related terms are connected to the $\sigma$-exchange if we write out the Lagrangians of $\sigma\pi\pi$ coupling,
  \begin{eqnarray}\label{eq:lagsigpipi}
\mathcal{L}_{\pi\sigma}&=&4\bar{c}_{d}\mathrm{Tr}(u\cdot u)\sigma+\bar{c}_{m}\mathrm{Tr}(\chi_{+})\sigma.%\qquad\bar{c}_{m,d}=\frac{1}{\sqrt{2}}c_{m,d}
  \end{eqnarray}
  Combining the vertices in Eqs.~\eqref{eq:laghsigma} and~\eqref{eq:lagsigpipi} and integrating out the $\sigma$ field one obtains that
  \begin{eqnarray}
  \tilde{c}_{1}=\frac{\bar{c}_{m}g_{\sigma}}{m_{\sigma}^{2}},~~\tilde{c}_{3}=\frac{8\bar{c}_{d}g_{\sigma}}{m_{\sigma}^{2}}=8\frac{\bar{c}_{d}}{\bar{c}_{m}}\tilde{c}_{1},~~\bar{c}_{m}\bar{c}_{d}>0,
  \end{eqnarray}
  in which $\bar{c}_{d,m}=\frac{1}{\sqrt{2}}c_{d,m}$~\cite{Bernard:1996gq}, and $|c_d|=26\pm7$ MeV, $|c_m|=80\pm21$ MeV~\cite{Guo:2009hi}.

  \item $\tilde{c}_2$---with the $D_0^\ast(2300)/D_1(2430)^0$-exchange

  \begin{eqnarray}
  \mathcal{L}_{\pi\mathcal{H}\mathcal{S}}&=&h\langle\mathcal{H}\gamma^{\mu}\gamma^{5}u_{\mu}\bar{\mathcal{S}}\rangle+\mathrm{H.c.},
  \end{eqnarray}
  where $\mathcal{S}=\frac{1+\slashed{v}}{2}\left[R^{*\mu}\gamma_{\mu}\gamma_{5}-R\right]$, and $\bar{\mathcal{S}}=\gamma^0\mathcal{S}^\dagger\gamma^0$. The $R^{*\mu}$ and $R$ denote the P-wave $1^+~[D_1(2430)^0]$ and $0^+~[D_0^\ast(2300)]$ charmed meson fields, respectively. The coupling constant $h$ can be extracted from the partial decay widthes of $D_0^\ast(2300)\to D\pi$ or $D_1(2430)\to D^\ast\pi$. We use $|h|=0.52$~\cite{Casalbuoni:1996pg} in our calculations.
  Considering both the $t$- and $u$-channel contributions one gets
  \begin{eqnarray}
  \tilde{c}_{2}&=&-\frac{h^{2}}{\Delta},
  \end{eqnarray}
  where $\Delta=m_{D_0^\ast(2300)}-m_D\simeq330$ MeV denotes the mass difference of $D_0^\ast(2300)$ and $D$ mesons.

  \item $\tilde{c}_4$---with the $\rho$-exchange

  In addition to the $\rho\mathcal{H}$ coupling in the second term of Eq.~\eqref{eq:vectorcoup}, we also need the $\rho\pi\pi$ Lagrangian, which reads~\cite{Casalbuoni:1996pg}
  \begin{eqnarray}\label{eq:lagrhopipi}
  \mathcal{L}_{\pi\rho}&=&-af_\pi^{2}\mathrm{Tr}[(\Gamma^{\mu}-\rho_{\mu})^2],\qquad a=2.
  \end{eqnarray}
  Combining the vertices in Eqs.~\eqref{eq:vectorcoup} and~\eqref{eq:lagrhopipi} and integrating out the $\rho$ field one gets
  \begin{eqnarray}
  \tilde{c}_{4}	= -\frac{2\lambda g_{V}^{2}f_\pi^{2}}{m_{\rho}^{2}}.
  \end{eqnarray}

  \item $\tilde{c}_5$---with the mass splittings of the neutral and charged $D^{(\ast)}$ mesons

  The $\tilde{c}_5$-term is related to the isospin breaking considering the $\hat{\chi}_+=2B_{0}\mathrm{diag}(m_{u}-m_{d},m_{d}-m_{u})$, with $m_{u,d}$ the masses of $u,d$ quarks. We first write out the relativistic Lagrangians of $D$ and $D^\ast$, which read
  \begin{eqnarray}
  \mathcal{L}_{\mathcal{H}}^{\mathrm{rel}}&=&\mathcal{D}_{\mu}P\mathcal{D}^{\mu}P^{\dagger}-m^{2}_0PP^{\dagger}\nonumber\\
  &&-\mathcal{D}_{\mu}P^{\ast\nu}\mathcal{D}^{\mu}P_{\nu}^{\ast\dagger}+m^{2}_{0^\ast}P^{\ast\nu}P_{\nu}^{\ast\dagger},
  \end{eqnarray}
  where $m_0$ and $m_{0^\ast}$ are the bare masses of $D$ and $D^\ast$, respectively. Here, we ignore the electromagnetic interactions and assume the mass splittings of the neutral and charged $D^{(\ast)}$ mesons come from the mass difference of $u,d$ quarks. Then we have
  \begin{eqnarray}
  -m_{D^{0}}^{2}	&=&	-m_{0}^{2}-4B_{0}\tilde{c}_{5}(m_{u}-m_{d})m_{D},\nonumber\\
-m_{D^{+}}^{2}	&=&	-m_{0}^{2}-4B_{0}\tilde{c}_{5}(m_{d}-m_{u})m_{D},\nonumber\\
-m_{D^{\ast0}}^{2}	&=&	-m_{0^{\ast}}^{2}-4B_{0}\tilde{c}_{5}(m_{u}-m_{d})m_{D^{\ast}},\nonumber\\
-m_{D^{\ast+}}^{2}	&=&	-m_{0^{\ast}}^{2}-4B_{0}\tilde{c}_{5}(m_{d}-m_{u})m_{D^{\ast}}.
  \end{eqnarray}
  With these equations we finally get
  \begin{eqnarray}
  \tilde{c}_5=\frac{m_{D^{0}}^{2}-m_{D^{+}}^{2}+m_{D^{\ast0}}^{2}-m_{D^{\ast+}}^{2}}{16\hat{m}_{D}(m_{K^{+}}^{2}-m_{K^{0}}^{2})},
  \end{eqnarray}
  where $\hat{m}_{D}=\frac{m_{D}+m_{D^{\ast}}}{2}$, and we have used $m_{K^{+}}^{2}=B_{0}(m_{u}+m_{s})$ and $m_{K^{0}}^{2}=B_{0}(m_{d}+m_{s})$.
\end{itemize}

\end{appendix}

\bibliography{ref}

\begin{thebibliography}{105}%
\makeatletter
\providecommand \@ifxundefined [1]{%
 \@ifx{#1\undefined}
}%
\providecommand \@ifnum [1]{%
 \ifnum #1\expandafter \@firstoftwo
 \else \expandafter \@secondoftwo
 \fi
}%
\providecommand \@ifx [1]{%
 \ifx #1\expandafter \@firstoftwo
 \else \expandafter \@secondoftwo
 \fi
}%
\providecommand \natexlab [1]{#1}%
\providecommand \enquote  [1]{``#1''}%
\providecommand \bibnamefont  [1]{#1}%
\providecommand \bibfnamefont [1]{#1}%
\providecommand \citenamefont [1]{#1}%
\providecommand \href@noop [0]{\@secondoftwo}%
\providecommand \href [0]{\begingroup \@sanitize@url \@href}%
\providecommand \@href[1]{\@@startlink{#1}\@@href}%
\providecommand \@@href[1]{\endgroup#1\@@endlink}%
\providecommand \@sanitize@url [0]{\catcode `\\12\catcode `\$12\catcode
  `\&12\catcode `\#12\catcode `\^12\catcode `\_12\catcode `\%12\relax}%
\providecommand \@@startlink[1]{}%
\providecommand \@@endlink[0]{}%
\providecommand \url  [0]{\begingroup\@sanitize@url \@url }%
\providecommand \@url [1]{\endgroup\@href {#1}{\urlprefix }}%
\providecommand \urlprefix  [0]{URL }%
\providecommand \Eprint [0]{\href }%
\providecommand \doibase [0]{http://dx.doi.org/}%
\providecommand \selectlanguage [0]{\@gobble}%
\providecommand \bibinfo  [0]{\@secondoftwo}%
\providecommand \bibfield  [0]{\@secondoftwo}%
\providecommand \translation [1]{[#1]}%
\providecommand \BibitemOpen [0]{}%
\providecommand \bibitemStop [0]{}%
\providecommand \bibitemNoStop [0]{.\EOS\space}%
\providecommand \EOS [0]{\spacefactor3000\relax}%
\providecommand \BibitemShut  [1]{\csname bibitem#1\endcsname}%
\let\auto@bib@innerbib\@empty
%</preamble>
\bibitem [{\citenamefont {Chen}\ \emph {et~al.}(2016)\citenamefont {Chen},
  \citenamefont {Chen}, \citenamefont {Liu},\ and\ \citenamefont
  {Zhu}}]{Chen:2016qju}%
  \BibitemOpen
  \bibfield  {author} {\bibinfo {author} {\bibfnamefont {H.-X.}\ \bibnamefont
  {Chen}}, \bibinfo {author} {\bibfnamefont {W.}~\bibnamefont {Chen}}, \bibinfo
  {author} {\bibfnamefont {X.}~\bibnamefont {Liu}}, \ and\ \bibinfo {author}
  {\bibfnamefont {S.-L.}\ \bibnamefont {Zhu}},\ }\href {\doibase
  10.1016/j.physrep.2016.05.004} {\bibfield  {journal} {\bibinfo  {journal}
  {Phys. Rept.}\ }\textbf {\bibinfo {volume} {639}},\ \bibinfo {pages} {1}
  (\bibinfo {year} {2016})},\ \Eprint {http://arxiv.org/abs/1601.02092}
  {arXiv:1601.02092 [hep-ph]} \BibitemShut {NoStop}%
\bibitem [{\citenamefont {Guo}\ \emph {et~al.}(2018)\citenamefont {Guo},
  \citenamefont {Hanhart}, \citenamefont {Mei\ss{}ner}, \citenamefont {Wang},
  \citenamefont {Zhao},\ and\ \citenamefont {Zou}}]{Guo:2017jvc}%
  \BibitemOpen
  \bibfield  {author} {\bibinfo {author} {\bibfnamefont {F.-K.}\ \bibnamefont
  {Guo}}, \bibinfo {author} {\bibfnamefont {C.}~\bibnamefont {Hanhart}},
  \bibinfo {author} {\bibfnamefont {U.-G.}\ \bibnamefont {Mei\ss{}ner}},
  \bibinfo {author} {\bibfnamefont {Q.}~\bibnamefont {Wang}}, \bibinfo {author}
  {\bibfnamefont {Q.}~\bibnamefont {Zhao}}, \ and\ \bibinfo {author}
  {\bibfnamefont {B.-S.}\ \bibnamefont {Zou}},\ }\href {\doibase
  10.1103/RevModPhys.90.015004} {\bibfield  {journal} {\bibinfo  {journal}
  {Rev. Mod. Phys.}\ }\textbf {\bibinfo {volume} {90}},\ \bibinfo {pages}
  {015004} (\bibinfo {year} {2018})},\ \Eprint
  {http://arxiv.org/abs/1705.00141} {arXiv:1705.00141 [hep-ph]} \BibitemShut
  {NoStop}%
\bibitem [{\citenamefont {Liu}\ \emph {et~al.}(2019)\citenamefont {Liu},
  \citenamefont {Chen}, \citenamefont {Chen}, \citenamefont {Liu},\ and\
  \citenamefont {Zhu}}]{Liu:2019zoy}%
  \BibitemOpen
  \bibfield  {author} {\bibinfo {author} {\bibfnamefont {Y.-R.}\ \bibnamefont
  {Liu}}, \bibinfo {author} {\bibfnamefont {H.-X.}\ \bibnamefont {Chen}},
  \bibinfo {author} {\bibfnamefont {W.}~\bibnamefont {Chen}}, \bibinfo {author}
  {\bibfnamefont {X.}~\bibnamefont {Liu}}, \ and\ \bibinfo {author}
  {\bibfnamefont {S.-L.}\ \bibnamefont {Zhu}},\ }\href {\doibase
  10.1016/j.ppnp.2019.04.003} {\bibfield  {journal} {\bibinfo  {journal} {Prog.
  Part. Nucl. Phys.}\ }\textbf {\bibinfo {volume} {107}},\ \bibinfo {pages}
  {237} (\bibinfo {year} {2019})},\ \Eprint {http://arxiv.org/abs/1903.11976}
  {arXiv:1903.11976 [hep-ph]} \BibitemShut {NoStop}%
\bibitem [{\citenamefont {Lebed}\ \emph {et~al.}(2017)\citenamefont {Lebed},
  \citenamefont {Mitchell},\ and\ \citenamefont {Swanson}}]{Lebed:2016hpi}%
  \BibitemOpen
  \bibfield  {author} {\bibinfo {author} {\bibfnamefont {R.~F.}\ \bibnamefont
  {Lebed}}, \bibinfo {author} {\bibfnamefont {R.~E.}\ \bibnamefont {Mitchell}},
  \ and\ \bibinfo {author} {\bibfnamefont {E.~S.}\ \bibnamefont {Swanson}},\
  }\href {\doibase 10.1016/j.ppnp.2016.11.003} {\bibfield  {journal} {\bibinfo
  {journal} {Prog. Part. Nucl. Phys.}\ }\textbf {\bibinfo {volume} {93}},\
  \bibinfo {pages} {143} (\bibinfo {year} {2017})},\ \Eprint
  {http://arxiv.org/abs/1610.04528} {arXiv:1610.04528 [hep-ph]} \BibitemShut
  {NoStop}%
\bibitem [{\citenamefont {Esposito}\ \emph {et~al.}(2017)\citenamefont
  {Esposito}, \citenamefont {Pilloni},\ and\ \citenamefont
  {Polosa}}]{Esposito:2016noz}%
  \BibitemOpen
  \bibfield  {author} {\bibinfo {author} {\bibfnamefont {A.}~\bibnamefont
  {Esposito}}, \bibinfo {author} {\bibfnamefont {A.}~\bibnamefont {Pilloni}}, \
  and\ \bibinfo {author} {\bibfnamefont {A.~D.}\ \bibnamefont {Polosa}},\
  }\href {\doibase 10.1016/j.physrep.2016.11.002} {\bibfield  {journal}
  {\bibinfo  {journal} {Phys. Rept.}\ }\textbf {\bibinfo {volume} {668}},\
  \bibinfo {pages} {1} (\bibinfo {year} {2017})},\ \Eprint
  {http://arxiv.org/abs/1611.07920} {arXiv:1611.07920 [hep-ph]} \BibitemShut
  {NoStop}%
\bibitem [{\citenamefont {Brambilla}\ \emph {et~al.}(2020)\citenamefont
  {Brambilla}, \citenamefont {Eidelman}, \citenamefont {Hanhart}, \citenamefont
  {Nefediev}, \citenamefont {Shen}, \citenamefont {Thomas}, \citenamefont
  {Vairo},\ and\ \citenamefont {Yuan}}]{Brambilla:2019esw}%
  \BibitemOpen
  \bibfield  {author} {\bibinfo {author} {\bibfnamefont {N.}~\bibnamefont
  {Brambilla}}, \bibinfo {author} {\bibfnamefont {S.}~\bibnamefont {Eidelman}},
  \bibinfo {author} {\bibfnamefont {C.}~\bibnamefont {Hanhart}}, \bibinfo
  {author} {\bibfnamefont {A.}~\bibnamefont {Nefediev}}, \bibinfo {author}
  {\bibfnamefont {C.-P.}\ \bibnamefont {Shen}}, \bibinfo {author}
  {\bibfnamefont {C.~E.}\ \bibnamefont {Thomas}}, \bibinfo {author}
  {\bibfnamefont {A.}~\bibnamefont {Vairo}}, \ and\ \bibinfo {author}
  {\bibfnamefont {C.-Z.}\ \bibnamefont {Yuan}},\ }\href {\doibase
  10.1016/j.physrep.2020.05.001} {\bibfield  {journal} {\bibinfo  {journal}
  {Phys. Rept.}\ }\textbf {\bibinfo {volume} {873}},\ \bibinfo {pages} {1}
  (\bibinfo {year} {2020})},\ \Eprint {http://arxiv.org/abs/1907.07583}
  {arXiv:1907.07583 [hep-ex]} \BibitemShut {NoStop}%
\bibitem [{\citenamefont {Chen}\ \emph
  {et~al.}(2021{\natexlab{a}})\citenamefont {Chen}, \citenamefont {Li},
  \citenamefont {Qian}, \citenamefont {Xie}, \citenamefont {Yang},
  \citenamefont {Zhang},\ and\ \citenamefont {Zhang}}]{Chen:2021ftn}%
  \BibitemOpen
  \bibfield  {author} {\bibinfo {author} {\bibfnamefont {S.}~\bibnamefont
  {Chen}}, \bibinfo {author} {\bibfnamefont {Y.}~\bibnamefont {Li}}, \bibinfo
  {author} {\bibfnamefont {W.}~\bibnamefont {Qian}}, \bibinfo {author}
  {\bibfnamefont {Y.}~\bibnamefont {Xie}}, \bibinfo {author} {\bibfnamefont
  {Z.}~\bibnamefont {Yang}}, \bibinfo {author} {\bibfnamefont {L.}~\bibnamefont
  {Zhang}}, \ and\ \bibinfo {author} {\bibfnamefont {Y.}~\bibnamefont
  {Zhang}},\ }\href@noop {} {\  (\bibinfo {year} {2021}{\natexlab{a}})},\
  \Eprint {http://arxiv.org/abs/2111.14360} {arXiv:2111.14360 [hep-ex]}
  \BibitemShut {NoStop}%
\bibitem [{\citenamefont {Chen}\ \emph
  {et~al.}(2022{\natexlab{a}})\citenamefont {Chen}, \citenamefont {Chen},
  \citenamefont {Liu}, \citenamefont {Liu},\ and\ \citenamefont
  {Zhu}}]{Chen:2022asf}%
  \BibitemOpen
  \bibfield  {author} {\bibinfo {author} {\bibfnamefont {H.-X.}\ \bibnamefont
  {Chen}}, \bibinfo {author} {\bibfnamefont {W.}~\bibnamefont {Chen}}, \bibinfo
  {author} {\bibfnamefont {X.}~\bibnamefont {Liu}}, \bibinfo {author}
  {\bibfnamefont {Y.-R.}\ \bibnamefont {Liu}}, \ and\ \bibinfo {author}
  {\bibfnamefont {S.-L.}\ \bibnamefont {Zhu}},\ }\href@noop {} {\  (\bibinfo
  {year} {2022}{\natexlab{a}})},\ \Eprint {http://arxiv.org/abs/2204.02649}
  {arXiv:2204.02649 [hep-ph]} \BibitemShut {NoStop}%
\bibitem [{\citenamefont {Meng}\ \emph {et~al.}(2022)\citenamefont {Meng},
  \citenamefont {Wang}, \citenamefont {Wang},\ and\ \citenamefont
  {Zhu}}]{Meng:2022ozq}%
  \BibitemOpen
  \bibfield  {author} {\bibinfo {author} {\bibfnamefont {L.}~\bibnamefont
  {Meng}}, \bibinfo {author} {\bibfnamefont {B.}~\bibnamefont {Wang}}, \bibinfo
  {author} {\bibfnamefont {G.-J.}\ \bibnamefont {Wang}}, \ and\ \bibinfo
  {author} {\bibfnamefont {S.-L.}\ \bibnamefont {Zhu}},\ }\href@noop {} {\
  (\bibinfo {year} {2022})},\ \Eprint {http://arxiv.org/abs/2204.08716}
  {arXiv:2204.08716 [hep-ph]} \BibitemShut {NoStop}%
\bibitem [{\citenamefont {Albuquerque}\ \emph {et~al.}(2022)\citenamefont
  {Albuquerque}, \citenamefont {Narison},\ and\ \citenamefont
  {Rabetiarivony}}]{Albuquerque:2022weq}%
  \BibitemOpen
  \bibfield  {author} {\bibinfo {author} {\bibfnamefont {R.}~\bibnamefont
  {Albuquerque}}, \bibinfo {author} {\bibfnamefont {S.}~\bibnamefont
  {Narison}}, \ and\ \bibinfo {author} {\bibfnamefont {D.}~\bibnamefont
  {Rabetiarivony}},\ }\href {\doibase 10.1016/j.nuclphysa.2022.122451}
  {\bibfield  {journal} {\bibinfo  {journal} {Nucl. Phys. A}\ }\textbf
  {\bibinfo {volume} {1023}},\ \bibinfo {pages} {122451} (\bibinfo {year}
  {2022})},\ \Eprint {http://arxiv.org/abs/2201.13449} {arXiv:2201.13449
  [hep-ph]} \BibitemShut {NoStop}%
\bibitem [{\citenamefont {Albuquerque}\ \emph {et~al.}(2023)\citenamefont
  {Albuquerque}, \citenamefont {Narison},\ and\ \citenamefont
  {Rabetiarivony}}]{Albuquerque:2023rrf}%
  \BibitemOpen
  \bibfield  {author} {\bibinfo {author} {\bibfnamefont {R.}~\bibnamefont
  {Albuquerque}}, \bibinfo {author} {\bibfnamefont {S.}~\bibnamefont
  {Narison}}, \ and\ \bibinfo {author} {\bibfnamefont {D.}~\bibnamefont
  {Rabetiarivony}},\ }\href {\doibase 10.1016/j.nuclphysa.2023.122637}
  {\bibfield  {journal} {\bibinfo  {journal} {Nucl. Phys. A}\ }\textbf
  {\bibinfo {volume} {1034}},\ \bibinfo {pages} {122637} (\bibinfo {year}
  {2023})},\ \Eprint {http://arxiv.org/abs/2301.08199} {arXiv:2301.08199
  [hep-ph]} \BibitemShut {NoStop}%
\bibitem [{\citenamefont {Weinberg}(1990)}]{Weinberg:1990rz}%
  \BibitemOpen
  \bibfield  {author} {\bibinfo {author} {\bibfnamefont {S.}~\bibnamefont
  {Weinberg}},\ }\href {\doibase 10.1016/0370-2693(90)90938-3} {\bibfield
  {journal} {\bibinfo  {journal} {Phys. Lett. B}\ }\textbf {\bibinfo {volume}
  {251}},\ \bibinfo {pages} {288} (\bibinfo {year} {1990})}\BibitemShut
  {NoStop}%
\bibitem [{\citenamefont {Weinberg}(1991)}]{Weinberg:1991um}%
  \BibitemOpen
  \bibfield  {author} {\bibinfo {author} {\bibfnamefont {S.}~\bibnamefont
  {Weinberg}},\ }\href {\doibase 10.1016/0550-3213(91)90231-L} {\bibfield
  {journal} {\bibinfo  {journal} {Nucl. Phys. B}\ }\textbf {\bibinfo {volume}
  {363}},\ \bibinfo {pages} {3} (\bibinfo {year} {1991})}\BibitemShut {NoStop}%
\bibitem [{\citenamefont {Epelbaum}\ \emph {et~al.}(2009)\citenamefont
  {Epelbaum}, \citenamefont {Hammer},\ and\ \citenamefont
  {Meissner}}]{Epelbaum:2008ga}%
  \BibitemOpen
  \bibfield  {author} {\bibinfo {author} {\bibfnamefont {E.}~\bibnamefont
  {Epelbaum}}, \bibinfo {author} {\bibfnamefont {H.-W.}\ \bibnamefont
  {Hammer}}, \ and\ \bibinfo {author} {\bibfnamefont {U.-G.}\ \bibnamefont
  {Meissner}},\ }\href {\doibase 10.1103/RevModPhys.81.1773} {\bibfield
  {journal} {\bibinfo  {journal} {Rev. Mod. Phys.}\ }\textbf {\bibinfo {volume}
  {81}},\ \bibinfo {pages} {1773} (\bibinfo {year} {2009})},\ \Eprint
  {http://arxiv.org/abs/0811.1338} {arXiv:0811.1338 [nucl-th]} \BibitemShut
  {NoStop}%
\bibitem [{\citenamefont {Machleidt}\ and\ \citenamefont
  {Entem}(2011)}]{Machleidt:2011zz}%
  \BibitemOpen
  \bibfield  {author} {\bibinfo {author} {\bibfnamefont {R.}~\bibnamefont
  {Machleidt}}\ and\ \bibinfo {author} {\bibfnamefont {D.~R.}\ \bibnamefont
  {Entem}},\ }\href {\doibase 10.1016/j.physrep.2011.02.001} {\bibfield
  {journal} {\bibinfo  {journal} {Phys. Rept.}\ }\textbf {\bibinfo {volume}
  {503}},\ \bibinfo {pages} {1} (\bibinfo {year} {2011})},\ \Eprint
  {http://arxiv.org/abs/1105.2919} {arXiv:1105.2919 [nucl-th]} \BibitemShut
  {NoStop}%
\bibitem [{\citenamefont {Ordonez}\ \emph {et~al.}(1996)\citenamefont
  {Ordonez}, \citenamefont {Ray},\ and\ \citenamefont {van
  Kolck}}]{Ordonez:1995rz}%
  \BibitemOpen
  \bibfield  {author} {\bibinfo {author} {\bibfnamefont {C.}~\bibnamefont
  {Ordonez}}, \bibinfo {author} {\bibfnamefont {L.}~\bibnamefont {Ray}}, \ and\
  \bibinfo {author} {\bibfnamefont {U.}~\bibnamefont {van Kolck}},\ }\href
  {\doibase 10.1103/PhysRevC.53.2086} {\bibfield  {journal} {\bibinfo
  {journal} {Phys. Rev. C}\ }\textbf {\bibinfo {volume} {53}},\ \bibinfo
  {pages} {2086} (\bibinfo {year} {1996})},\ \Eprint
  {http://arxiv.org/abs/hep-ph/9511380} {arXiv:hep-ph/9511380} \BibitemShut
  {NoStop}%
\bibitem [{\citenamefont {Kaiser}\ \emph {et~al.}(1997)\citenamefont {Kaiser},
  \citenamefont {Brockmann},\ and\ \citenamefont {Weise}}]{Kaiser:1997mw}%
  \BibitemOpen
  \bibfield  {author} {\bibinfo {author} {\bibfnamefont {N.}~\bibnamefont
  {Kaiser}}, \bibinfo {author} {\bibfnamefont {R.}~\bibnamefont {Brockmann}}, \
  and\ \bibinfo {author} {\bibfnamefont {W.}~\bibnamefont {Weise}},\ }\href
  {\doibase 10.1016/S0375-9474(97)00586-1} {\bibfield  {journal} {\bibinfo
  {journal} {Nucl. Phys. A}\ }\textbf {\bibinfo {volume} {625}},\ \bibinfo
  {pages} {758} (\bibinfo {year} {1997})},\ \Eprint
  {http://arxiv.org/abs/nucl-th/9706045} {arXiv:nucl-th/9706045} \BibitemShut
  {NoStop}%
\bibitem [{\citenamefont {Epelbaum}\ \emph {et~al.}(2000)\citenamefont
  {Epelbaum}, \citenamefont {Gloeckle},\ and\ \citenamefont
  {Meissner}}]{Epelbaum:1999dj}%
  \BibitemOpen
  \bibfield  {author} {\bibinfo {author} {\bibfnamefont {E.}~\bibnamefont
  {Epelbaum}}, \bibinfo {author} {\bibfnamefont {W.}~\bibnamefont {Gloeckle}},
  \ and\ \bibinfo {author} {\bibfnamefont {U.-G.}\ \bibnamefont {Meissner}},\
  }\href {\doibase 10.1016/S0375-9474(99)00821-0} {\bibfield  {journal}
  {\bibinfo  {journal} {Nucl. Phys. A}\ }\textbf {\bibinfo {volume} {671}},\
  \bibinfo {pages} {295} (\bibinfo {year} {2000})},\ \Eprint
  {http://arxiv.org/abs/nucl-th/9910064} {arXiv:nucl-th/9910064} \BibitemShut
  {NoStop}%
\bibitem [{\citenamefont {Epelbaum}(2006)}]{Epelbaum:2005pn}%
  \BibitemOpen
  \bibfield  {author} {\bibinfo {author} {\bibfnamefont {E.}~\bibnamefont
  {Epelbaum}},\ }\href {\doibase 10.1016/j.ppnp.2005.09.002} {\bibfield
  {journal} {\bibinfo  {journal} {Prog. Part. Nucl. Phys.}\ }\textbf {\bibinfo
  {volume} {57}},\ \bibinfo {pages} {654} (\bibinfo {year} {2006})},\ \Eprint
  {http://arxiv.org/abs/nucl-th/0509032} {arXiv:nucl-th/0509032} \BibitemShut
  {NoStop}%
\bibitem [{\citenamefont {Bernard}\ \emph {et~al.}(1997)\citenamefont
  {Bernard}, \citenamefont {Kaiser},\ and\ \citenamefont
  {Mei\ss~ner}}]{Bernard:1996gq}%
  \BibitemOpen
  \bibfield  {author} {\bibinfo {author} {\bibfnamefont {V.}~\bibnamefont
  {Bernard}}, \bibinfo {author} {\bibfnamefont {N.}~\bibnamefont {Kaiser}}, \
  and\ \bibinfo {author} {\bibfnamefont {U.-G.}\ \bibnamefont {Mei\ss~ner}},\
  }\href {\doibase 10.1016/S0375-9474(97)00021-3} {\bibfield  {journal}
  {\bibinfo  {journal} {Nucl. Phys. A}\ }\textbf {\bibinfo {volume} {615}},\
  \bibinfo {pages} {483} (\bibinfo {year} {1997})},\ \Eprint
  {http://arxiv.org/abs/hep-ph/9611253} {arXiv:hep-ph/9611253} \BibitemShut
  {NoStop}%
\bibitem [{\citenamefont {Epelbaum}\ \emph
  {et~al.}(2002{\natexlab{a}})\citenamefont {Epelbaum}, \citenamefont {Nogga},
  \citenamefont {Gloeckle}, \citenamefont {Kamada}, \citenamefont {Meissner},\
  and\ \citenamefont {Witala}}]{Epelbaum:2002ji}%
  \BibitemOpen
  \bibfield  {author} {\bibinfo {author} {\bibfnamefont {E.}~\bibnamefont
  {Epelbaum}}, \bibinfo {author} {\bibfnamefont {A.}~\bibnamefont {Nogga}},
  \bibinfo {author} {\bibfnamefont {W.}~\bibnamefont {Gloeckle}}, \bibinfo
  {author} {\bibfnamefont {H.}~\bibnamefont {Kamada}}, \bibinfo {author}
  {\bibfnamefont {U.~G.}\ \bibnamefont {Meissner}}, \ and\ \bibinfo {author}
  {\bibfnamefont {H.}~\bibnamefont {Witala}},\ }\href {\doibase
  10.1140/epja/i2002-10048-2} {\bibfield  {journal} {\bibinfo  {journal} {Eur.
  Phys. J. A}\ }\textbf {\bibinfo {volume} {15}},\ \bibinfo {pages} {543}
  (\bibinfo {year} {2002}{\natexlab{a}})},\ \Eprint
  {http://arxiv.org/abs/nucl-th/0201064} {arXiv:nucl-th/0201064} \BibitemShut
  {NoStop}%
\bibitem [{\citenamefont {Epelbaum}\ \emph
  {et~al.}(2004{\natexlab{a}})\citenamefont {Epelbaum}, \citenamefont
  {Gloeckle},\ and\ \citenamefont {Meissner}}]{Epelbaum:2003gr}%
  \BibitemOpen
  \bibfield  {author} {\bibinfo {author} {\bibfnamefont {E.}~\bibnamefont
  {Epelbaum}}, \bibinfo {author} {\bibfnamefont {W.}~\bibnamefont {Gloeckle}},
  \ and\ \bibinfo {author} {\bibfnamefont {U.-G.}\ \bibnamefont {Meissner}},\
  }\href {\doibase 10.1140/epja/i2003-10096-0} {\bibfield  {journal} {\bibinfo
  {journal} {Eur. Phys. J. A}\ }\textbf {\bibinfo {volume} {19}},\ \bibinfo
  {pages} {125} (\bibinfo {year} {2004}{\natexlab{a}})},\ \Eprint
  {http://arxiv.org/abs/nucl-th/0304037} {arXiv:nucl-th/0304037} \BibitemShut
  {NoStop}%
\bibitem [{\citenamefont {Fettes}\ \emph {et~al.}(1998)\citenamefont {Fettes},
  \citenamefont {Meissner},\ and\ \citenamefont {Steininger}}]{Fettes:1998ud}%
  \BibitemOpen
  \bibfield  {author} {\bibinfo {author} {\bibfnamefont {N.}~\bibnamefont
  {Fettes}}, \bibinfo {author} {\bibfnamefont {U.-G.}\ \bibnamefont
  {Meissner}}, \ and\ \bibinfo {author} {\bibfnamefont {S.}~\bibnamefont
  {Steininger}},\ }\href {\doibase 10.1016/S0375-9474(98)00452-7} {\bibfield
  {journal} {\bibinfo  {journal} {Nucl. Phys. A}\ }\textbf {\bibinfo {volume}
  {640}},\ \bibinfo {pages} {199} (\bibinfo {year} {1998})},\ \Eprint
  {http://arxiv.org/abs/hep-ph/9803266} {arXiv:hep-ph/9803266} \BibitemShut
  {NoStop}%
\bibitem [{\citenamefont {Krebs}\ \emph {et~al.}(2007)\citenamefont {Krebs},
  \citenamefont {Epelbaum},\ and\ \citenamefont {Meissner}}]{Krebs:2007rh}%
  \BibitemOpen
  \bibfield  {author} {\bibinfo {author} {\bibfnamefont {H.}~\bibnamefont
  {Krebs}}, \bibinfo {author} {\bibfnamefont {E.}~\bibnamefont {Epelbaum}}, \
  and\ \bibinfo {author} {\bibfnamefont {U.-G.}\ \bibnamefont {Meissner}},\
  }\href {\doibase 10.1140/epja/i2007-10372-y} {\bibfield  {journal} {\bibinfo
  {journal} {Eur. Phys. J. A}\ }\textbf {\bibinfo {volume} {32}},\ \bibinfo
  {pages} {127} (\bibinfo {year} {2007})},\ \Eprint
  {http://arxiv.org/abs/nucl-th/0703087} {arXiv:nucl-th/0703087} \BibitemShut
  {NoStop}%
\bibitem [{\citenamefont {Epelbaum}\ \emph
  {et~al.}(2004{\natexlab{b}})\citenamefont {Epelbaum}, \citenamefont
  {Gloeckle},\ and\ \citenamefont {Meissner}}]{Epelbaum:2003xx}%
  \BibitemOpen
  \bibfield  {author} {\bibinfo {author} {\bibfnamefont {E.}~\bibnamefont
  {Epelbaum}}, \bibinfo {author} {\bibfnamefont {W.}~\bibnamefont {Gloeckle}},
  \ and\ \bibinfo {author} {\bibfnamefont {U.-G.}\ \bibnamefont {Meissner}},\
  }\href {\doibase 10.1140/epja/i2003-10129-8} {\bibfield  {journal} {\bibinfo
  {journal} {Eur. Phys. J. A}\ }\textbf {\bibinfo {volume} {19}},\ \bibinfo
  {pages} {401} (\bibinfo {year} {2004}{\natexlab{b}})},\ \Eprint
  {http://arxiv.org/abs/nucl-th/0308010} {arXiv:nucl-th/0308010} \BibitemShut
  {NoStop}%
\bibitem [{\citenamefont {Epelbaum}\ \emph {et~al.}(2005)\citenamefont
  {Epelbaum}, \citenamefont {Glockle},\ and\ \citenamefont
  {Meissner}}]{Epelbaum:2004fk}%
  \BibitemOpen
  \bibfield  {author} {\bibinfo {author} {\bibfnamefont {E.}~\bibnamefont
  {Epelbaum}}, \bibinfo {author} {\bibfnamefont {W.}~\bibnamefont {Glockle}}, \
  and\ \bibinfo {author} {\bibfnamefont {U.-G.}\ \bibnamefont {Meissner}},\
  }\href {\doibase 10.1016/j.nuclphysa.2004.09.107} {\bibfield  {journal}
  {\bibinfo  {journal} {Nucl. Phys. A}\ }\textbf {\bibinfo {volume} {747}},\
  \bibinfo {pages} {362} (\bibinfo {year} {2005})},\ \Eprint
  {http://arxiv.org/abs/nucl-th/0405048} {arXiv:nucl-th/0405048} \BibitemShut
  {NoStop}%
\bibitem [{\citenamefont {Epelbaum}\ \emph
  {et~al.}(2015{\natexlab{a}})\citenamefont {Epelbaum}, \citenamefont {Krebs},\
  and\ \citenamefont {Mei\ss{}ner}}]{Epelbaum:2014efa}%
  \BibitemOpen
  \bibfield  {author} {\bibinfo {author} {\bibfnamefont {E.}~\bibnamefont
  {Epelbaum}}, \bibinfo {author} {\bibfnamefont {H.}~\bibnamefont {Krebs}}, \
  and\ \bibinfo {author} {\bibfnamefont {U.~G.}\ \bibnamefont {Mei\ss{}ner}},\
  }\href {\doibase 10.1140/epja/i2015-15053-8} {\bibfield  {journal} {\bibinfo
  {journal} {Eur. Phys. J. A}\ }\textbf {\bibinfo {volume} {51}},\ \bibinfo
  {pages} {53} (\bibinfo {year} {2015}{\natexlab{a}})},\ \Eprint
  {http://arxiv.org/abs/1412.0142} {arXiv:1412.0142 [nucl-th]} \BibitemShut
  {NoStop}%
\bibitem [{\citenamefont {Epelbaum}\ \emph
  {et~al.}(2015{\natexlab{b}})\citenamefont {Epelbaum}, \citenamefont {Krebs},\
  and\ \citenamefont {Mei\ss{}ner}}]{Epelbaum:2014sza}%
  \BibitemOpen
  \bibfield  {author} {\bibinfo {author} {\bibfnamefont {E.}~\bibnamefont
  {Epelbaum}}, \bibinfo {author} {\bibfnamefont {H.}~\bibnamefont {Krebs}}, \
  and\ \bibinfo {author} {\bibfnamefont {U.~G.}\ \bibnamefont {Mei\ss{}ner}},\
  }\href {\doibase 10.1103/PhysRevLett.115.122301} {\bibfield  {journal}
  {\bibinfo  {journal} {Phys. Rev. Lett.}\ }\textbf {\bibinfo {volume} {115}},\
  \bibinfo {pages} {122301} (\bibinfo {year} {2015}{\natexlab{b}})},\ \Eprint
  {http://arxiv.org/abs/1412.4623} {arXiv:1412.4623 [nucl-th]} \BibitemShut
  {NoStop}%
\bibitem [{\citenamefont {Reinert}\ \emph {et~al.}(2018)\citenamefont
  {Reinert}, \citenamefont {Krebs},\ and\ \citenamefont
  {Epelbaum}}]{Reinert:2017usi}%
  \BibitemOpen
  \bibfield  {author} {\bibinfo {author} {\bibfnamefont {P.}~\bibnamefont
  {Reinert}}, \bibinfo {author} {\bibfnamefont {H.}~\bibnamefont {Krebs}}, \
  and\ \bibinfo {author} {\bibfnamefont {E.}~\bibnamefont {Epelbaum}},\ }\href
  {\doibase 10.1140/epja/i2018-12516-4} {\bibfield  {journal} {\bibinfo
  {journal} {Eur. Phys. J. A}\ }\textbf {\bibinfo {volume} {54}},\ \bibinfo
  {pages} {86} (\bibinfo {year} {2018})},\ \Eprint
  {http://arxiv.org/abs/1711.08821} {arXiv:1711.08821 [nucl-th]} \BibitemShut
  {NoStop}%
\bibitem [{\citenamefont {Epelbaum}\ \emph {et~al.}(2020)\citenamefont
  {Epelbaum}, \citenamefont {Krebs},\ and\ \citenamefont
  {Reinert}}]{Epelbaum:2019kcf}%
  \BibitemOpen
  \bibfield  {author} {\bibinfo {author} {\bibfnamefont {E.}~\bibnamefont
  {Epelbaum}}, \bibinfo {author} {\bibfnamefont {H.}~\bibnamefont {Krebs}}, \
  and\ \bibinfo {author} {\bibfnamefont {P.}~\bibnamefont {Reinert}},\ }\href
  {\doibase 10.3389/fphy.2020.00098} {\bibfield  {journal} {\bibinfo  {journal}
  {Front. in Phys.}\ }\textbf {\bibinfo {volume} {8}},\ \bibinfo {pages} {98}
  (\bibinfo {year} {2020})},\ \Eprint {http://arxiv.org/abs/1911.11875}
  {arXiv:1911.11875 [nucl-th]} \BibitemShut {NoStop}%
\bibitem [{\citenamefont {Donoghue}\ and\ \citenamefont
  {Holstein}(1998)}]{Donoghue:1998krd}%
  \BibitemOpen
  \bibfield  {author} {\bibinfo {author} {\bibfnamefont {J.~F.}\ \bibnamefont
  {Donoghue}}\ and\ \bibinfo {author} {\bibfnamefont {B.~R.}\ \bibnamefont
  {Holstein}},\ }\href {\doibase 10.1016/S0370-2693(98)00859-4} {\bibfield
  {journal} {\bibinfo  {journal} {Phys. Lett. B}\ }\textbf {\bibinfo {volume}
  {436}},\ \bibinfo {pages} {331} (\bibinfo {year} {1998})}\BibitemShut
  {NoStop}%
\bibitem [{\citenamefont {Donoghue}\ \emph {et~al.}(1999)\citenamefont
  {Donoghue}, \citenamefont {Holstein},\ and\ \citenamefont
  {Borasoy}}]{Donoghue:1998bs}%
  \BibitemOpen
  \bibfield  {author} {\bibinfo {author} {\bibfnamefont {J.~F.}\ \bibnamefont
  {Donoghue}}, \bibinfo {author} {\bibfnamefont {B.~R.}\ \bibnamefont
  {Holstein}}, \ and\ \bibinfo {author} {\bibfnamefont {B.}~\bibnamefont
  {Borasoy}},\ }\href {\doibase 10.1103/PhysRevD.59.036002} {\bibfield
  {journal} {\bibinfo  {journal} {Phys. Rev. D}\ }\textbf {\bibinfo {volume}
  {59}},\ \bibinfo {pages} {036002} (\bibinfo {year} {1999})},\ \Eprint
  {http://arxiv.org/abs/hep-ph/9804281} {arXiv:hep-ph/9804281} \BibitemShut
  {NoStop}%
\bibitem [{\citenamefont {Xiao}\ \emph {et~al.}(2020)\citenamefont {Xiao},
  \citenamefont {Wang}, \citenamefont {Lu},\ and\ \citenamefont
  {Geng}}]{Xiao:2020ozd}%
  \BibitemOpen
  \bibfield  {author} {\bibinfo {author} {\bibfnamefont {Y.}~\bibnamefont
  {Xiao}}, \bibinfo {author} {\bibfnamefont {C.-X.}\ \bibnamefont {Wang}},
  \bibinfo {author} {\bibfnamefont {J.-X.}\ \bibnamefont {Lu}}, \ and\ \bibinfo
  {author} {\bibfnamefont {L.-S.}\ \bibnamefont {Geng}},\ }\href {\doibase
  10.1103/PhysRevC.102.054001} {\bibfield  {journal} {\bibinfo  {journal}
  {Phys. Rev. C}\ }\textbf {\bibinfo {volume} {102}},\ \bibinfo {pages}
  {054001} (\bibinfo {year} {2020})},\ \Eprint
  {http://arxiv.org/abs/2007.13675} {arXiv:2007.13675 [nucl-th]} \BibitemShut
  {NoStop}%
\bibitem [{\citenamefont {Wang}\ \emph {et~al.}(2022)\citenamefont {Wang},
  \citenamefont {Lu}, \citenamefont {Xiao},\ and\ \citenamefont
  {Geng}}]{Wang:2021kos}%
  \BibitemOpen
  \bibfield  {author} {\bibinfo {author} {\bibfnamefont {C.-X.}\ \bibnamefont
  {Wang}}, \bibinfo {author} {\bibfnamefont {J.-X.}\ \bibnamefont {Lu}},
  \bibinfo {author} {\bibfnamefont {Y.}~\bibnamefont {Xiao}}, \ and\ \bibinfo
  {author} {\bibfnamefont {L.-S.}\ \bibnamefont {Geng}},\ }\href {\doibase
  10.1103/PhysRevC.105.014003} {\bibfield  {journal} {\bibinfo  {journal}
  {Phys. Rev. C}\ }\textbf {\bibinfo {volume} {105}},\ \bibinfo {pages}
  {014003} (\bibinfo {year} {2022})},\ \Eprint
  {http://arxiv.org/abs/2110.05278} {arXiv:2110.05278 [nucl-th]} \BibitemShut
  {NoStop}%
\bibitem [{\citenamefont {Lu}\ \emph {et~al.}(2022)\citenamefont {Lu},
  \citenamefont {Wang}, \citenamefont {Xiao}, \citenamefont {Geng},
  \citenamefont {Meng},\ and\ \citenamefont {Ring}}]{Lu:2021gsb}%
  \BibitemOpen
  \bibfield  {author} {\bibinfo {author} {\bibfnamefont {J.-X.}\ \bibnamefont
  {Lu}}, \bibinfo {author} {\bibfnamefont {C.-X.}\ \bibnamefont {Wang}},
  \bibinfo {author} {\bibfnamefont {Y.}~\bibnamefont {Xiao}}, \bibinfo {author}
  {\bibfnamefont {L.-S.}\ \bibnamefont {Geng}}, \bibinfo {author}
  {\bibfnamefont {J.}~\bibnamefont {Meng}}, \ and\ \bibinfo {author}
  {\bibfnamefont {P.}~\bibnamefont {Ring}},\ }\href {\doibase
  10.1103/PhysRevLett.128.142002} {\bibfield  {journal} {\bibinfo  {journal}
  {Phys. Rev. Lett.}\ }\textbf {\bibinfo {volume} {128}},\ \bibinfo {pages}
  {142002} (\bibinfo {year} {2022})},\ \Eprint
  {http://arxiv.org/abs/2111.07766} {arXiv:2111.07766 [nucl-th]} \BibitemShut
  {NoStop}%
\bibitem [{\citenamefont {Liu}\ \emph {et~al.}(2014)\citenamefont {Liu},
  \citenamefont {Li},\ and\ \citenamefont {Zhu}}]{Liu:2012vd}%
  \BibitemOpen
  \bibfield  {author} {\bibinfo {author} {\bibfnamefont {Z.-W.}\ \bibnamefont
  {Liu}}, \bibinfo {author} {\bibfnamefont {N.}~\bibnamefont {Li}}, \ and\
  \bibinfo {author} {\bibfnamefont {S.-L.}\ \bibnamefont {Zhu}},\ }\href
  {\doibase 10.1103/PhysRevD.89.074015} {\bibfield  {journal} {\bibinfo
  {journal} {Phys. Rev. D}\ }\textbf {\bibinfo {volume} {89}},\ \bibinfo
  {pages} {074015} (\bibinfo {year} {2014})},\ \Eprint
  {http://arxiv.org/abs/1211.3578} {arXiv:1211.3578 [hep-ph]} \BibitemShut
  {NoStop}%
\bibitem [{\citenamefont {Xu}\ \emph {et~al.}(2019)\citenamefont {Xu},
  \citenamefont {Wang}, \citenamefont {Liu},\ and\ \citenamefont
  {Liu}}]{Xu:2017tsr}%
  \BibitemOpen
  \bibfield  {author} {\bibinfo {author} {\bibfnamefont {H.}~\bibnamefont
  {Xu}}, \bibinfo {author} {\bibfnamefont {B.}~\bibnamefont {Wang}}, \bibinfo
  {author} {\bibfnamefont {Z.-W.}\ \bibnamefont {Liu}}, \ and\ \bibinfo
  {author} {\bibfnamefont {X.}~\bibnamefont {Liu}},\ }\href {\doibase
  10.1103/PhysRevD.99.014027} {\bibfield  {journal} {\bibinfo  {journal} {Phys.
  Rev. D}\ }\textbf {\bibinfo {volume} {99}},\ \bibinfo {pages} {014027}
  (\bibinfo {year} {2019})},\ \bibinfo {note} {[Erratum: Phys.Rev.D 104, 119903
  (2021)]},\ \Eprint {http://arxiv.org/abs/1708.06918} {arXiv:1708.06918
  [hep-ph]} \BibitemShut {NoStop}%
\bibitem [{\citenamefont {Aaij}\ \emph
  {et~al.}(2022{\natexlab{a}})\citenamefont {Aaij} \emph
  {et~al.}}]{LHCb:2021vvq}%
  \BibitemOpen
  \bibfield  {author} {\bibinfo {author} {\bibfnamefont {R.}~\bibnamefont
  {Aaij}} \emph {et~al.} (\bibinfo {collaboration} {LHCb}),\ }\href {\doibase
  10.1038/s41567-022-01614-y} {\bibfield  {journal} {\bibinfo  {journal}
  {Nature Phys.}\ }\textbf {\bibinfo {volume} {18}},\ \bibinfo {pages} {751}
  (\bibinfo {year} {2022}{\natexlab{a}})},\ \Eprint
  {http://arxiv.org/abs/2109.01038} {arXiv:2109.01038 [hep-ex]} \BibitemShut
  {NoStop}%
\bibitem [{\citenamefont {Aaij}\ \emph
  {et~al.}(2022{\natexlab{b}})\citenamefont {Aaij} \emph
  {et~al.}}]{LHCb:2021auc}%
  \BibitemOpen
  \bibfield  {author} {\bibinfo {author} {\bibfnamefont {R.}~\bibnamefont
  {Aaij}} \emph {et~al.} (\bibinfo {collaboration} {LHCb}),\ }\href {\doibase
  10.1038/s41467-022-30206-w} {\bibfield  {journal} {\bibinfo  {journal}
  {Nature Commun.}\ }\textbf {\bibinfo {volume} {13}},\ \bibinfo {pages} {3351}
  (\bibinfo {year} {2022}{\natexlab{b}})},\ \Eprint
  {http://arxiv.org/abs/2109.01056} {arXiv:2109.01056 [hep-ex]} \BibitemShut
  {NoStop}%
\bibitem [{\citenamefont {Wang}\ \emph
  {et~al.}(2019{\natexlab{a}})\citenamefont {Wang}, \citenamefont {Liu},\ and\
  \citenamefont {Liu}}]{Wang:2018atz}%
  \BibitemOpen
  \bibfield  {author} {\bibinfo {author} {\bibfnamefont {B.}~\bibnamefont
  {Wang}}, \bibinfo {author} {\bibfnamefont {Z.-W.}\ \bibnamefont {Liu}}, \
  and\ \bibinfo {author} {\bibfnamefont {X.}~\bibnamefont {Liu}},\ }\href
  {\doibase 10.1103/PhysRevD.99.036007} {\bibfield  {journal} {\bibinfo
  {journal} {Phys. Rev. D}\ }\textbf {\bibinfo {volume} {99}},\ \bibinfo
  {pages} {036007} (\bibinfo {year} {2019}{\natexlab{a}})},\ \Eprint
  {http://arxiv.org/abs/1812.04457} {arXiv:1812.04457 [hep-ph]} \BibitemShut
  {NoStop}%
\bibitem [{\citenamefont {Meng}\ \emph {et~al.}(2019)\citenamefont {Meng},
  \citenamefont {Wang}, \citenamefont {Wang},\ and\ \citenamefont
  {Zhu}}]{Meng:2019ilv}%
  \BibitemOpen
  \bibfield  {author} {\bibinfo {author} {\bibfnamefont {L.}~\bibnamefont
  {Meng}}, \bibinfo {author} {\bibfnamefont {B.}~\bibnamefont {Wang}}, \bibinfo
  {author} {\bibfnamefont {G.-J.}\ \bibnamefont {Wang}}, \ and\ \bibinfo
  {author} {\bibfnamefont {S.-L.}\ \bibnamefont {Zhu}},\ }\href {\doibase
  10.1103/PhysRevD.100.014031} {\bibfield  {journal} {\bibinfo  {journal}
  {Phys. Rev. D}\ }\textbf {\bibinfo {volume} {100}},\ \bibinfo {pages}
  {014031} (\bibinfo {year} {2019})},\ \Eprint
  {http://arxiv.org/abs/1905.04113} {arXiv:1905.04113 [hep-ph]} \BibitemShut
  {NoStop}%
\bibitem [{\citenamefont {Wang}\ \emph
  {et~al.}(2019{\natexlab{b}})\citenamefont {Wang}, \citenamefont {Meng},\ and\
  \citenamefont {Zhu}}]{Wang:2019ato}%
  \BibitemOpen
  \bibfield  {author} {\bibinfo {author} {\bibfnamefont {B.}~\bibnamefont
  {Wang}}, \bibinfo {author} {\bibfnamefont {L.}~\bibnamefont {Meng}}, \ and\
  \bibinfo {author} {\bibfnamefont {S.-L.}\ \bibnamefont {Zhu}},\ }\href
  {\doibase 10.1007/JHEP11(2019)108} {\bibfield  {journal} {\bibinfo  {journal}
  {JHEP}\ }\textbf {\bibinfo {volume} {11}},\ \bibinfo {pages} {108} (\bibinfo
  {year} {2019}{\natexlab{b}})},\ \Eprint {http://arxiv.org/abs/1909.13054}
  {arXiv:1909.13054 [hep-ph]} \BibitemShut {NoStop}%
\bibitem [{\citenamefont {Gershon}(2022)}]{Gershon:2022xnn}%
  \BibitemOpen
  \bibfield  {author} {\bibinfo {author} {\bibfnamefont {T.}~\bibnamefont
  {Gershon}} (\bibinfo {collaboration} {LHCb}),\ }\href@noop {} {\  (\bibinfo
  {year} {2022})},\ \Eprint {http://arxiv.org/abs/2206.15233} {arXiv:2206.15233
  [hep-ex]} \BibitemShut {NoStop}%
\bibitem [{\citenamefont {Wang}\ \emph {et~al.}(2020)\citenamefont {Wang},
  \citenamefont {Meng},\ and\ \citenamefont {Zhu}}]{Wang:2019nvm}%
  \BibitemOpen
  \bibfield  {author} {\bibinfo {author} {\bibfnamefont {B.}~\bibnamefont
  {Wang}}, \bibinfo {author} {\bibfnamefont {L.}~\bibnamefont {Meng}}, \ and\
  \bibinfo {author} {\bibfnamefont {S.-L.}\ \bibnamefont {Zhu}},\ }\href
  {\doibase 10.1103/PhysRevD.101.034018} {\bibfield  {journal} {\bibinfo
  {journal} {Phys. Rev. D}\ }\textbf {\bibinfo {volume} {101}},\ \bibinfo
  {pages} {034018} (\bibinfo {year} {2020})},\ \Eprint
  {http://arxiv.org/abs/1912.12592} {arXiv:1912.12592 [hep-ph]} \BibitemShut
  {NoStop}%
\bibitem [{\citenamefont {Aaij}\ \emph {et~al.}(2021)\citenamefont {Aaij} \emph
  {et~al.}}]{LHCb:2020jpq}%
  \BibitemOpen
  \bibfield  {author} {\bibinfo {author} {\bibfnamefont {R.}~\bibnamefont
  {Aaij}} \emph {et~al.} (\bibinfo {collaboration} {LHCb}),\ }\href {\doibase
  10.1016/j.scib.2021.02.030} {\bibfield  {journal} {\bibinfo  {journal} {Sci.
  Bull.}\ }\textbf {\bibinfo {volume} {66}},\ \bibinfo {pages} {1278} (\bibinfo
  {year} {2021})},\ \Eprint {http://arxiv.org/abs/2012.10380} {arXiv:2012.10380
  [hep-ex]} \BibitemShut {NoStop}%
\bibitem [{LHC(2022)}]{LHCb:2022jad}%
  \BibitemOpen
  \href@noop {} {\  (\bibinfo {year} {2022})},\ \Eprint
  {http://arxiv.org/abs/2210.10346} {arXiv:2210.10346 [hep-ex]} \BibitemShut
  {NoStop}%
\bibitem [{\citenamefont {Chen}\ \emph
  {et~al.}(2021{\natexlab{b}})\citenamefont {Chen}, \citenamefont {Wang},\ and\
  \citenamefont {Zhu}}]{Chen:2021htr}%
  \BibitemOpen
  \bibfield  {author} {\bibinfo {author} {\bibfnamefont {K.}~\bibnamefont
  {Chen}}, \bibinfo {author} {\bibfnamefont {B.}~\bibnamefont {Wang}}, \ and\
  \bibinfo {author} {\bibfnamefont {S.-L.}\ \bibnamefont {Zhu}},\ }\href
  {\doibase 10.1103/PhysRevD.103.116017} {\bibfield  {journal} {\bibinfo
  {journal} {Phys. Rev. D}\ }\textbf {\bibinfo {volume} {103}},\ \bibinfo
  {pages} {116017} (\bibinfo {year} {2021}{\natexlab{b}})},\ \Eprint
  {http://arxiv.org/abs/2102.05868} {arXiv:2102.05868 [hep-ph]} \BibitemShut
  {NoStop}%
\bibitem [{\citenamefont {Chen}\ \emph
  {et~al.}(2022{\natexlab{b}})\citenamefont {Chen}, \citenamefont {Huang},
  \citenamefont {Wang},\ and\ \citenamefont {Zhu}}]{Chen:2022iil}%
  \BibitemOpen
  \bibfield  {author} {\bibinfo {author} {\bibfnamefont {K.}~\bibnamefont
  {Chen}}, \bibinfo {author} {\bibfnamefont {B.-L.}\ \bibnamefont {Huang}},
  \bibinfo {author} {\bibfnamefont {B.}~\bibnamefont {Wang}}, \ and\ \bibinfo
  {author} {\bibfnamefont {S.-L.}\ \bibnamefont {Zhu}},\ }\href@noop {} {\
  (\bibinfo {year} {2022}{\natexlab{b}})},\ \Eprint
  {http://arxiv.org/abs/2204.13316} {arXiv:2204.13316 [hep-ph]} \BibitemShut
  {NoStop}%
\bibitem [{\citenamefont {Lyu}\ \emph {et~al.}(2023)\citenamefont {Lyu},
  \citenamefont {Aoki}, \citenamefont {Doi}, \citenamefont {Hatsuda},
  \citenamefont {Ikeda},\ and\ \citenamefont {Meng}}]{Lyu:2023xro}%
  \BibitemOpen
  \bibfield  {author} {\bibinfo {author} {\bibfnamefont {Y.}~\bibnamefont
  {Lyu}}, \bibinfo {author} {\bibfnamefont {S.}~\bibnamefont {Aoki}}, \bibinfo
  {author} {\bibfnamefont {T.}~\bibnamefont {Doi}}, \bibinfo {author}
  {\bibfnamefont {T.}~\bibnamefont {Hatsuda}}, \bibinfo {author} {\bibfnamefont
  {Y.}~\bibnamefont {Ikeda}}, \ and\ \bibinfo {author} {\bibfnamefont
  {J.}~\bibnamefont {Meng}},\ }\href@noop {} {\  (\bibinfo {year} {2023})},\
  \Eprint {http://arxiv.org/abs/2302.04505} {arXiv:2302.04505 [hep-lat]}
  \BibitemShut {NoStop}%
\bibitem [{\citenamefont {Meng}\ \emph {et~al.}(2021)\citenamefont {Meng},
  \citenamefont {Wang}, \citenamefont {Wang},\ and\ \citenamefont
  {Zhu}}]{Meng:2021jnw}%
  \BibitemOpen
  \bibfield  {author} {\bibinfo {author} {\bibfnamefont {L.}~\bibnamefont
  {Meng}}, \bibinfo {author} {\bibfnamefont {G.-J.}\ \bibnamefont {Wang}},
  \bibinfo {author} {\bibfnamefont {B.}~\bibnamefont {Wang}}, \ and\ \bibinfo
  {author} {\bibfnamefont {S.-L.}\ \bibnamefont {Zhu}},\ }\href {\doibase
  10.1103/PhysRevD.104.L051502} {\bibfield  {journal} {\bibinfo  {journal}
  {Phys. Rev. D}\ }\textbf {\bibinfo {volume} {104}},\ \bibinfo {pages}
  {051502} (\bibinfo {year} {2021})},\ \Eprint
  {http://arxiv.org/abs/2107.14784} {arXiv:2107.14784 [hep-ph]} \BibitemShut
  {NoStop}%
\bibitem [{\citenamefont {Agaev}\ \emph {et~al.}(2022)\citenamefont {Agaev},
  \citenamefont {Azizi},\ and\ \citenamefont {Sundu}}]{Agaev:2021vur}%
  \BibitemOpen
  \bibfield  {author} {\bibinfo {author} {\bibfnamefont {S.~S.}\ \bibnamefont
  {Agaev}}, \bibinfo {author} {\bibfnamefont {K.}~\bibnamefont {Azizi}}, \ and\
  \bibinfo {author} {\bibfnamefont {H.}~\bibnamefont {Sundu}},\ }\href
  {\doibase 10.1016/j.nuclphysb.2022.115650} {\bibfield  {journal} {\bibinfo
  {journal} {Nucl. Phys. B}\ }\textbf {\bibinfo {volume} {975}},\ \bibinfo
  {pages} {115650} (\bibinfo {year} {2022})},\ \Eprint
  {http://arxiv.org/abs/2108.00188} {arXiv:2108.00188 [hep-ph]} \BibitemShut
  {NoStop}%
\bibitem [{\citenamefont {Ling}\ \emph {et~al.}(2022)\citenamefont {Ling},
  \citenamefont {Liu}, \citenamefont {Geng}, \citenamefont {Wang},\ and\
  \citenamefont {Xie}}]{Ling:2021bir}%
  \BibitemOpen
  \bibfield  {author} {\bibinfo {author} {\bibfnamefont {X.-Z.}\ \bibnamefont
  {Ling}}, \bibinfo {author} {\bibfnamefont {M.-Z.}\ \bibnamefont {Liu}},
  \bibinfo {author} {\bibfnamefont {L.-S.}\ \bibnamefont {Geng}}, \bibinfo
  {author} {\bibfnamefont {E.}~\bibnamefont {Wang}}, \ and\ \bibinfo {author}
  {\bibfnamefont {J.-J.}\ \bibnamefont {Xie}},\ }\href {\doibase
  10.1016/j.physletb.2022.136897} {\bibfield  {journal} {\bibinfo  {journal}
  {Phys. Lett. B}\ }\textbf {\bibinfo {volume} {826}},\ \bibinfo {pages}
  {136897} (\bibinfo {year} {2022})},\ \Eprint
  {http://arxiv.org/abs/2108.00947} {arXiv:2108.00947 [hep-ph]} \BibitemShut
  {NoStop}%
\bibitem [{\citenamefont {Yan}\ and\ \citenamefont
  {Valderrama}(2022)}]{Yan:2021wdl}%
  \BibitemOpen
  \bibfield  {author} {\bibinfo {author} {\bibfnamefont {M.-J.}\ \bibnamefont
  {Yan}}\ and\ \bibinfo {author} {\bibfnamefont {M.~P.}\ \bibnamefont
  {Valderrama}},\ }\href {\doibase 10.1103/PhysRevD.105.014007} {\bibfield
  {journal} {\bibinfo  {journal} {Phys. Rev. D}\ }\textbf {\bibinfo {volume}
  {105}},\ \bibinfo {pages} {014007} (\bibinfo {year} {2022})},\ \Eprint
  {http://arxiv.org/abs/2108.04785} {arXiv:2108.04785 [hep-ph]} \BibitemShut
  {NoStop}%
\bibitem [{\citenamefont {Feijoo}\ \emph {et~al.}(2021)\citenamefont {Feijoo},
  \citenamefont {Liang},\ and\ \citenamefont {Oset}}]{Feijoo:2021ppq}%
  \BibitemOpen
  \bibfield  {author} {\bibinfo {author} {\bibfnamefont {A.}~\bibnamefont
  {Feijoo}}, \bibinfo {author} {\bibfnamefont {W.~H.}\ \bibnamefont {Liang}}, \
  and\ \bibinfo {author} {\bibfnamefont {E.}~\bibnamefont {Oset}},\ }\href
  {\doibase 10.1103/PhysRevD.104.114015} {\bibfield  {journal} {\bibinfo
  {journal} {Phys. Rev. D}\ }\textbf {\bibinfo {volume} {104}},\ \bibinfo
  {pages} {114015} (\bibinfo {year} {2021})},\ \Eprint
  {http://arxiv.org/abs/2108.02730} {arXiv:2108.02730 [hep-ph]} \BibitemShut
  {NoStop}%
\bibitem [{\citenamefont {Ren}\ \emph {et~al.}(2022)\citenamefont {Ren},
  \citenamefont {Wu},\ and\ \citenamefont {Zhu}}]{Ren:2021dsi}%
  \BibitemOpen
  \bibfield  {author} {\bibinfo {author} {\bibfnamefont {H.}~\bibnamefont
  {Ren}}, \bibinfo {author} {\bibfnamefont {F.}~\bibnamefont {Wu}}, \ and\
  \bibinfo {author} {\bibfnamefont {R.}~\bibnamefont {Zhu}},\ }\href {\doibase
  10.1155/2022/9103031} {\bibfield  {journal} {\bibinfo  {journal} {Adv. High
  Energy Phys.}\ }\textbf {\bibinfo {volume} {2022}},\ \bibinfo {pages}
  {9103031} (\bibinfo {year} {2022})},\ \Eprint
  {http://arxiv.org/abs/2109.02531} {arXiv:2109.02531 [hep-ph]} \BibitemShut
  {NoStop}%
\bibitem [{\citenamefont {Dong}\ \emph {et~al.}(2021)\citenamefont {Dong},
  \citenamefont {Guo},\ and\ \citenamefont {Zou}}]{Dong:2021bvy}%
  \BibitemOpen
  \bibfield  {author} {\bibinfo {author} {\bibfnamefont {X.-K.}\ \bibnamefont
  {Dong}}, \bibinfo {author} {\bibfnamefont {F.-K.}\ \bibnamefont {Guo}}, \
  and\ \bibinfo {author} {\bibfnamefont {B.-S.}\ \bibnamefont {Zou}},\ }\href
  {\doibase 10.1088/1572-9494/ac27a2} {\bibfield  {journal} {\bibinfo
  {journal} {Commun. Theor. Phys.}\ }\textbf {\bibinfo {volume} {73}},\
  \bibinfo {pages} {125201} (\bibinfo {year} {2021})},\ \Eprint
  {http://arxiv.org/abs/2108.02673} {arXiv:2108.02673 [hep-ph]} \BibitemShut
  {NoStop}%
\bibitem [{\citenamefont {Chen}\ \emph
  {et~al.}(2021{\natexlab{c}})\citenamefont {Chen}, \citenamefont {Huang},
  \citenamefont {Liu},\ and\ \citenamefont {Zhu}}]{Chen:2021vhg}%
  \BibitemOpen
  \bibfield  {author} {\bibinfo {author} {\bibfnamefont {R.}~\bibnamefont
  {Chen}}, \bibinfo {author} {\bibfnamefont {Q.}~\bibnamefont {Huang}},
  \bibinfo {author} {\bibfnamefont {X.}~\bibnamefont {Liu}}, \ and\ \bibinfo
  {author} {\bibfnamefont {S.-L.}\ \bibnamefont {Zhu}},\ }\href {\doibase
  10.1103/PhysRevD.104.114042} {\bibfield  {journal} {\bibinfo  {journal}
  {Phys. Rev. D}\ }\textbf {\bibinfo {volume} {104}},\ \bibinfo {pages}
  {114042} (\bibinfo {year} {2021}{\natexlab{c}})},\ \Eprint
  {http://arxiv.org/abs/2108.01911} {arXiv:2108.01911 [hep-ph]} \BibitemShut
  {NoStop}%
\bibitem [{\citenamefont {Weng}\ \emph {et~al.}(2022)\citenamefont {Weng},
  \citenamefont {Deng},\ and\ \citenamefont {Zhu}}]{Weng:2021hje}%
  \BibitemOpen
  \bibfield  {author} {\bibinfo {author} {\bibfnamefont {X.-Z.}\ \bibnamefont
  {Weng}}, \bibinfo {author} {\bibfnamefont {W.-Z.}\ \bibnamefont {Deng}}, \
  and\ \bibinfo {author} {\bibfnamefont {S.-L.}\ \bibnamefont {Zhu}},\ }\href
  {\doibase 10.1088/1674-1137/ac2ed0} {\bibfield  {journal} {\bibinfo
  {journal} {Chin. Phys. C}\ }\textbf {\bibinfo {volume} {46}},\ \bibinfo
  {pages} {013102} (\bibinfo {year} {2022})},\ \Eprint
  {http://arxiv.org/abs/2108.07242} {arXiv:2108.07242 [hep-ph]} \BibitemShut
  {NoStop}%
\bibitem [{\citenamefont {Xin}\ and\ \citenamefont {Wang}(2022)}]{Xin:2021wcr}%
  \BibitemOpen
  \bibfield  {author} {\bibinfo {author} {\bibfnamefont {Q.}~\bibnamefont
  {Xin}}\ and\ \bibinfo {author} {\bibfnamefont {Z.-G.}\ \bibnamefont {Wang}},\
  }\href {\doibase 10.1140/epja/s10050-022-00752-4} {\bibfield  {journal}
  {\bibinfo  {journal} {Eur. Phys. J. A}\ }\textbf {\bibinfo {volume} {58}},\
  \bibinfo {pages} {110} (\bibinfo {year} {2022})},\ \Eprint
  {http://arxiv.org/abs/2108.12597} {arXiv:2108.12597 [hep-ph]} \BibitemShut
  {NoStop}%
\bibitem [{\citenamefont {Chen}\ and\ \citenamefont
  {Yang}(2022)}]{Chen:2021tnn}%
  \BibitemOpen
  \bibfield  {author} {\bibinfo {author} {\bibfnamefont {X.}~\bibnamefont
  {Chen}}\ and\ \bibinfo {author} {\bibfnamefont {Y.}~\bibnamefont {Yang}},\
  }\href {\doibase 10.1088/1674-1137/ac4ee8} {\bibfield  {journal} {\bibinfo
  {journal} {Chin. Phys. C}\ }\textbf {\bibinfo {volume} {46}},\ \bibinfo
  {pages} {054103} (\bibinfo {year} {2022})},\ \Eprint
  {http://arxiv.org/abs/2109.02828} {arXiv:2109.02828 [hep-ph]} \BibitemShut
  {NoStop}%
\bibitem [{\citenamefont {Chen}\ \emph
  {et~al.}(2022{\natexlab{c}})\citenamefont {Chen}, \citenamefont {Chen},
  \citenamefont {Meng}, \citenamefont {Wang},\ and\ \citenamefont
  {Zhu}}]{Chen:2021cfl}%
  \BibitemOpen
  \bibfield  {author} {\bibinfo {author} {\bibfnamefont {K.}~\bibnamefont
  {Chen}}, \bibinfo {author} {\bibfnamefont {R.}~\bibnamefont {Chen}}, \bibinfo
  {author} {\bibfnamefont {L.}~\bibnamefont {Meng}}, \bibinfo {author}
  {\bibfnamefont {B.}~\bibnamefont {Wang}}, \ and\ \bibinfo {author}
  {\bibfnamefont {S.-L.}\ \bibnamefont {Zhu}},\ }\href {\doibase
  10.1140/epjc/s10052-022-10540-5} {\bibfield  {journal} {\bibinfo  {journal}
  {Eur. Phys. J. C}\ }\textbf {\bibinfo {volume} {82}},\ \bibinfo {pages} {581}
  (\bibinfo {year} {2022}{\natexlab{c}})},\ \Eprint
  {http://arxiv.org/abs/2109.13057} {arXiv:2109.13057 [hep-ph]} \BibitemShut
  {NoStop}%
\bibitem [{\citenamefont {Deng}\ and\ \citenamefont
  {Zhu}(2022)}]{Deng:2021gnb}%
  \BibitemOpen
  \bibfield  {author} {\bibinfo {author} {\bibfnamefont {C.}~\bibnamefont
  {Deng}}\ and\ \bibinfo {author} {\bibfnamefont {S.-L.}\ \bibnamefont {Zhu}},\
  }\href {\doibase 10.1103/PhysRevD.105.054015} {\bibfield  {journal} {\bibinfo
   {journal} {Phys. Rev. D}\ }\textbf {\bibinfo {volume} {105}},\ \bibinfo
  {pages} {054015} (\bibinfo {year} {2022})},\ \Eprint
  {http://arxiv.org/abs/2112.12472} {arXiv:2112.12472 [hep-ph]} \BibitemShut
  {NoStop}%
\bibitem [{\citenamefont {Ke}\ \emph {et~al.}(2022)\citenamefont {Ke},
  \citenamefont {Liu},\ and\ \citenamefont {Li}}]{Ke:2021rxd}%
  \BibitemOpen
  \bibfield  {author} {\bibinfo {author} {\bibfnamefont {H.-W.}\ \bibnamefont
  {Ke}}, \bibinfo {author} {\bibfnamefont {X.-H.}\ \bibnamefont {Liu}}, \ and\
  \bibinfo {author} {\bibfnamefont {X.-Q.}\ \bibnamefont {Li}},\ }\href
  {\doibase 10.1140/epjc/s10052-022-10092-8} {\bibfield  {journal} {\bibinfo
  {journal} {Eur. Phys. J. C}\ }\textbf {\bibinfo {volume} {82}},\ \bibinfo
  {pages} {144} (\bibinfo {year} {2022})},\ \Eprint
  {http://arxiv.org/abs/2112.14142} {arXiv:2112.14142 [hep-ph]} \BibitemShut
  {NoStop}%
\bibitem [{\citenamefont {Padmanath}\ and\ \citenamefont
  {Prelovsek}(2022)}]{Padmanath:2022cvl}%
  \BibitemOpen
  \bibfield  {author} {\bibinfo {author} {\bibfnamefont {M.}~\bibnamefont
  {Padmanath}}\ and\ \bibinfo {author} {\bibfnamefont {S.}~\bibnamefont
  {Prelovsek}},\ }\href {\doibase 10.1103/PhysRevLett.129.032002} {\bibfield
  {journal} {\bibinfo  {journal} {Phys. Rev. Lett.}\ }\textbf {\bibinfo
  {volume} {129}},\ \bibinfo {pages} {032002} (\bibinfo {year} {2022})},\
  \Eprint {http://arxiv.org/abs/2202.10110} {arXiv:2202.10110 [hep-lat]}
  \BibitemShut {NoStop}%
\bibitem [{\citenamefont {Lin}\ \emph {et~al.}(2022)\citenamefont {Lin},
  \citenamefont {Cheng},\ and\ \citenamefont {Zhu}}]{Lin:2022wmj}%
  \BibitemOpen
  \bibfield  {author} {\bibinfo {author} {\bibfnamefont {Z.-Y.}\ \bibnamefont
  {Lin}}, \bibinfo {author} {\bibfnamefont {J.-B.}\ \bibnamefont {Cheng}}, \
  and\ \bibinfo {author} {\bibfnamefont {S.-L.}\ \bibnamefont {Zhu}},\
  }\href@noop {} {\  (\bibinfo {year} {2022})},\ \Eprint
  {http://arxiv.org/abs/2205.14628} {arXiv:2205.14628 [hep-ph]} \BibitemShut
  {NoStop}%
\bibitem [{\citenamefont {Kim}\ \emph {et~al.}(2022)\citenamefont {Kim},
  \citenamefont {Oka},\ and\ \citenamefont {Suzuki}}]{Kim:2022mpa}%
  \BibitemOpen
  \bibfield  {author} {\bibinfo {author} {\bibfnamefont {Y.}~\bibnamefont
  {Kim}}, \bibinfo {author} {\bibfnamefont {M.}~\bibnamefont {Oka}}, \ and\
  \bibinfo {author} {\bibfnamefont {K.}~\bibnamefont {Suzuki}},\ }\href
  {\doibase 10.1103/PhysRevD.105.074021} {\bibfield  {journal} {\bibinfo
  {journal} {Phys. Rev. D}\ }\textbf {\bibinfo {volume} {105}},\ \bibinfo
  {pages} {074021} (\bibinfo {year} {2022})},\ \Eprint
  {http://arxiv.org/abs/2202.06520} {arXiv:2202.06520 [hep-ph]} \BibitemShut
  {NoStop}%
\bibitem [{\citenamefont {Cheng}\ \emph {et~al.}(2022)\citenamefont {Cheng},
  \citenamefont {Lin},\ and\ \citenamefont {Zhu}}]{Cheng:2022qcm}%
  \BibitemOpen
  \bibfield  {author} {\bibinfo {author} {\bibfnamefont {J.-B.}\ \bibnamefont
  {Cheng}}, \bibinfo {author} {\bibfnamefont {Z.-Y.}\ \bibnamefont {Lin}}, \
  and\ \bibinfo {author} {\bibfnamefont {S.-L.}\ \bibnamefont {Zhu}},\ }\href
  {\doibase 10.1103/PhysRevD.106.016012} {\bibfield  {journal} {\bibinfo
  {journal} {Phys. Rev. D}\ }\textbf {\bibinfo {volume} {106}},\ \bibinfo
  {pages} {016012} (\bibinfo {year} {2022})},\ \Eprint
  {http://arxiv.org/abs/2205.13354} {arXiv:2205.13354 [hep-ph]} \BibitemShut
  {NoStop}%
\bibitem [{\citenamefont {Qin}\ \emph {et~al.}(2021)\citenamefont {Qin},
  \citenamefont {Shen},\ and\ \citenamefont {Yu}}]{Qin:2020zlg}%
  \BibitemOpen
  \bibfield  {author} {\bibinfo {author} {\bibfnamefont {Q.}~\bibnamefont
  {Qin}}, \bibinfo {author} {\bibfnamefont {Y.-F.}\ \bibnamefont {Shen}}, \
  and\ \bibinfo {author} {\bibfnamefont {F.-S.}\ \bibnamefont {Yu}},\ }\href
  {\doibase 10.1088/1674-1137/ac1b97} {\bibfield  {journal} {\bibinfo
  {journal} {Chin. Phys. C}\ }\textbf {\bibinfo {volume} {45}},\ \bibinfo
  {pages} {103106} (\bibinfo {year} {2021})},\ \Eprint
  {http://arxiv.org/abs/2008.08026} {arXiv:2008.08026 [hep-ph]} \BibitemShut
  {NoStop}%
\bibitem [{\citenamefont {Huang}\ \emph {et~al.}(2021)\citenamefont {Huang},
  \citenamefont {Zhu}, \citenamefont {Geng},\ and\ \citenamefont
  {Wang}}]{Huang:2021urd}%
  \BibitemOpen
  \bibfield  {author} {\bibinfo {author} {\bibfnamefont {Y.}~\bibnamefont
  {Huang}}, \bibinfo {author} {\bibfnamefont {H.~Q.}\ \bibnamefont {Zhu}},
  \bibinfo {author} {\bibfnamefont {L.-S.}\ \bibnamefont {Geng}}, \ and\
  \bibinfo {author} {\bibfnamefont {R.}~\bibnamefont {Wang}},\ }\href {\doibase
  10.1103/PhysRevD.104.116008} {\bibfield  {journal} {\bibinfo  {journal}
  {Phys. Rev. D}\ }\textbf {\bibinfo {volume} {104}},\ \bibinfo {pages}
  {116008} (\bibinfo {year} {2021})},\ \Eprint
  {http://arxiv.org/abs/2108.13028} {arXiv:2108.13028 [hep-ph]} \BibitemShut
  {NoStop}%
\bibitem [{\citenamefont {Jin}\ \emph {et~al.}(2021)\citenamefont {Jin},
  \citenamefont {Li}, \citenamefont {Liu}, \citenamefont {Qin}, \citenamefont
  {Si},\ and\ \citenamefont {Yu}}]{Jin:2021cxj}%
  \BibitemOpen
  \bibfield  {author} {\bibinfo {author} {\bibfnamefont {Y.}~\bibnamefont
  {Jin}}, \bibinfo {author} {\bibfnamefont {S.-Y.}\ \bibnamefont {Li}},
  \bibinfo {author} {\bibfnamefont {Y.-R.}\ \bibnamefont {Liu}}, \bibinfo
  {author} {\bibfnamefont {Q.}~\bibnamefont {Qin}}, \bibinfo {author}
  {\bibfnamefont {Z.-G.}\ \bibnamefont {Si}}, \ and\ \bibinfo {author}
  {\bibfnamefont {F.-S.}\ \bibnamefont {Yu}},\ }\href {\doibase
  10.1103/PhysRevD.104.114009} {\bibfield  {journal} {\bibinfo  {journal}
  {Phys. Rev. D}\ }\textbf {\bibinfo {volume} {104}},\ \bibinfo {pages}
  {114009} (\bibinfo {year} {2021})},\ \Eprint
  {http://arxiv.org/abs/2109.05678} {arXiv:2109.05678 [hep-ph]} \BibitemShut
  {NoStop}%
\bibitem [{\citenamefont {Hu}\ \emph {et~al.}(2021)\citenamefont {Hu},
  \citenamefont {Liao}, \citenamefont {Wang}, \citenamefont {Wang},
  \citenamefont {Xing},\ and\ \citenamefont {Zhang}}]{Hu:2021gdg}%
  \BibitemOpen
  \bibfield  {author} {\bibinfo {author} {\bibfnamefont {Y.}~\bibnamefont
  {Hu}}, \bibinfo {author} {\bibfnamefont {J.}~\bibnamefont {Liao}}, \bibinfo
  {author} {\bibfnamefont {E.}~\bibnamefont {Wang}}, \bibinfo {author}
  {\bibfnamefont {Q.}~\bibnamefont {Wang}}, \bibinfo {author} {\bibfnamefont
  {H.}~\bibnamefont {Xing}}, \ and\ \bibinfo {author} {\bibfnamefont
  {H.}~\bibnamefont {Zhang}},\ }\href {\doibase 10.1103/PhysRevD.104.L111502}
  {\bibfield  {journal} {\bibinfo  {journal} {Phys. Rev. D}\ }\textbf {\bibinfo
  {volume} {104}},\ \bibinfo {pages} {L111502} (\bibinfo {year} {2021})},\
  \Eprint {http://arxiv.org/abs/2109.07733} {arXiv:2109.07733 [hep-ph]}
  \BibitemShut {NoStop}%
\bibitem [{\citenamefont {Abreu}\ \emph {et~al.}(2022)\citenamefont {Abreu},
  \citenamefont {Navarra},\ and\ \citenamefont {Vieira}}]{Abreu:2022lfy}%
  \BibitemOpen
  \bibfield  {author} {\bibinfo {author} {\bibfnamefont {L.~M.}\ \bibnamefont
  {Abreu}}, \bibinfo {author} {\bibfnamefont {F.~S.}\ \bibnamefont {Navarra}},
  \ and\ \bibinfo {author} {\bibfnamefont {H.~P.~L.}\ \bibnamefont {Vieira}},\
  }\href {\doibase 10.1103/PhysRevD.105.116029} {\bibfield  {journal} {\bibinfo
   {journal} {Phys. Rev. D}\ }\textbf {\bibinfo {volume} {105}},\ \bibinfo
  {pages} {116029} (\bibinfo {year} {2022})},\ \Eprint
  {http://arxiv.org/abs/2202.10882} {arXiv:2202.10882 [hep-ph]} \BibitemShut
  {NoStop}%
\bibitem [{\citenamefont {Braaten}\ \emph {et~al.}(2022)\citenamefont
  {Braaten}, \citenamefont {He}, \citenamefont {Ingles},\ and\ \citenamefont
  {Jiang}}]{Braaten:2022elw}%
  \BibitemOpen
  \bibfield  {author} {\bibinfo {author} {\bibfnamefont {E.}~\bibnamefont
  {Braaten}}, \bibinfo {author} {\bibfnamefont {L.-P.}\ \bibnamefont {He}},
  \bibinfo {author} {\bibfnamefont {K.}~\bibnamefont {Ingles}}, \ and\ \bibinfo
  {author} {\bibfnamefont {J.}~\bibnamefont {Jiang}},\ }\href@noop {} {\
  (\bibinfo {year} {2022})},\ \Eprint {http://arxiv.org/abs/2202.03900}
  {arXiv:2202.03900 [hep-ph]} \BibitemShut {NoStop}%
\bibitem [{\citenamefont {Dai}\ \emph {et~al.}(2022)\citenamefont {Dai},
  \citenamefont {Sun}, \citenamefont {Kang}, \citenamefont {Szczepaniak},\ and\
  \citenamefont {Yu}}]{Dai:2021wxi}%
  \BibitemOpen
  \bibfield  {author} {\bibinfo {author} {\bibfnamefont {L.-Y.}\ \bibnamefont
  {Dai}}, \bibinfo {author} {\bibfnamefont {X.}~\bibnamefont {Sun}}, \bibinfo
  {author} {\bibfnamefont {X.-W.}\ \bibnamefont {Kang}}, \bibinfo {author}
  {\bibfnamefont {A.~P.}\ \bibnamefont {Szczepaniak}}, \ and\ \bibinfo {author}
  {\bibfnamefont {J.-S.}\ \bibnamefont {Yu}},\ }\href {\doibase
  10.1103/PhysRevD.105.L051507} {\bibfield  {journal} {\bibinfo  {journal}
  {Phys. Rev. D}\ }\textbf {\bibinfo {volume} {105}},\ \bibinfo {pages}
  {L051507} (\bibinfo {year} {2022})},\ \Eprint
  {http://arxiv.org/abs/2108.06002} {arXiv:2108.06002 [hep-ph]} \BibitemShut
  {NoStop}%
\bibitem [{\citenamefont {Fleming}\ \emph {et~al.}(2021)\citenamefont
  {Fleming}, \citenamefont {Hodges},\ and\ \citenamefont
  {Mehen}}]{Fleming:2021wmk}%
  \BibitemOpen
  \bibfield  {author} {\bibinfo {author} {\bibfnamefont {S.}~\bibnamefont
  {Fleming}}, \bibinfo {author} {\bibfnamefont {R.}~\bibnamefont {Hodges}}, \
  and\ \bibinfo {author} {\bibfnamefont {T.}~\bibnamefont {Mehen}},\ }\href
  {\doibase 10.1103/PhysRevD.104.116010} {\bibfield  {journal} {\bibinfo
  {journal} {Phys. Rev. D}\ }\textbf {\bibinfo {volume} {104}},\ \bibinfo
  {pages} {116010} (\bibinfo {year} {2021})},\ \Eprint
  {http://arxiv.org/abs/2109.02188} {arXiv:2109.02188 [hep-ph]} \BibitemShut
  {NoStop}%
\bibitem [{\citenamefont {Du}\ \emph {et~al.}(2022)\citenamefont {Du},
  \citenamefont {Baru}, \citenamefont {Dong}, \citenamefont {Filin},
  \citenamefont {Guo}, \citenamefont {Hanhart}, \citenamefont {Nefediev},
  \citenamefont {Nieves},\ and\ \citenamefont {Wang}}]{Du:2021zzh}%
  \BibitemOpen
  \bibfield  {author} {\bibinfo {author} {\bibfnamefont {M.-L.}\ \bibnamefont
  {Du}}, \bibinfo {author} {\bibfnamefont {V.}~\bibnamefont {Baru}}, \bibinfo
  {author} {\bibfnamefont {X.-K.}\ \bibnamefont {Dong}}, \bibinfo {author}
  {\bibfnamefont {A.}~\bibnamefont {Filin}}, \bibinfo {author} {\bibfnamefont
  {F.-K.}\ \bibnamefont {Guo}}, \bibinfo {author} {\bibfnamefont
  {C.}~\bibnamefont {Hanhart}}, \bibinfo {author} {\bibfnamefont
  {A.}~\bibnamefont {Nefediev}}, \bibinfo {author} {\bibfnamefont
  {J.}~\bibnamefont {Nieves}}, \ and\ \bibinfo {author} {\bibfnamefont
  {Q.}~\bibnamefont {Wang}},\ }\href {\doibase 10.1103/PhysRevD.105.014024}
  {\bibfield  {journal} {\bibinfo  {journal} {Phys. Rev. D}\ }\textbf {\bibinfo
  {volume} {105}},\ \bibinfo {pages} {014024} (\bibinfo {year} {2022})},\
  \Eprint {http://arxiv.org/abs/2110.13765} {arXiv:2110.13765 [hep-ph]}
  \BibitemShut {NoStop}%
\bibitem [{\citenamefont {Azizi}\ and\ \citenamefont
  {\"Ozdem}(2021)}]{Azizi:2021aib}%
  \BibitemOpen
  \bibfield  {author} {\bibinfo {author} {\bibfnamefont {K.}~\bibnamefont
  {Azizi}}\ and\ \bibinfo {author} {\bibfnamefont {U.}~\bibnamefont
  {\"Ozdem}},\ }\href {\doibase 10.1103/PhysRevD.104.114002} {\bibfield
  {journal} {\bibinfo  {journal} {Phys. Rev. D}\ }\textbf {\bibinfo {volume}
  {104}},\ \bibinfo {pages} {114002} (\bibinfo {year} {2021})},\ \Eprint
  {http://arxiv.org/abs/2109.02390} {arXiv:2109.02390 [hep-ph]} \BibitemShut
  {NoStop}%
\bibitem [{\citenamefont {Machleidt}\ \emph {et~al.}(1987)\citenamefont
  {Machleidt}, \citenamefont {Holinde},\ and\ \citenamefont
  {Elster}}]{Machleidt:1987hj}%
  \BibitemOpen
  \bibfield  {author} {\bibinfo {author} {\bibfnamefont {R.}~\bibnamefont
  {Machleidt}}, \bibinfo {author} {\bibfnamefont {K.}~\bibnamefont {Holinde}},
  \ and\ \bibinfo {author} {\bibfnamefont {C.}~\bibnamefont {Elster}},\ }\href
  {\doibase 10.1016/S0370-1573(87)80002-9} {\bibfield  {journal} {\bibinfo
  {journal} {Phys. Rept.}\ }\textbf {\bibinfo {volume} {149}},\ \bibinfo
  {pages} {1} (\bibinfo {year} {1987})}\BibitemShut {NoStop}%
\bibitem [{\citenamefont {Ecker}\ \emph {et~al.}(1989)\citenamefont {Ecker},
  \citenamefont {Gasser}, \citenamefont {Pich},\ and\ \citenamefont
  {de~Rafael}}]{Ecker:1988te}%
  \BibitemOpen
  \bibfield  {author} {\bibinfo {author} {\bibfnamefont {G.}~\bibnamefont
  {Ecker}}, \bibinfo {author} {\bibfnamefont {J.}~\bibnamefont {Gasser}},
  \bibinfo {author} {\bibfnamefont {A.}~\bibnamefont {Pich}}, \ and\ \bibinfo
  {author} {\bibfnamefont {E.}~\bibnamefont {de~Rafael}},\ }\href {\doibase
  10.1016/0550-3213(89)90346-5} {\bibfield  {journal} {\bibinfo  {journal}
  {Nucl. Phys. B}\ }\textbf {\bibinfo {volume} {321}},\ \bibinfo {pages} {311}
  (\bibinfo {year} {1989})}\BibitemShut {NoStop}%
\bibitem [{\citenamefont {Epelbaum}\ \emph
  {et~al.}(2002{\natexlab{b}})\citenamefont {Epelbaum}, \citenamefont
  {Meissner}, \citenamefont {Gloeckle},\ and\ \citenamefont
  {Elster}}]{Epelbaum:2001fm}%
  \BibitemOpen
  \bibfield  {author} {\bibinfo {author} {\bibfnamefont {E.}~\bibnamefont
  {Epelbaum}}, \bibinfo {author} {\bibfnamefont {U.~G.}\ \bibnamefont
  {Meissner}}, \bibinfo {author} {\bibfnamefont {W.}~\bibnamefont {Gloeckle}},
  \ and\ \bibinfo {author} {\bibfnamefont {C.}~\bibnamefont {Elster}},\ }\href
  {\doibase 10.1103/PhysRevC.65.044001} {\bibfield  {journal} {\bibinfo
  {journal} {Phys. Rev. C}\ }\textbf {\bibinfo {volume} {65}},\ \bibinfo
  {pages} {044001} (\bibinfo {year} {2002}{\natexlab{b}})},\ \Eprint
  {http://arxiv.org/abs/nucl-th/0106007} {arXiv:nucl-th/0106007} \BibitemShut
  {NoStop}%
\bibitem [{\citenamefont {Du}\ \emph {et~al.}(2016)\citenamefont {Du},
  \citenamefont {Guo}, \citenamefont {Mei\ss{}ner},\ and\ \citenamefont
  {Yao}}]{Du:2016tgp}%
  \BibitemOpen
  \bibfield  {author} {\bibinfo {author} {\bibfnamefont {M.-L.}\ \bibnamefont
  {Du}}, \bibinfo {author} {\bibfnamefont {F.-K.}\ \bibnamefont {Guo}},
  \bibinfo {author} {\bibfnamefont {U.-G.}\ \bibnamefont {Mei\ss{}ner}}, \ and\
  \bibinfo {author} {\bibfnamefont {D.-L.}\ \bibnamefont {Yao}},\ }\href
  {\doibase 10.1103/PhysRevD.94.094037} {\bibfield  {journal} {\bibinfo
  {journal} {Phys. Rev. D}\ }\textbf {\bibinfo {volume} {94}},\ \bibinfo
  {pages} {094037} (\bibinfo {year} {2016})},\ \Eprint
  {http://arxiv.org/abs/1610.02963} {arXiv:1610.02963 [hep-ph]} \BibitemShut
  {NoStop}%
\bibitem [{\citenamefont {Xu}(2022)}]{Xu:2021vsi}%
  \BibitemOpen
  \bibfield  {author} {\bibinfo {author} {\bibfnamefont {H.}~\bibnamefont
  {Xu}},\ }\href {\doibase 10.1103/PhysRevD.105.034013} {\bibfield  {journal}
  {\bibinfo  {journal} {Phys. Rev. D}\ }\textbf {\bibinfo {volume} {105}},\
  \bibinfo {pages} {034013} (\bibinfo {year} {2022})},\ \Eprint
  {http://arxiv.org/abs/2112.10722} {arXiv:2112.10722 [hep-ph]} \BibitemShut
  {NoStop}%
\bibitem [{\citenamefont {Peng}\ \emph {et~al.}(2022)\citenamefont {Peng},
  \citenamefont {S\'anchez~S\'anchez}, \citenamefont {Yan},\ and\ \citenamefont
  {Pavon~Valderrama}}]{Peng:2021hkr}%
  \BibitemOpen
  \bibfield  {author} {\bibinfo {author} {\bibfnamefont {F.-Z.}\ \bibnamefont
  {Peng}}, \bibinfo {author} {\bibfnamefont {M.}~\bibnamefont
  {S\'anchez~S\'anchez}}, \bibinfo {author} {\bibfnamefont {M.-J.}\
  \bibnamefont {Yan}}, \ and\ \bibinfo {author} {\bibfnamefont
  {M.}~\bibnamefont {Pavon~Valderrama}},\ }\href {\doibase
  10.1103/PhysRevD.105.034028} {\bibfield  {journal} {\bibinfo  {journal}
  {Phys. Rev. D}\ }\textbf {\bibinfo {volume} {105}},\ \bibinfo {pages}
  {034028} (\bibinfo {year} {2022})},\ \Eprint
  {http://arxiv.org/abs/2101.07213} {arXiv:2101.07213 [hep-ph]} \BibitemShut
  {NoStop}%
\bibitem [{\citenamefont {Wise}(1992)}]{Wise:1992hn}%
  \BibitemOpen
  \bibfield  {author} {\bibinfo {author} {\bibfnamefont {M.~B.}\ \bibnamefont
  {Wise}},\ }\href {\doibase 10.1103/PhysRevD.45.R2188} {\bibfield  {journal}
  {\bibinfo  {journal} {Phys. Rev. D}\ }\textbf {\bibinfo {volume} {45}},\
  \bibinfo {pages} {R2188} (\bibinfo {year} {1992})}\BibitemShut {NoStop}%
\bibitem [{\citenamefont {Manohar}\ and\ \citenamefont
  {Wise}(2000)}]{Manohar:2000dt}%
  \BibitemOpen
  \bibfield  {author} {\bibinfo {author} {\bibfnamefont {A.~V.}\ \bibnamefont
  {Manohar}}\ and\ \bibinfo {author} {\bibfnamefont {M.~B.}\ \bibnamefont
  {Wise}},\ }\href@noop {} {\emph {\bibinfo {title} {{Heavy quark physics}}}},\
  Vol.~\bibinfo {volume} {10}\ (\bibinfo {year} {2000})\BibitemShut {NoStop}%
\bibitem [{\citenamefont {Ericson}\ and\ \citenamefont
  {Weise}(1988)}]{Ericson:1988gk}%
  \BibitemOpen
  \bibfield  {author} {\bibinfo {author} {\bibfnamefont {T.~E.~O.}\
  \bibnamefont {Ericson}}\ and\ \bibinfo {author} {\bibfnamefont
  {W.}~\bibnamefont {Weise}},\ }\href@noop {} {\emph {\bibinfo {title} {{Pions
  and Nuclei}}}}\ (\bibinfo  {publisher} {Clarendon Press},\ \bibinfo {address}
  {Oxford, UK},\ \bibinfo {year} {1988})\BibitemShut {NoStop}%
\bibitem [{\citenamefont {Epelbaum}\ \emph {et~al.}(2022)\citenamefont
  {Epelbaum}, \citenamefont {Krebs},\ and\ \citenamefont
  {Reinert}}]{Epelbaum:2022cyo}%
  \BibitemOpen
  \bibfield  {author} {\bibinfo {author} {\bibfnamefont {E.}~\bibnamefont
  {Epelbaum}}, \bibinfo {author} {\bibfnamefont {H.}~\bibnamefont {Krebs}}, \
  and\ \bibinfo {author} {\bibfnamefont {P.}~\bibnamefont {Reinert}},\
  }\href@noop {} {\  (\bibinfo {year} {2022})},\ \Eprint
  {http://arxiv.org/abs/2206.07072} {arXiv:2206.07072 [nucl-th]} \BibitemShut
  {NoStop}%
\bibitem [{\citenamefont {Bernard}\ \emph {et~al.}(1995)\citenamefont
  {Bernard}, \citenamefont {Kaiser},\ and\ \citenamefont
  {Mei\ss~ner}}]{Bernard:1995dp}%
  \BibitemOpen
  \bibfield  {author} {\bibinfo {author} {\bibfnamefont {V.}~\bibnamefont
  {Bernard}}, \bibinfo {author} {\bibfnamefont {N.}~\bibnamefont {Kaiser}}, \
  and\ \bibinfo {author} {\bibfnamefont {U.-G.}\ \bibnamefont {Mei\ss~ner}},\
  }\href {\doibase 10.1142/S0218301395000092} {\bibfield  {journal} {\bibinfo
  {journal} {Int. J. Mod. Phys. E}\ }\textbf {\bibinfo {volume} {4}},\ \bibinfo
  {pages} {193} (\bibinfo {year} {1995})},\ \Eprint
  {http://arxiv.org/abs/hep-ph/9501384} {arXiv:hep-ph/9501384} \BibitemShut
  {NoStop}%
\bibitem [{\citenamefont {Lyu}\ \emph {et~al.}(2022)\citenamefont {Lyu},
  \citenamefont {Doi}, \citenamefont {Hatsuda}, \citenamefont {Ikeda},
  \citenamefont {Meng}, \citenamefont {Sasaki},\ and\ \citenamefont
  {Sugiura}}]{Lyu:2022imf}%
  \BibitemOpen
  \bibfield  {author} {\bibinfo {author} {\bibfnamefont {Y.}~\bibnamefont
  {Lyu}}, \bibinfo {author} {\bibfnamefont {T.}~\bibnamefont {Doi}}, \bibinfo
  {author} {\bibfnamefont {T.}~\bibnamefont {Hatsuda}}, \bibinfo {author}
  {\bibfnamefont {Y.}~\bibnamefont {Ikeda}}, \bibinfo {author} {\bibfnamefont
  {J.}~\bibnamefont {Meng}}, \bibinfo {author} {\bibfnamefont {K.}~\bibnamefont
  {Sasaki}}, \ and\ \bibinfo {author} {\bibfnamefont {T.}~\bibnamefont
  {Sugiura}},\ }\href {\doibase 10.1103/PhysRevD.106.074507} {\bibfield
  {journal} {\bibinfo  {journal} {Phys. Rev. D}\ }\textbf {\bibinfo {volume}
  {106}},\ \bibinfo {pages} {074507} (\bibinfo {year} {2022})},\ \Eprint
  {http://arxiv.org/abs/2205.10544} {arXiv:2205.10544 [hep-lat]} \BibitemShut
  {NoStop}%
\bibitem [{\citenamefont {Fleming}\ \emph {et~al.}(2007)\citenamefont
  {Fleming}, \citenamefont {Kusunoki}, \citenamefont {Mehen},\ and\
  \citenamefont {van Kolck}}]{Fleming:2007rp}%
  \BibitemOpen
  \bibfield  {author} {\bibinfo {author} {\bibfnamefont {S.}~\bibnamefont
  {Fleming}}, \bibinfo {author} {\bibfnamefont {M.}~\bibnamefont {Kusunoki}},
  \bibinfo {author} {\bibfnamefont {T.}~\bibnamefont {Mehen}}, \ and\ \bibinfo
  {author} {\bibfnamefont {U.}~\bibnamefont {van Kolck}},\ }\href {\doibase
  10.1103/PhysRevD.76.034006} {\bibfield  {journal} {\bibinfo  {journal} {Phys.
  Rev. D}\ }\textbf {\bibinfo {volume} {76}},\ \bibinfo {pages} {034006}
  (\bibinfo {year} {2007})},\ \Eprint {http://arxiv.org/abs/hep-ph/0703168}
  {arXiv:hep-ph/0703168} \BibitemShut {NoStop}%
\bibitem [{\citenamefont {Golak}\ \emph {et~al.}(2010)\citenamefont {Golak}
  \emph {et~al.}}]{Golak:2009ri}%
  \BibitemOpen
  \bibfield  {author} {\bibinfo {author} {\bibfnamefont {J.}~\bibnamefont
  {Golak}} \emph {et~al.},\ }\href {\doibase 10.1140/epja/i2009-10903-6}
  {\bibfield  {journal} {\bibinfo  {journal} {Eur. Phys. J. A}\ }\textbf
  {\bibinfo {volume} {43}},\ \bibinfo {pages} {241} (\bibinfo {year} {2010})},\
  \Eprint {http://arxiv.org/abs/0911.4173} {arXiv:0911.4173 [nucl-th]}
  \BibitemShut {NoStop}%
\bibitem [{\citenamefont {Workman}\ and\ \citenamefont
  {Others}(2022)}]{Workman:2022ynf}%
  \BibitemOpen
  \bibfield  {author} {\bibinfo {author} {\bibfnamefont {R.~L.}\ \bibnamefont
  {Workman}}\ and\ \bibinfo {author} {\bibnamefont {Others}} (\bibinfo
  {collaboration} {Particle Data Group}),\ }\href {\doibase
  10.1093/ptep/ptac097} {\bibfield  {journal} {\bibinfo  {journal} {PTEP}\
  }\textbf {\bibinfo {volume} {2022}},\ \bibinfo {pages} {083C01} (\bibinfo
  {year} {2022})}\BibitemShut {NoStop}%
\bibitem [{\citenamefont {Bardeen}\ \emph {et~al.}(2003)\citenamefont
  {Bardeen}, \citenamefont {Eichten},\ and\ \citenamefont
  {Hill}}]{Bardeen:2003kt}%
  \BibitemOpen
  \bibfield  {author} {\bibinfo {author} {\bibfnamefont {W.~A.}\ \bibnamefont
  {Bardeen}}, \bibinfo {author} {\bibfnamefont {E.~J.}\ \bibnamefont
  {Eichten}}, \ and\ \bibinfo {author} {\bibfnamefont {C.~T.}\ \bibnamefont
  {Hill}},\ }\href {\doibase 10.1103/PhysRevD.68.054024} {\bibfield  {journal}
  {\bibinfo  {journal} {Phys. Rev. D}\ }\textbf {\bibinfo {volume} {68}},\
  \bibinfo {pages} {054024} (\bibinfo {year} {2003})},\ \Eprint
  {http://arxiv.org/abs/hep-ph/0305049} {arXiv:hep-ph/0305049} \BibitemShut
  {NoStop}%
\bibitem [{\citenamefont {Moir}\ \emph {et~al.}(2016)\citenamefont {Moir},
  \citenamefont {Peardon}, \citenamefont {Ryan}, \citenamefont {Thomas},\ and\
  \citenamefont {Wilson}}]{Moir:2016srx}%
  \BibitemOpen
  \bibfield  {author} {\bibinfo {author} {\bibfnamefont {G.}~\bibnamefont
  {Moir}}, \bibinfo {author} {\bibfnamefont {M.}~\bibnamefont {Peardon}},
  \bibinfo {author} {\bibfnamefont {S.~M.}\ \bibnamefont {Ryan}}, \bibinfo
  {author} {\bibfnamefont {C.~E.}\ \bibnamefont {Thomas}}, \ and\ \bibinfo
  {author} {\bibfnamefont {D.~J.}\ \bibnamefont {Wilson}},\ }\href {\doibase
  10.1007/JHEP10(2016)011} {\bibfield  {journal} {\bibinfo  {journal} {JHEP}\
  }\textbf {\bibinfo {volume} {10}},\ \bibinfo {pages} {011} (\bibinfo {year}
  {2016})},\ \Eprint {http://arxiv.org/abs/1607.07093} {arXiv:1607.07093
  [hep-lat]} \BibitemShut {NoStop}%
\bibitem [{\citenamefont {Du}\ \emph {et~al.}(2021)\citenamefont {Du},
  \citenamefont {Guo}, \citenamefont {Hanhart}, \citenamefont {Kubis},\ and\
  \citenamefont {Mei\ss{}ner}}]{Du:2020pui}%
  \BibitemOpen
  \bibfield  {author} {\bibinfo {author} {\bibfnamefont {M.-L.}\ \bibnamefont
  {Du}}, \bibinfo {author} {\bibfnamefont {F.-K.}\ \bibnamefont {Guo}},
  \bibinfo {author} {\bibfnamefont {C.}~\bibnamefont {Hanhart}}, \bibinfo
  {author} {\bibfnamefont {B.}~\bibnamefont {Kubis}}, \ and\ \bibinfo {author}
  {\bibfnamefont {U.-G.}\ \bibnamefont {Mei\ss{}ner}},\ }\href {\doibase
  10.1103/PhysRevLett.126.192001} {\bibfield  {journal} {\bibinfo  {journal}
  {Phys. Rev. Lett.}\ }\textbf {\bibinfo {volume} {126}},\ \bibinfo {pages}
  {192001} (\bibinfo {year} {2021})},\ \Eprint
  {http://arxiv.org/abs/2012.04599} {arXiv:2012.04599 [hep-ph]} \BibitemShut
  {NoStop}%
\bibitem [{\citenamefont {Gayer}\ \emph {et~al.}(2021)\citenamefont {Gayer},
  \citenamefont {Lang}, \citenamefont {Ryan}, \citenamefont {Tims},
  \citenamefont {Thomas},\ and\ \citenamefont {Wilson}}]{Gayer:2021xzv}%
  \BibitemOpen
  \bibfield  {author} {\bibinfo {author} {\bibfnamefont {L.}~\bibnamefont
  {Gayer}}, \bibinfo {author} {\bibfnamefont {N.}~\bibnamefont {Lang}},
  \bibinfo {author} {\bibfnamefont {S.~M.}\ \bibnamefont {Ryan}}, \bibinfo
  {author} {\bibfnamefont {D.}~\bibnamefont {Tims}}, \bibinfo {author}
  {\bibfnamefont {C.~E.}\ \bibnamefont {Thomas}}, \ and\ \bibinfo {author}
  {\bibfnamefont {D.~J.}\ \bibnamefont {Wilson}} (\bibinfo {collaboration}
  {Hadron Spectrum}),\ }\href {\doibase 10.1007/JHEP07(2021)123} {\bibfield
  {journal} {\bibinfo  {journal} {JHEP}\ }\textbf {\bibinfo {volume} {07}},\
  \bibinfo {pages} {123} (\bibinfo {year} {2021})},\ \Eprint
  {http://arxiv.org/abs/2102.04973} {arXiv:2102.04973 [hep-lat]} \BibitemShut
  {NoStop}%
\bibitem [{\citenamefont {Caprini}\ \emph {et~al.}(2006)\citenamefont
  {Caprini}, \citenamefont {Colangelo},\ and\ \citenamefont
  {Leutwyler}}]{Caprini:2005zr}%
  \BibitemOpen
  \bibfield  {author} {\bibinfo {author} {\bibfnamefont {I.}~\bibnamefont
  {Caprini}}, \bibinfo {author} {\bibfnamefont {G.}~\bibnamefont {Colangelo}},
  \ and\ \bibinfo {author} {\bibfnamefont {H.}~\bibnamefont {Leutwyler}},\
  }\href {\doibase 10.1103/PhysRevLett.96.132001} {\bibfield  {journal}
  {\bibinfo  {journal} {Phys. Rev. Lett.}\ }\textbf {\bibinfo {volume} {96}},\
  \bibinfo {pages} {132001} (\bibinfo {year} {2006})},\ \Eprint
  {http://arxiv.org/abs/hep-ph/0512364} {arXiv:hep-ph/0512364} \BibitemShut
  {NoStop}%
\bibitem [{\citenamefont {Dai}\ and\ \citenamefont
  {Pennington}(2014)}]{Dai:2014zta}%
  \BibitemOpen
  \bibfield  {author} {\bibinfo {author} {\bibfnamefont {L.-Y.}\ \bibnamefont
  {Dai}}\ and\ \bibinfo {author} {\bibfnamefont {M.~R.}\ \bibnamefont
  {Pennington}},\ }\href {\doibase 10.1103/PhysRevD.90.036004} {\bibfield
  {journal} {\bibinfo  {journal} {Phys. Rev. D}\ }\textbf {\bibinfo {volume}
  {90}},\ \bibinfo {pages} {036004} (\bibinfo {year} {2014})},\ \Eprint
  {http://arxiv.org/abs/1404.7524} {arXiv:1404.7524 [hep-ph]} \BibitemShut
  {NoStop}%
\bibitem [{\citenamefont {Yndurain}\ \emph {et~al.}(2007)\citenamefont
  {Yndurain}, \citenamefont {Garcia-Martin},\ and\ \citenamefont
  {Pelaez}}]{Yndurain:2007qm}%
  \BibitemOpen
  \bibfield  {author} {\bibinfo {author} {\bibfnamefont {F.~J.}\ \bibnamefont
  {Yndurain}}, \bibinfo {author} {\bibfnamefont {R.}~\bibnamefont
  {Garcia-Martin}}, \ and\ \bibinfo {author} {\bibfnamefont {J.~R.}\
  \bibnamefont {Pelaez}},\ }\href {\doibase 10.1103/PhysRevD.76.074034}
  {\bibfield  {journal} {\bibinfo  {journal} {Phys. Rev. D}\ }\textbf {\bibinfo
  {volume} {76}},\ \bibinfo {pages} {074034} (\bibinfo {year} {2007})},\
  \Eprint {http://arxiv.org/abs/hep-ph/0701025} {arXiv:hep-ph/0701025}
  \BibitemShut {NoStop}%
\bibitem [{\citenamefont {Mennessier}\ \emph {et~al.}(2008)\citenamefont
  {Mennessier}, \citenamefont {Narison},\ and\ \citenamefont
  {Ochs}}]{Mennessier:2008kk}%
  \BibitemOpen
  \bibfield  {author} {\bibinfo {author} {\bibfnamefont {G.}~\bibnamefont
  {Mennessier}}, \bibinfo {author} {\bibfnamefont {S.}~\bibnamefont {Narison}},
  \ and\ \bibinfo {author} {\bibfnamefont {W.}~\bibnamefont {Ochs}},\ }\href
  {\doibase 10.1016/j.physletb.2008.06.018} {\bibfield  {journal} {\bibinfo
  {journal} {Phys. Lett. B}\ }\textbf {\bibinfo {volume} {665}},\ \bibinfo
  {pages} {205} (\bibinfo {year} {2008})},\ \Eprint
  {http://arxiv.org/abs/0804.4452} {arXiv:0804.4452 [hep-ph]} \BibitemShut
  {NoStop}%
\bibitem [{\citenamefont {Mennessier}\ \emph {et~al.}(2010)\citenamefont
  {Mennessier}, \citenamefont {Narison},\ and\ \citenamefont
  {Wang}}]{Mennessier:2010xg}%
  \BibitemOpen
  \bibfield  {author} {\bibinfo {author} {\bibfnamefont {G.}~\bibnamefont
  {Mennessier}}, \bibinfo {author} {\bibfnamefont {S.}~\bibnamefont {Narison}},
  \ and\ \bibinfo {author} {\bibfnamefont {X.~G.}\ \bibnamefont {Wang}},\
  }\href {\doibase 10.1016/j.physletb.2010.03.031} {\bibfield  {journal}
  {\bibinfo  {journal} {Phys. Lett. B}\ }\textbf {\bibinfo {volume} {688}},\
  \bibinfo {pages} {59} (\bibinfo {year} {2010})},\ \Eprint
  {http://arxiv.org/abs/1002.1402} {arXiv:1002.1402 [hep-ph]} \BibitemShut
  {NoStop}%
\bibitem [{\citenamefont {Casalbuoni}\ \emph {et~al.}(1997)\citenamefont
  {Casalbuoni}, \citenamefont {Deandrea}, \citenamefont {Di~Bartolomeo},
  \citenamefont {Gatto}, \citenamefont {Feruglio},\ and\ \citenamefont
  {Nardulli}}]{Casalbuoni:1996pg}%
  \BibitemOpen
  \bibfield  {author} {\bibinfo {author} {\bibfnamefont {R.}~\bibnamefont
  {Casalbuoni}}, \bibinfo {author} {\bibfnamefont {A.}~\bibnamefont
  {Deandrea}}, \bibinfo {author} {\bibfnamefont {N.}~\bibnamefont
  {Di~Bartolomeo}}, \bibinfo {author} {\bibfnamefont {R.}~\bibnamefont
  {Gatto}}, \bibinfo {author} {\bibfnamefont {F.}~\bibnamefont {Feruglio}}, \
  and\ \bibinfo {author} {\bibfnamefont {G.}~\bibnamefont {Nardulli}},\ }\href
  {\doibase 10.1016/S0370-1573(96)00027-0} {\bibfield  {journal} {\bibinfo
  {journal} {Phys. Rept.}\ }\textbf {\bibinfo {volume} {281}},\ \bibinfo
  {pages} {145} (\bibinfo {year} {1997})},\ \Eprint
  {http://arxiv.org/abs/hep-ph/9605342} {arXiv:hep-ph/9605342} \BibitemShut
  {NoStop}%
\bibitem [{\citenamefont {Casalbuoni}\ \emph {et~al.}(1992)\citenamefont
  {Casalbuoni}, \citenamefont {Deandrea}, \citenamefont {Di~Bartolomeo},
  \citenamefont {Gatto}, \citenamefont {Feruglio},\ and\ \citenamefont
  {Nardulli}}]{Casalbuoni:1992gi}%
  \BibitemOpen
  \bibfield  {author} {\bibinfo {author} {\bibfnamefont {R.}~\bibnamefont
  {Casalbuoni}}, \bibinfo {author} {\bibfnamefont {A.}~\bibnamefont
  {Deandrea}}, \bibinfo {author} {\bibfnamefont {N.}~\bibnamefont
  {Di~Bartolomeo}}, \bibinfo {author} {\bibfnamefont {R.}~\bibnamefont
  {Gatto}}, \bibinfo {author} {\bibfnamefont {F.}~\bibnamefont {Feruglio}}, \
  and\ \bibinfo {author} {\bibfnamefont {G.}~\bibnamefont {Nardulli}},\ }\href
  {\doibase 10.1016/0370-2693(92)91189-G} {\bibfield  {journal} {\bibinfo
  {journal} {Phys. Lett. B}\ }\textbf {\bibinfo {volume} {292}},\ \bibinfo
  {pages} {371} (\bibinfo {year} {1992})},\ \Eprint
  {http://arxiv.org/abs/hep-ph/9209248} {arXiv:hep-ph/9209248} \BibitemShut
  {NoStop}%
\bibitem [{\citenamefont {Yan}\ \emph {et~al.}(2021)\citenamefont {Yan},
  \citenamefont {Peng}, \citenamefont {S\'anchez~S\'anchez},\ and\
  \citenamefont {Pavon~Valderrama}}]{Yan:2021tcp}%
  \BibitemOpen
  \bibfield  {author} {\bibinfo {author} {\bibfnamefont {M.-J.}\ \bibnamefont
  {Yan}}, \bibinfo {author} {\bibfnamefont {F.-Z.}\ \bibnamefont {Peng}},
  \bibinfo {author} {\bibfnamefont {M.}~\bibnamefont {S\'anchez~S\'anchez}}, \
  and\ \bibinfo {author} {\bibfnamefont {M.}~\bibnamefont {Pavon~Valderrama}},\
  }\href {\doibase 10.1103/PhysRevD.104.114025} {\bibfield  {journal} {\bibinfo
   {journal} {Phys. Rev. D}\ }\textbf {\bibinfo {volume} {104}},\ \bibinfo
  {pages} {114025} (\bibinfo {year} {2021})},\ \Eprint
  {http://arxiv.org/abs/2102.13058} {arXiv:2102.13058 [hep-ph]} \BibitemShut
  {NoStop}%
\bibitem [{\citenamefont {Guo}\ and\ \citenamefont
  {Sanz-Cillero}(2009)}]{Guo:2009hi}%
  \BibitemOpen
  \bibfield  {author} {\bibinfo {author} {\bibfnamefont {Z.-H.}\ \bibnamefont
  {Guo}}\ and\ \bibinfo {author} {\bibfnamefont {J.~J.}\ \bibnamefont
  {Sanz-Cillero}},\ }\href {\doibase 10.1103/PhysRevD.79.096006} {\bibfield
  {journal} {\bibinfo  {journal} {Phys. Rev. D}\ }\textbf {\bibinfo {volume}
  {79}},\ \bibinfo {pages} {096006} (\bibinfo {year} {2009})},\ \Eprint
  {http://arxiv.org/abs/0903.0782} {arXiv:0903.0782 [hep-ph]} \BibitemShut
  {NoStop}%
\end{thebibliography}%
\end{document}